\documentclass[a4paper,fleqn]{cas-sc}

\definecolor{cite_color}{rgb}{0.0, 0.58, 0.71}
\usepackage{natbib}
\setcitestyle{authoryear,open={},close={}} 
\usepackage{hyperref}
\hypersetup{colorlinks=true,citecolor=cite_color,linkcolor=cite_color}
\usepackage{amsthm}
\usepackage{adjustbox} 
\usepackage{array} 
\usepackage{geometry} 
\usepackage{subcaption}
\usepackage{lineno}
\usepackage{booktabs}
\usepackage[linesnumbered,ruled,vlined]{algorithm2e}
\usepackage{graphicx}
\usepackage{threeparttable}
\usepackage{pifont}  
\usepackage{amssymb} 
\usepackage{makecell}
\usepackage{multirow}
\usepackage{float}
\usepackage{longtable}
\usepackage{lscape}
\usepackage{adjustbox}
\usepackage{pifont}
\usepackage{array}
\usepackage[most]{tcolorbox}
\usepackage{tabularx}
\usepackage{threeparttable}
\usepackage[T1]{fontenc}    
\usepackage{tikz}
\definecolor{db}{rgb}{0.0, 0.2, 0.7}
\newcolumntype{Z}{>{\raggedright\arraybackslash}X}
\renewcommand\arraystretch{1.4}
\renewcommand{\figurename}{Fig.}

\makeatletter
\renewcommand*{\fnum@figure}[1]{\figurename~\thefigure.}
\makeatother
\def\tsc#1{\csdef{#1}{\textsc{\lowercase{#1}}\xspace}}
\tsc{WGM}
\tsc{QE}
\tsc{EP}
\tsc{PMS}
\tsc{BEC}
\tsc{DE}

\begin{document}
\let\WriteBookmarks\relax
\def\floatpagepagefraction{1}
\let\printorcid\relax 

\def\textpagefraction{.001}
\shorttitle{}
\shortauthors{Mohammad Anis et~al.}

\title [mode = title]{Learning from geometry-aware near misses to real-time COR: A corridor-wide grouped random parameters GEV framework}

\author[1]{\textcolor{black}{Mohammad Anis}}[]

\credit{Conceptualization, Methodology, Writing – original draft,  Software, Writing – review \& editing}

\address[1]{Zachry Department of Civil $\&$ Environmental Engineering, Texas A$\&$M University, College Station, TX 77843, USA}

\author%
[1]
{\textcolor{black}{Yang Zhou}}
\cormark[1]
\ead{yangzhou295@tamu.edu}
\credit{Conceptualization, Methodology, Writing – review \& editing, Supervision}

\author%
[1]
{\textcolor{black}{Dominique Lord}}
\credit{Conceptualization, Methodology, Writing – review \& editing, Supervision}

\cortext[cor1]{Corresponding author}

\begin{abstract}
Real-time, corridor-wide prediction of crash-occurrence risk (COR) is challenging because existing near-miss EVT models oversimplify collision geometry, neglect vehicle–infrastructure (V–I) interactions, and fail to adequately account for spatial heterogeneity in traffic and roadway conditions. To address these gaps, this study develops a geometry-aware 2D-TTC near-miss extraction and integrates it with a hierarchical Bayesian structure grouped random parameters (HBSGRP–UGEV) to estimate short-term COR in urban corridors. Building on prior grouped EVT formulations while explicitly accommodating both V–V and V–I near-miss processes within a unified corridor-wide modeling framework. High-resolution trajectories from the Argoverse-2 dataset were analyzed across 28 sites on Miami’s Biscayne Boulevard to extract extreme near-miss events. The model incorporates vehicle dynamics and roadway features as covariates, with partial pooling across segments and intersections to capture corridor-wide heterogeneity. Results show that the HBSGRP–UGEV framework outperforms fixed-parameter models, reducing DIC by up to 7.5\% (V–V) and 3.1\% (V–I). Predictive validation using ROC–AUC confirms strong accuracy (0.89 for V–V segments, 0.82 for intersections, 0.79 for V–I segments, and 0.75 for intersections). Grouped random-parameter results indicate that relative (speed, distance, and deceleration) dominate V–V near-miss risk on segments, whereas V–I segment risk is primarily associated with relative distance; at intersections, V–V risk is driven by relative (speed and distance), while V–I dynamics exhibit no statistically significant effects. These findings demonstrate the value of a geometry-aware, spatially adaptive framework for proactive corridor safety management, supporting both real-time interventions and long-term Vision Zero goals.

\end{abstract}

\begin{keywords}
Crash occurrence risk \sep
Corridor-wide safety \sep
Near-miss \sep 
Geometry-aware 2D-TTC \sep
Hierarchical Bayesian modeling \sep
Grouped random parameters \sep
\end{keywords}

\maketitle

\section{Introduction}

The increasing complexity of urban corridors, rising traffic volumes, and the coexistence of human-driven vehicles (HDVs) and autonomous vehicles (AVs) require innovative safety assessment approaches. Urban corridors are particularly vulnerable, as even a single crash can disrupt mobility, trigger cascading congestion, and result in substantial economic losses. In 2023, traffic crashes in the US claimed 40901 lives, with a fatality rate of 1.26 per 100 million vehicle miles traveled (VMT)  (\citep{NCSA2025}). The total annual cost of crashes exceeds \$1.85 trillion, including \$460 billion in direct economic losses and \$1.4 trillion in reduced quality of life (\citep{Blincoe2023}). Despite decades of research, corridor-wide crash risk prediction remains methodologically limited.

Crash risk prediction has traditionally relied on retrospective crash data (\citep{arun2021systematic,lee2017intersection,pei2011joint}). Although such approaches have advanced understanding of crash patterns, they face well-documented limitations (\citep{lord2010statistical, lord2021highway, mannering2020big}), including the rarity and stochastic nature of crashes, underreporting (\citep{arun2021systematic, tarko2018estimating}), temporal delays, and spatial mismatches with real-time traffic conditions. These drawbacks highlight the need for proactive approaches that anticipate rather than react to crash risk.

To address these challenges, researchers have shifted toward proactive trajectory-based surrogate safety measures (SSMs) (\citep{ali2023assessing,arun2021systematic,mahmud2017application}). SSMs use frequent near-miss events as proxies for crashes, enabling the study of traffic conflict dynamics at a much higher temporal resolution (e.g., \citep{li2024beyond, tarko2018estimating, davis2011outline}). Among these, time-based indicators such as Time-to-Collision (TTC), Post-Encroachment Time (PET), and Modified TTC (MTTC) are widely used (\citep{hayward1971near, perkins1967traffic, minderhoud2001extended, ozbay2008derivation, allen1978analysis, venthuruthiyil2022anticipated}). However, their assumptions of constant velocity and linear paths can bias near-miss characterization, especially in turning and boundary-constrained settings. Recent two-dimensional (2D) extensions of TTC (\citep{li2024beyond, anis2025real}) that incorporate position, heading, yaw rate, speed, and acceleration often still rely on simplified vehicle shape approximations (e.g., circular). Although \citet{hou2014new} compared alternative vehicle-shape representations (e.g., circle, rectangle, and hybrid forms) for simplified TTC computation in simulation, a corridor-scale, high-fidelity 2D-TTC that propagates geometry-aware footprints and jointly extracts V–V and V–I near-misses from trajectories remains limited.

Recent studies have integrated computed SSMs with the extreme value theory (EVT) framework to estimate crash occurrence risk (COR) in urban corridors (\citep{lanzaro2023comparison, kamel2023real, kamel2024real, kamel2024transferability, singh2024bayesian, ghoul2025cyclist, singh2025autonomous, anis2025real}). Since EVT can extrapolate from frequent near misses or SSMs to rare crash-prone extremes, it offers a statistically principled framework for proactive safety analysis. Yet most near-miss and EVT frameworks remain limited, focusing narrowly on V–V interactions and excluding V–I near-misses such as lane edges, curbs, and barriers. This omission is critical because V–I events account for a large share of severe roadway departure crashes (\citep{islam2021crash}). Recent advances in AV sensing and high-definition (HD) mapping, such as the Argoverse-2 dataset (\citep{wilson2023argoverse}), now enable precise alignment of roadway boundaries with vehicle trajectories, creating new opportunities to model both V–V and V–I near misses with geometric fidelity.

While SSMs capture frequent near misses, robust COR estimation requires extrapolating extreme near misses to rare crash outcomes. EVT (\citep{fu2021comparison,fu2021multivariate,zheng2021modeling,ali2023assessing, zheng2021validating, songchitruksa2006extreme, zheng2014freeway, tarko2012use}) has become the most widely used method for this task, with two dominant formulations: the Generalized Extreme Value (GEV) distribution using Block Maxima (BM) sampling, and the Generalized Pareto Distribution (GPD) using Peaks-over-Threshold (POT) sampling (\citep{coles2001introduction}). The BM approach is particularly well-suited to real-time applications due to its short block structure, reduced threshold sensitivity, and compatibility with discrete-time samples (\citep{fu2022bayesian}).

Early applications of EVT in traffic safety used stationary GEV models (\citep{wang2018combined, guo2019comparison, orsini2019collision, alozi2022evaluating, hussain2022hybrid}), which assumed constant parameters across the study area. Over time, nonstationary GEV formulations have been developed, allowing covariates to influence distribution parameters (\citep{tahir2024non,ali2022extreme,fu2022bayesian,zheng2019bayesian,nazir2023car,kar2024crash,ali2023estimating,alozi2022evaluating,zheng2014freeway,kar2023non}) and thereby represent variation across traffic states and locations. (\citep{lanzaro2023comparison, kamel2023real, kamel2024real, kamel2024transferability, singh2024bayesian, ghoul2025cyclist, singh2025autonomous, anis2025real}). Bayesian hierarchical extensions further address data sparsity and unobserved heterogeneity (\citep{zheng2019univariate, fu2022random}). A growing body of EVT-based safety studies now incorporates vehicle dynamics and roadway context (\citep{fu2022random, kumar2024risk, anis2025real, singh2025autonomous}) and, in some cases, distinguishes traffic-flow directions (\citep{kamel2024real}). However, direction-wise parametrization of vehicle-dynamics effects remains limited, particularly for corridor segments where interaction mechanisms can differ systematically by travel direction and access density. Moreover, most EVT applications focus on intersection settings (\citep{fu2023identification, fu2021random, fu2022random, ghoul2023dynamic, kumar2024risk}), only a small subset extends to corridor or network-scale inference (\citep{ghoul2023real, kamel2023real, kamel2024real, kamel2024transferability, singh2024bayesian, anis2025real, singh2025autonomous}).

To address these gaps, this paper makes two contributions. First, it develops a corridor-wide COR framework that incorporates both V–V and V–I interactions by implementing a geometry-aware 2D-TTC formulation that forward-projects rectangular vehicle footprints and uses HD-map boundaries to define V–I exposure, enabling joint extraction of V–V and V–I extremes from AV trajectories. Second, it adapts and extends a hierarchical Bayesian structure grouped random-parameters (HBSGRP) UGEV framework for corridor-scale modeling, integrating fixed roadway and time-varying vehicle-dynamics covariates within an interaction-window BM framework to capture multilevel heterogeneity across intersections and directional segments. Together, these contributions establish an infrastructure-aware and spatially adaptive framework for proactive corridor-wide crash-risk estimation that links high-fidelity near-miss geometry with hierarchical EVT inference, supporting both real-time intervention and longer-term safety planning.

The remainder of this paper introduces the proposed 2D-TTC indicators, the HBSGRP–UGEV modeling framework, and the data preparation used in the analysis. This is followed by the presentation of model estimation, validation, and comparative results. The paper concludes with key findings, limitations, and directions for future research.

\section{Near-miss detection framework}\label{2}

This section presents the computational framework for estimating a geometry-aware, high-fidelity 2D-TTC to support real-time COR inference. The proposed workflow follows projection-based TTC concepts that forward-simulate oriented rectangular footprints, but is implemented directly on high-frequency AV trajectories and HD-map boundary primitives. The framework integrates a continuous-time kinematic bicycle model, explicit vehicle footprint geometry, and a spatial proximity algorithm to capture both V–V and V–I interactions. Vehicles are represented as oriented bounding boxes, and V–I exposure is defined using the curb/median boundaries of the vehicle’s current carriageway extracted from the HD map, enabling consistent joint detection of near misses along the corridor. Together, these components provide a deterministic, high-resolution basis for extracting extreme near-miss severity used in the subsequent EVT-based COR model.

\subsection{Vehicle dynamics model}

The vehicle dynamics module provides the predictive foundation of the framework. Which adopts the kinematic bicycle model for short-horizon interaction forecasting, where steering and acceleration dominate, and lateral tire slip can be neglected. The motion is represented by a four-dimensional (4D) dynamic state vector. At any time \(t\), the state of vehicle \(i\) is given by Eq.~(\ref{eq:1}):

\begin{align}
\mathbf{x}_i(t)= \begin{bmatrix} \label{eq:1}
x_i(t) \\[1mm]
y_i(t) \\[1mm]
v_i(t)
\end{bmatrix}
\end{align}

Here, \( x_i(t), y_i(t) \) denote the vehicle’s global center coordinates, \( \theta_i(t) \) its heading angle, and \( v_i(t) \) its speed. Over the prediction window, each vehicle is assumed to maintain constant acceleration \( a_i \) and steering angle \( \delta_i \), simplifying short-horizon trajectory integration. The motion then follows the standard kinematic bicycle Eqs.~(\ref{eq:2}-\ref{eq:5}):

\begin{align}
\dot{x}_i(t) &= v_i(t)\cos\theta_i(t) \label{eq:2}\\[1mm]
\dot{y}_i(t) &= v_i(t)\sin\theta_i(t) \label{eq:3}\\[1mm]
\dot{\theta}_i(t) &= \frac{v_i(t)}{L_i}\tan\delta_i \label{eq:4}\\[1mm]
\dot{v}_i(t) &= a_i \label{eq:5}
\end{align}

Here, \(L_i\) is the vehicle wheelbase, influencing the turning radius and path curvature of vehicle \(i\). The steering angle  \(\delta_i\) and acceleration \(a_i\) are treated as fixed inputs over the lookahead period. Eqs. (\ref{eq:2}-\ref{eq:5}) describe the rates of change in position, orientation, and velocity, determining the vehicle’s trajectory. To model interactions, individual states are concatenated into an eight-dimensional (8D) joint state vector, defined as:

\begin{align}\label{eq:6}
X(t) = \begin{bmatrix} x_A(t) & y_A(t) & \theta_A(t) & v_A(t) & x_B(t) & y_B(t) & \theta_B(t) & v_B(t) \end{bmatrix}^\top
\end{align}

The vector field that governs the evolution of this combined state is \( {F} \) defined as:

\begin{align}\label{eq:7}
\dot{X}(t) = F\bigl(X(t)\bigr) = \begin{bmatrix} v_A\cos\theta_A & v_A\sin\theta_A  & \frac{v_A}{L_A}\tan\delta_A & a_A & v_B\cos\theta_B & v_B\sin\theta_B & \frac{v_B}{L_B}\tan\delta_B & a_B \end{bmatrix}^\top
\end{align}

The state-space vector field in Eq.~(\ref{eq:7}) gives the instantaneous rate of change of the joint state \( \dot{X}(t) \). Its forward evolution is computed using the fourth-order Runge–Kutta (RK4) method, which balances accuracy and efficiency for nonlinear vehicle dynamics. The time horizon is discretized into \( N \) steps of size \( \Delta t \) (\( t_n = n \Delta t \), \( n = 0, 1, \dots, N \)), and the RK4 update for state \( {X}(t_{n+1}) \) from \({X}(t_n) \) is:

\begin{align} \label{eq:8}
X_{n+1} = X_n + \frac{\Delta t}{6} \left( k_1 + 2k_2 + 2k_3 + k_4 \right)
\end{align}

where the intermediate slope evaluations \( {k}_1, \dots, {k}_4 \in \mathbb{R}^8 \) are defined as:

\begin{align} \label{eq:9}
k_1 &= F(X_n) \notag \\
k_2 &= F\left(X_n + \frac{\Delta t}{2} k_1 \right) \notag \\
k_3 &= F\left(X_n + \frac{\Delta t}{2} k_2 \right) \notag \\
k_4 &= F\left(X_n + \Delta t \cdot k_3 \right)
\end{align}

This integration is repeated for each timestep \( n \in \{0, \dots, N-1\} \), producing a sequence of predicted states \( \{ {X}_n \} \) over the horizon. These predicted states are then converted to oriented footprints for geometric proximity checks against other vehicles and HD-map boundaries.

\subsection{Vehicle geometry representation}

Although the dynamic model predicts vehicle positions and headings, accurate near-miss detection requires a realistic representation of the vehicle footprint. Simplifying vehicles as points or circles often misrepresents interactions, especially during lane changes, swerving, or angled approaches. To address this, each vehicle is modeled as a rigid rectangle defined by its actual length $L_i^{\text{veh}}$ and width $W_i^{\text{veh}}$  (vehicle $i \in \{A, B\}$). The vehicle is centered at its center of gravity, with four corners specified in the body-fixed coordinate system as:

\begin{align}\label{eq:10}
{b}_i^{(j)} =
\begin{bmatrix}
\xi_i^{(j)} \\
\eta_i^{(j)}
\end{bmatrix}, \quad j = 1, 2, 3, 4
\end{align}

where 
\( \xi_i^{(j)} \in \left\{ +\frac{L_i^{\text{veh}}}{2},\ -\frac{L_i^{\text{veh}}}{2} \right\} \)
and
\( \eta_i^{(j)} \in \left\{ +\frac{W_i^{\text{veh}}}{2},\ -\frac{W_i^{\text{veh}}}{2} \right\} \). 
This results in four corners: front-left, front-right, rear-left, and rear-right, expressed relative to the vehicle’s center in the local frame.

To compute the position of each corner in the global coordinate frame at time \( t \), the body-fixed corner coordinates are first rotated by the vehicle’s heading angle \( \theta_i(t) \), and then translated by the vehicle’s center position \( (x_i(t), y_i(t)) \). This transformation is written as:

\begin{align}\label{eq:11}
{p}_i^{(j)}(t) = 
\begin{bmatrix}
x_i(t) \\
y_i(t)
\end{bmatrix}
+
R(\theta_i(t)) \cdot {b}_i^{(j)}
\end{align}

where $R(\theta_i(t))$ is the 2D rotation matrix defined as:

\begin{align}\label{eq:12}
R(\theta_i(t)) =
\begin{bmatrix}
\cos \theta_i(t) & -\sin \theta_i(t) \\
\sin \theta_i(t) & \cos \theta_i(t)
\end{bmatrix}
\end{align}

This transformation rotates the local corner coordinates by the heading angle and translates them into the global frame. Repeating it for all corners at each timestep produces the vehicle’s time-varying geometric footprint, which serves as input for near-miss detection. This footprint construction is applied at every RK4 step, yielding a time-varying oriented bounding box suitable for both V–V and V–I proximity evaluation.

\subsection{V-V near-miss detection}
Vehicles are modeled as rigid rectangles, and predicted states and orientations determine their interactions. At each RK4 integration timestep, the global positions of all four corners are computed using the kinematic and rotational transformations. A V–V near-miss is detected when the predicted footprints exhibit near-contact under a numerical tolerance $\epsilon$, implemented through corner-based proximity checks across all corner pairs. Let ${p}_A^{(j)}(t)$ and ${p}_B^{(k)}(t)$ denote the global coordinates of corners 
$j$ and $k$ of vehicles A and B at time $t$. A V–V near-miss is flagged if any pair of these corners satisfies the proximity condition.

\begin{align}
{p}_A^{(j)}(t) = {p}_B^{(k)}(t)
\label{eq:13}
\end{align}

An exact match between the corners of two vehicles is highly improbable, given floating-point arithmetic and continuous motion. Instead, a practical criterion is applied: a collision is flagged if the difference in both the x and y coordinates of any corner pair is less than or equal to a small threshold $\epsilon$:

\begin{align}
|x_A^{(j)} - x_B^{(k)}| &\leq \varepsilon \label{eq:14}\\[1mm]
|y_A^{(j)} - y_B^{(k)}| &\leq \varepsilon \label{eq:15}
\end{align}

The rotation–translation transformation from each vehicle’s center and heading angle gives the global corner coordinates, for vehicles $A$ and $B$:
\begin{align}
x_A^{(j)} &= x_A + \xi_A^{(j)} \cos \theta_A - \eta_A^{(j)} \sin \theta_A \\
y_A^{(j)} &= y_A + \xi_A^{(j)} \sin \theta_A + \eta_A^{(j)} \cos \theta_A\\
x_B^{(k)} &= x_B + \xi_B^{(k)} \cos \theta_B - \eta_B^{(k)} \sin \theta_B \\
y_B^{(k)} &= y_B + \xi_B^{(k)} \sin \theta_B + \eta_B^{(k)} \cos \theta_B
\label{eq:16-19}
\end{align}

At each timestep, all 16 corner-pair combinations are evaluated. A V–V near-miss is flagged if any pair satisfies the proximity condition, and the corresponding collision time $t_c$ is recorded for 2D-TTC computation. This discrete-time footprint-based check aligns with high-frequency trajectory sampling and avoids the geometric bias introduced by circular approximations in oblique and turning interactions.

\subsection{V-I near-miss detection}

In addition to V-V interactions, the framework also detects near-misses between vehicles and roadway infrastructure, including curbs, medians, and barriers. Capturing these V–I events is critical for identifying edge encroachments or departures that often precede run-off-road crashes.

In this study, the scope is defined as the curb and median boundaries of the vehicle’s current carriageway as encoded in the HD map; opposing-direction curbs, minor side-street sidewalks, and internal lane markings are beyond the scope. To do so, curb and boundaries are represented as static polylines of $M$ discrete points $(X_\ell, Y_\ell)$ consistent with HD map formats in datasets such as Argoverse-2. At each timestep, the global positions of the four vehicle corners ${p}_i^{(j)}(t)$ are obtained from the predicted state. The distance between a vehicle corner $j$ and boundary point $\ell$, the distance at time $t$ is computed as:

\begin{align}\label{eq:20}
d_{j,\ell}(t) = \sqrt{(x^{(j)}(t) - X_\ell)^2 + (y^{(j)}(t) - Y_\ell)^2}
\end{align}

A near-miss is flagged if $d_{j,\ell}(t) \leq \varepsilon$ for any corner and boundary point, where is a proximity threshold accounting for numerical tolerances. When this occurs, the corresponding time $t_c$ is recorded as the instant of boundary contact. These events are then passed to the 2D-TTC framework to support extreme risk estimation.

\subsection{COR estimation}

\begin{figure}[h]
    \centering
    \includegraphics[width=0.7\textwidth]{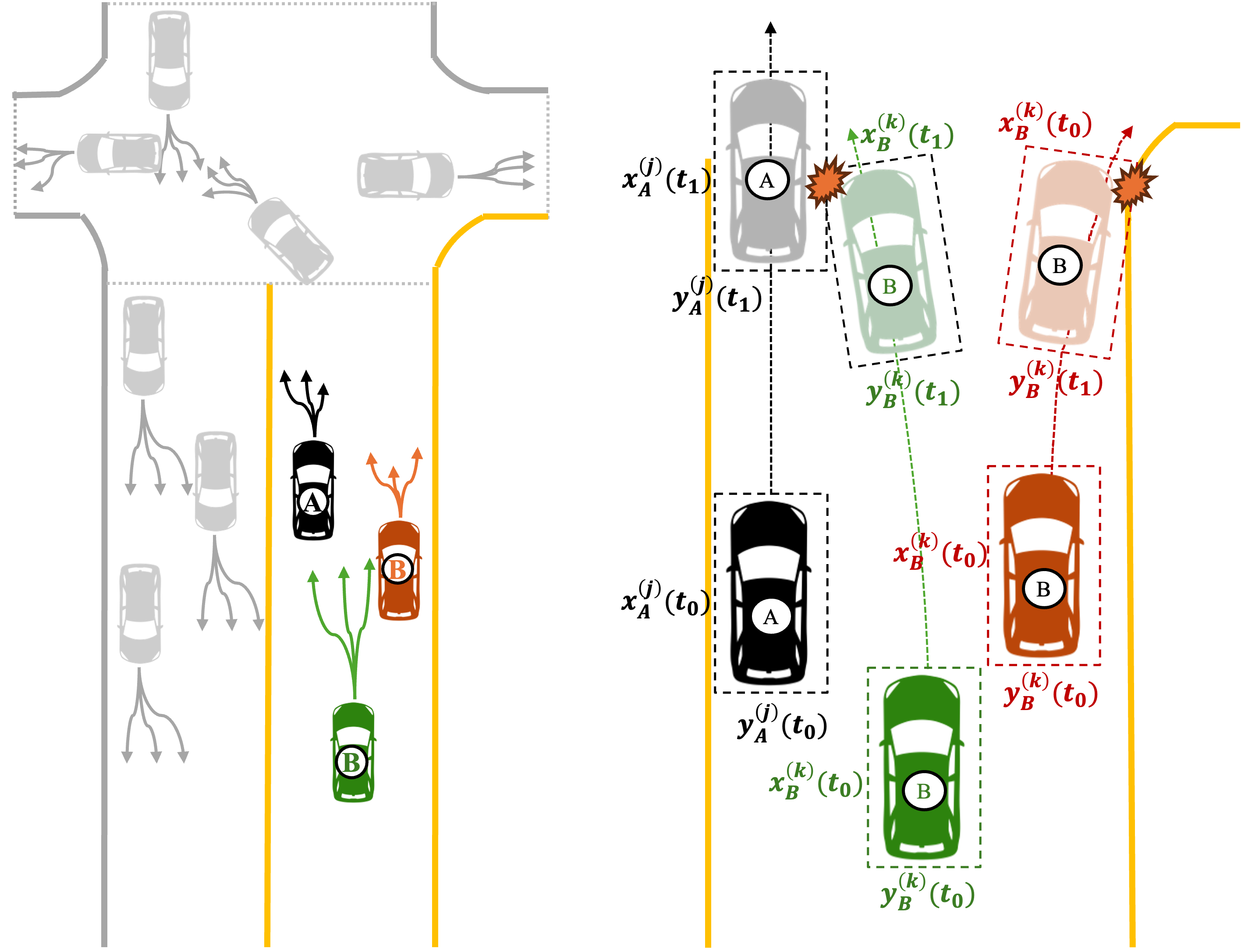}
    \caption{Computation of 2D-TTC for near-miss scenarios}
    \label{fig:1}
\end{figure}

The near-miss detection framework integrates vehicle dynamics, geometric transformations, and spatial proximity checks to identify both V–V and V–I events. Forward trajectories are simulated using RK4 integration, vehicle corners are transformed into global coordinates, and geometric overlaps are evaluated to detect potential near-misses across the prediction horizon. The procedure is summarized in Algorithm~\ref{alg:collision}, and a conceptual illustration is provided in Fig.~\ref{fig:1}. These steps yield a time series of 2D-TTC values for each interaction block.

\begin{algorithm}[t]
\caption{Near-miss detection based on state integration and geometric overlap}
\label{alg:collision}
\KwIn{Initial state $\mathbf{X}(0)$, vehicle parameters $l_i$, $L_i^{\text{veh}}, W_i^{\text{veh}}$, control inputs $a_i$, $\delta_i$, and boundary points $\{(X_\ell, Y_\ell)\}$}
\KwOut{Earliest collision time $t_c$}

\For{each timestep $n = 0$ to $N-1$}{
    Simulate $\mathbf{X}_n \rightarrow \mathbf{X}_{n+1}$ using RK4 integration\;
    Compute global corner positions $\mathbf{p}_i^{(j)}$ for each vehicle\;

    \For{each corner pair $(j, k) \in \{1,2,3,4\}^2$}{
        \If{$|x_A^{(j)} - x_B^{(k)}| \leq \varepsilon$ \textbf{or} $|y_A^{(j)} - y_B^{(k)}| \leq \varepsilon$}{
            Record V-V near-miss, set $t_c$\;
        }
    }

    \For{each corner $j$ and boundary point $\ell$}{
        Compute $d_{j,\ell}(t) = \sqrt{(x^{(j)} - X_\ell)^2 + (y^{(j)} - Y_\ell)^2}$\;
        \If{$d_{j,\ell}(t) \leq \varepsilon$}{
            Record V-I near-miss, set $t_c$\;
        }
    }
}
\Return{the earliest collision time $t_c$ as 2D-TTC}
\end{algorithm}

\begin{figure}[h]
    \centering
    \includegraphics[width=0.9\textwidth]{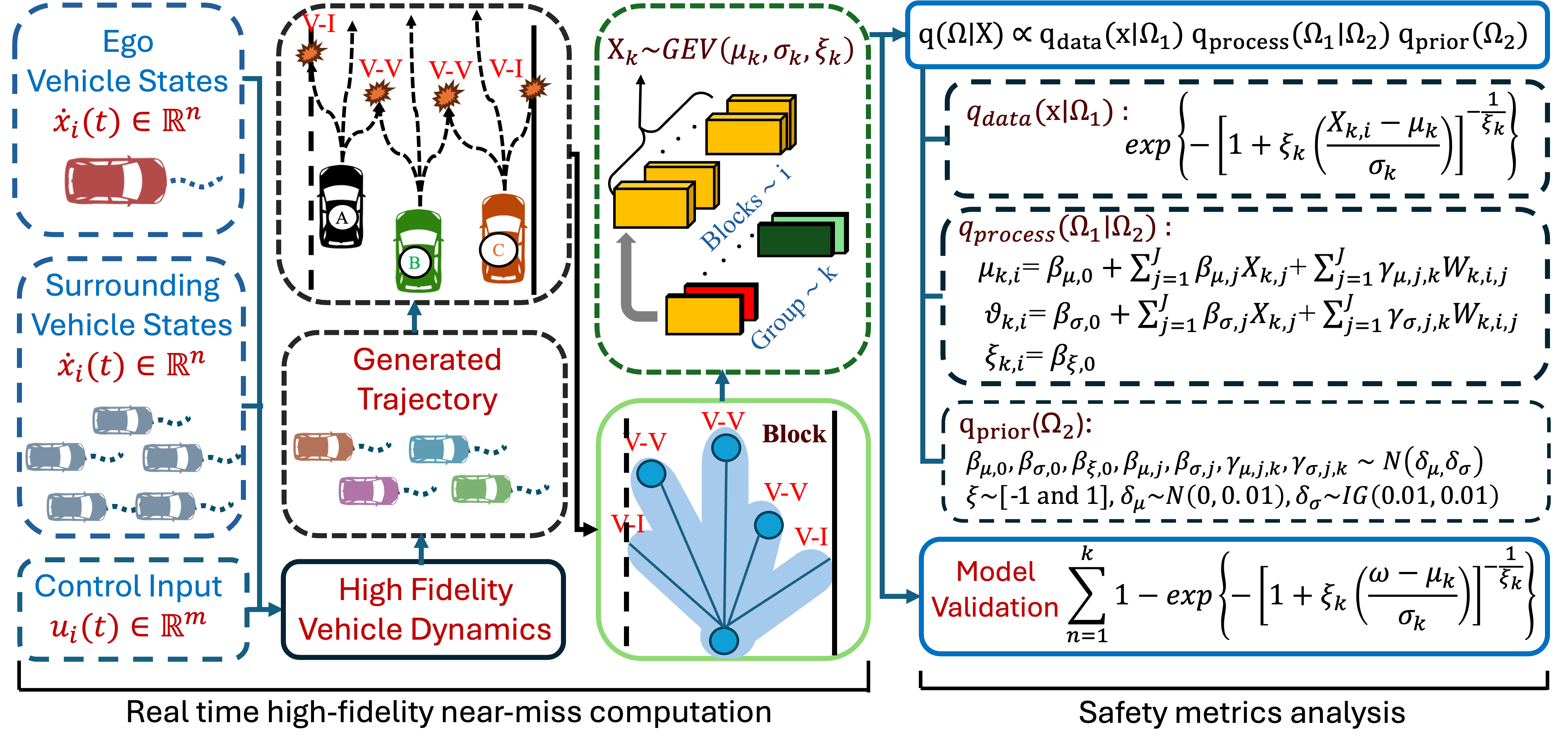}
    \caption{COR time series generation}
    \label{fig:2}
\end{figure}

While the above procedure deterministically detects and localizes near-misses, COR requires a probabilistic mapping from near-miss severity to short-term crash risk. Accordingly, BM, defined as the minimum 2D-TTC observed across the selected window, is modelled using EVT. COR is then defined as the estimated probability that 2D-TTC falls below a pre-specified critical threshold. Here, this block definition preserves short-lived maneuver dynamics at the native trajectory resolution while yielding a sufficient set of approximately independent extremes for stable inference across locations. The end-to-end pipeline from trajectory simulation to COR estimation is summarized in Fig.~\ref{fig:2}.

\section{Extreme value theory (EVT)}\label{3}

Extreme Value Theory (EVT) provides the statistical foundation for estimating COR by extrapolating rare, high-severity events from frequent near-misses. In this study, extreme events are represented by block maxima of 2D-TTC derived from high-resolution vehicle trajectories (Algorithm~\ref{alg:collision}). EVT enables risk estimation without requiring large volumes of historical crash data (e.g., \citep{zheng2020novel, fu2022bayesian}). Two EVT approaches are commonly applied: the Block Maxima (BM) method, which fits maxima or minima from fixed-duration blocks using the Generalized Extreme Value (GEV) distribution (\citep{coles2001introduction}), and the Peaks Over Threshold (POT) method, which models threshold exceedances using the Generalized Pareto Distribution (GPD). Although POT can be data-efficient, it is highly sensitive to threshold selection and typically requires declustering and handling of dependencies when applied to dense trajectory data. By contrast, BM aligns naturally with real-time AV unique-segmented data, avoids subjective thresholds, and supports scalable time-series analysis (\citep{fu2022bayesian, songchitruksa2006extreme}). Consequently, this study adopts the BM–GEV framework (see Fig.~\ref{fig:2}).

The GEV has been widely applied in traffic safety research. Early studies modeled near-misses using univariate and multivariate GEV formulations (e.g., citep{orsini2020large, zhang2019real, zheng2014freeway}). Recent advances incorporate nonstationary structures, linking GEV parameters to covariates such as vehicle dynamics (\citep{anis2025real, fu2021random, fu2022random}). Bayesian hierarchical models have also addressed sparse data and unobserved heterogeneity across vehicles, sites, and traffic contexts. With increasing availability of AV trajectories and HD maps, there is growing motivation to combine geometry-aware near-miss extraction (including V–I exposure) with hierarchical EVT to support corridor-scale risk inference.

Despite this progress, most UGEV models still neglect the joint influence of fixed and time-variant parameters, which are critical but distinct sources of heterogeneity. For instance, (e.g., lane count, median type) is relatively stable within segments or intersections, whereas vehicle dynamics (e.g., speed, deceleration, spacing) vary within and across interaction windows and drive short-term volatility. Capturing this heterogeneity is crucial for estimating corridor or network-level crash risk. To address this gap, this study employs a hierarchical Bayesian structure grouped random parameters (HBSGRP) UGEV model. Unlike standard random-parameter models, which assign variation independently to each unit, grouped random parameters capture shared variation across clusters (e.g., segments, intersections) while allowing for within-group variability. This structure provides both flexibility and parsimony, making it well-suited for high-resolution, corridor-scale crash risk modeling (\citep{washington2020statistical, cai2018developing, islam2023grouped}).

\subsection{Univariate generalized extreme value (UGEV)}\label{}

The BM-based UGEV model is used to characterize the tail behavior of extreme (e.g., 2D-TTC), which represents the most critical events (near-miss: V–V and V–I interactions). To implement this framework, continuous vehicle trajectories are separated into fixed-duration windows. Within each window, the minimum 2D-TTC is extracted and negated to align with the upper-tail modeling convention (\citep{coles2001introduction}). The resulting BM sequence serves as input for parameter estimation. Throughout this study, we focus on maxima, denoted by \(M_n\), under the assumption that each block consists of \(n\) independent and identically distributed (i.i.d., \(X_1, X_2, \ldots, X_n\)). To align with conventional UGEV formulations, we consider the negative values of the original variables derived from Algorithm~\ref{alg:collision}, thereby defining the block maximum as $M_n = \max\{-X_1, -X_2, \ldots, -X_n\}$. Under suitable regularity conditions, if there exist sequences of constants \( a_n \in \mathbb{R} \) and \( b_n > 0 \), the normalized block maxima \( M_n^* = (M_n - b_n)/a_n \) converge in distribution to a non-degenerate limit \( G(x) \) as \( n \to \infty \), then: $\Pr\left(\frac{M_n - a_n}{b_n} \leq X\right) \xrightarrow[n \to \infty]{} f(x)$, where \(G(x)\) is a non-degenerate distribution function. According to EVT, the limiting distribution \( G(x) \) must belong to one of three families: Gumbel, Fréchet, or Weibull, each of which can be captured by the parametric form of the GEV distribution (\citep{coles2001introduction}). Let the UGEV be defined as:

\begin{equation} \label{eq:22} 
G(x; \mu_x, \sigma_x, \xi_x) = \exp\left\{ - \left[1 + \xi_x \left(\frac{x - \mu_x}{\sigma_x}\right)\right]^{-1/\xi_x} \right\}, \quad \text{where } 1 + \xi_x \left(\frac{x - \mu_x}{\sigma_x}\right) > 0
\end{equation}

Here, \( \mu_x \in \mathbb{R} \) is the location parameter, \( \sigma_x > 0 \) the scale, and \( \xi_x \in \mathbb{R} \) the shape. The value of \( \xi_x \) determines tail heaviness: \( \xi_x > 0 \) corresponds to the heavy-tailed Fréchet, \( \xi_x < 0 \) to the bounded Weibull, and \( \xi_x = 0 \) to the light-tailed Gumbel case.

\subsubsection{Hierarchical Bayesian Structure Grouped Random Parameters (HBSGRP) model \label{sec3.3.1}}

To capture structured heterogeneity in extreme outcomes across the corridor, we employ the HBSGRP model. In this study, groups \(k=1,\dots,K\) correspond to corridor elements (intersections or directional segments), so each block extreme \(X_{k,i}\) is associated with a specific spatial context. Covariates are divided into (i) group-level fixed effects \(Y\), which describe relatively stable roadway context (e.g., lane configuration, median type), and (ii) interaction-level dynamics \(W\), which vary across blocks (e.g., relative speed, relative distance). The key feature of HBSGRP is that selected dynamic effects are allowed to vary by group, enabling systematic differences in interactions across corridor elements while maintaining corridor-wide coherence through partial pooling.

The model is formulated as a three-layer hierarchical Bayesian structure: a data layer that models block extremes using a UGEV distribution, a process layer that introduces grouped random parameters (partial pooling across corridor elements), and a prior layer that regularizes the model and quantifies uncertainty. Under a latent Gaussian formulation, the UGEV parameters are linked to covariates through identity link functions that combine global fixed effects with group-specific random effects, allowing the model to represent corridor-wide structure while preserving localized heterogeneity.

Bayesian inference proceeds by updating prior beliefs with observed data to obtain the posterior distribution. Formally, the joint posterior of parameters \( \Theta \), given data \( X \), is:

\begin{equation}
q(\Theta \mid X) \propto q_{\text{data}}(X \mid \Theta) \cdot q_{\text{process}}(\Theta \mid \Psi) \cdot q_{\text{prior}}(\Psi)
\label{eq:23}
\end{equation}

In this formulation, \( q(\Theta \mid X) \) denotes the posterior distribution, integrating both the data likelihood and prior knowledge. The term \( q_{\text{data}}(X \mid \Theta) \) corresponds to the likelihood function defined by the EVT assumptions. The component \( q_{\text{process}}(\Theta \mid \Psi) \) captures the hierarchical dependency structure, including the group-varying parameters. Lastly, \( q_{\text{prior}}(\Psi) \) encompasses the prior distributions of hyperparameters that control the model's flexibility and penalize over-complexity. Eqs.~(\ref{eq:24}–\ref{eq:30}) therefore summarize the full HBSGRP–UGEV hierarchy used in this study. 

The likelihood function for the data layer is specified as each extreme value sample \(X_{k,i}\), observed for event \(i\) within group \(k\), is modeled as:

\begin{equation}
q_{\text{data}}(X_{k,i} \mid \Theta) = \prod_{k=1}^{K} \prod_{i=1}^{N_k} 
\frac{1}{\sigma_{k}} 
\exp \left\{ - \left[1 + \xi_{k,i} \left(\frac{X_{k,i} - \mu_{k,i}}{\sigma_{k,i}}\right)\right]^{-1/\xi_{k,i}} \right\} 
\left[1 + \xi_{k,i} \left(\frac{X_{k,i} - \mu_{k,i}}{\sigma_{k,i}}\right)\right]^{-1-1/\xi_{k}}
\label{eq:24}
\end{equation}

where $\mu_{k,i}$, $\sigma_{k,i}$, and $\xi_{k,i}$ are the location, scale, and shape parameters of the UGEV distribution for observation $i$ in group $k$. This layer expresses how block maxima of 2D-TTC are modeled as extreme outcomes that differ across groups and blocks.

Parameters are expressed as functions of group-level fixed-effect covariates \(Y_{k,j}\) (e.g., lane no, lane width, median type, etc.) and sample-specific random-effect covariates \(W_{k,i,j}\) (e.g., speed, acceleration, distance, etc.) as follows:

\begin{equation}
\mu_{k,i} = \beta_{\mu,0} + \sum_{j=1}^{J} \beta_{\mu,j} Y_{k,j} + \sum_{j=1}^{J'} \gamma_{\mu,j,k} W_{k,i,j}
\label{eq:25}
\end{equation}

\begin{equation}
\vartheta_{k,i} = \beta_{\sigma,0} + \sum_{j=1}^{J} \beta_{\sigma,j} Y_{k,j} + \sum_{j=1}^{J'} \gamma_{\sigma,j,k} W_{k,i,j}, \quad \sigma_{k,i} = \exp(\vartheta_{k,i})
\label{eq:26}
\end{equation}

\begin{equation}
\xi_{k,i} = \beta_{\xi,0}
\label{eq:27}
\end{equation}

In these expressions, $Y_{k,j}$ denotes the $j$-th group-level fixed-effect covariate for group $k$, and $W_{k,i,j}$ denotes the $j$-th observation-level covariate for event $i$ within group $k$. The parameters $\beta_{\mu,0}$, $\beta_{\sigma,0}$, and $\beta_{\xi,0}$ are global intercepts for the location, log-scale, and shape parameters, respectively. The coefficients $\beta_{\mu,j}$, $\beta_{\sigma,j}$, and $\beta_{\xi,j}$ are global fixed-effect coefficients associated with geometric and static traffic variables, whereas $\gamma_{\mu,j,k}$ and $\gamma_{\sigma,j,k}$ are grouped random-effect coefficients that vary across groups $k$ and are associated with dynamic, time-varying covariates. Taken together, Eqs.~\eqref{eq:24}--\eqref{eq:27} show how the HBSGRP model combines corridor-level geometric characteristics (through fixed effects on $Y_{k,j}$) with time-varying driving conditions (through group-specific random effects on $W_{k,i,j}$), making the formulation well suited for corridor-level risk analysis.

Therefore, the contribution of the process layer, modeling within-group variability, is defined as

\begin{multline}
q_{\text{process}}(\Theta \mid \Psi) = 
\prod_{k=1}^K \prod_{i=1}^{N_k} 
\frac{1}{\sqrt{2\pi \tau_\mu^2}} \exp\left\{ -\frac{1}{2\tau_\mu^2} (\mu_{k,i} - \bar{\mu}_{k})^2 \right\} \\
\times \frac{1}{\sqrt{2\pi \tau_\vartheta^2}} \exp\left\{ -\frac{1}{2\tau_\vartheta^2} (\vartheta_{k,i} - \bar{\vartheta}_{k})^2 \right\}
\times \frac{1}{\sqrt{2\pi \tau_\xi^2}} \exp\left\{ -\frac{1}{2\tau_\xi^2} (\xi_{k,i} - \bar{\xi}_{k})^2 \right\}
\label{eq:28}
\end{multline}

where $\bar{\mu}_k$, $\bar{\vartheta}_k$, and $\bar{\xi}_k$ are group-specific average parameters. The variance components $\tau_\mu^2$, $\tau_\vartheta^2$, and $\tau_\xi^2$ control the amount of between-group heterogeneity and induce shrinkage toward group means, thereby encoding the ``grouped random parameters'' aspect of the HBSGRP model and allowing different corridor elements to share information while retaining distinct risk profiles.

Prior distributions are assigned to all model parameters. The priors for fixed and random effect coefficients are as follows:

\begin{multline}
q_{\text{prior}}(\Psi) = q_{\beta_{\mu,0}}(\beta_{\mu,0}) \cdot \prod_{j=1}^{J} q_{\beta_{\mu,j}}(\beta_{\mu,j}) \cdot \prod_{k=1}^{K}\prod_{j=1}^{J'} q_{\gamma_{\mu,j,k}}(\gamma_{\mu,j,k}) \\
\times q_{\beta_{\sigma,0}}(\beta_{\sigma,0}) \cdot \prod_{j=1}^{J} q_{\beta_{\sigma,j}}(\beta_{\sigma,j}) \cdot \prod_{k=1}^{K}\prod_{j=1}^{J'} q_{\gamma_{\sigma,j,k}}(\gamma_{\sigma,j,k}) \
\times q_{\beta_{\xi,0}}(\beta_{\xi,0}) \cdot \prod_{j=1}^{J} q_{\beta_{\xi,j}}(\beta_{\xi,j})
\label{eq:29}
\end{multline}

Hyperpriors for the variance parameters \(\tau_\mu^2, \tau_\vartheta^2, \tau_\xi^2\) are modeled using inverse gamma distributions:

\begin{equation}
\tau_\mu^2 \sim \text{IG}(\alpha_\mu, \beta_\mu), \quad 
\tau_\vartheta^2 \sim \text{IG}(\alpha_\vartheta, \beta_\vartheta), \quad 
\tau_\xi^2 \sim \text{IG}(\alpha_\xi, \beta_\xi)
\label{eq:30}
\end{equation}

where \(\text{IG}(\alpha, \beta)\) denotes an inverse gamma prior distribution ensuring positivity constraints for variance parameters.These prior and hyperprior specifications provide regularization for both fixed and grouped random effects, thereby completing the hierarchical definition of the HBSGRP model.

\subsection{COR estimation}

Building on the HBSGRP--UGEV model, COR is quantified by mapping block-level extreme near-miss severity to the probability of a crash-level outcome. For each corridor element $k$ (intersection or directional segment), the trajectory stream is partitioned into fixed-duration interaction blocks, and the \emph{minimum} 2D--TTC in each block is extracted and transformed to the UGEV scale (Section~\ref{sec3.3.1}). These block extremes form a sequence $X_{k,i}$,  which is modeled using a nonstationary UGEV distribution with parameters $\mu_{k,i}$, $\sigma_{k,i}$, and $\xi_{k,i}$. On the transformed scale, a crash-level event corresponds to the physical boundary at which the $\omega = 0$ (i.e., 2D--TTC $=0$). Therefore, the probability of a crash occurring within the  interaction window $i$ in group $k$ is defined as the upper-tail probability of the fitted UGEV distribution evaluated at $\omega=0$:

\begin{equation}
P_{k,i}^{\text{crash}} 
= \Pr(X_{k,i} \geq \omega) 
= 1 - G(\omega; \mu_{k,i}, \sigma_{k,i}, \xi_{k,i})
= 1 - \exp\left\{ -\left[1 + \xi_{k,i} \left( \frac{\omega - \mu_{k,i}}{\sigma_{k,i}} \right) \right]^{-1/\xi_{k,i}} \right\}, \omega=0
\label{eq:31}
\end{equation}

Aggregating these block-level crash probabilities over an exposure period $T$ yields the expected COR for corridor element $k$:

\begin{equation}
COR_k = \frac{1}{T} \sum_{i=1}^{N_k} P_{k,i}^{\text{crash}}
\label{eq:32}
\end{equation}

where $N_k$ is the number of blocks observed for group $k$. The total expected crash frequency across the corridor is then obtained by summing over all groups:
\begin{equation}
CF_{\text{total}} = \sum_{k=1}^{K} COR_k
\label{eq:33}
\end{equation}

Negative TTC thresholds (e.g., $-0.1$~s to $-3$~s) are not interpreted as crash boundaries. Instead, they are used solely to probe increasingly severe regions of the near-miss distribution and to compare relative crash-prone risk across locations. Accordingly, COR should be interpreted as a model-based measure of crash-prone extreme conflict risk extrapolated from near-miss interactions, rather than as a direct count or prediction of observed crashes.

\section{Data description}\label{4}
The effectiveness of real-time COR estimation depends on the spatiotemporal fidelity and contextual coverage of trajectory data. Traditional sensing technologies (e.g., roadside cameras, loop detectors, fixed LiDAR) can support safety analysis by capturing speed, spacing, and acceleration patterns (\citep{islam2021crash, yuan2019real, li2020real}). However, these systems often face limited spatial coverage, occlusion, overlapping detections, and deployment/maintenance burdens, which constrain scalability and data quality (\citep{st2013automated}). In contrast, AV datasets provide high-frequency, multi-agent trajectories with precise localization, enabling consistent reconstruction of short-horizon kinematics and interaction geometry needed for near-miss-based COR modeling. Despite this potential, AV data remain underutilized in generalized real-time COR frameworks. For example, studies using the Waymo Open Dataset have mainly analyzed localized conflicts and often omit HD-map-aligned roadway boundaries needed to represent V–I exposure, limiting transferability to corridor-scale inference (\citep{anis2025real}).

To address these challenges, this study uses the Argoverse-2 dataset (\citep{wilson2023argoverse}), collected in 2021 across six U.S. cities (Austin, Detroit, Miami, Palo Alto, Pittsburgh, and Washington, D.C.) using SAE Level 4 AVs. Argoverse-2 contains $>$250k driving scenarios spanning 113 roadway segments, with each 11-second scenario sampled at 10 Hz. Compared to Waymo Open (\citep{ettinger2021large}), Lyft Level-5 (\citep{houston2021one}), and nuScenes (\citep{caesar2020nuscenes}), Argoverse-2 is curated to emphasize interaction-rich episodes (e.g., intersection encounters, lane changes, and complex crossings) that are informative for extreme-conflict modeling. A defining feature is its scene-adaptive sampling, which prioritizes high-interaction moments rather than routine free-flow conditions. Because this design intentionally over-samples kinematically and socially unusual behavior (\citep{wilson2023argoverse}), the extracted near-misses and inferred COR should be interpreted as conditional on safety-critical traffic states, rather than as unconditional averages over typical corridor operations. Scenarios include agent identifiers, AV status, and map-based behavioral annotations. The sensor suite comprises seven ring cameras and two front-facing stereo cameras (20 Hz), supported by dual 64-beam VLP-32C LiDAR sensors, which produce ~107k 3D points per frame. HD maps contain lane centerlines, crosswalks, medians, and drivable areas, all registered to local coordinates.

\begin{table}[htbp]
  \centering
  \begin{threeparttable}
    \caption{Representative open-source AV datasets with applications in safety and crash-risk analysis}
    \label{tab:1}
    \scriptsize
    \begin{tabularx}{\linewidth}{@{}lZZZZZ@{}}
      \toprule
      \textbf{Dataset} & \textbf{Trajectory coverage\tnote{a}} & 
      \textbf{Safety relevance\tnote{b}} & 
      \textbf{Duration / Frequency} & 
      \textbf{HD map} & 
      \textbf{Notes} \\
      \midrule
      Argoverse-2 & 250k scenarios; 113 segments; 6 cities & Near-miss analysis, surrogate safety measures & 11s per scenario, 10 Hz & Yes & Rich lane-level geometry; 10 object classes (\cite{wilson2023argoverse}) \\
      nuScenes & 40k scenes; $\sim$1000 segments; 2 cities & Multi-agent interactions, trajectory risk indicators & 20s per scene, 2 Hz & No & 23 object classes; widely used urban dataset (\cite{caesar2020nuscenes}) \\
      Waymo & 390k scenarios; $\sim$1950 segments; 6 cities & Motion forecasting, crash-prone interaction detection & Up to 20s, 10 Hz & Yes & 12.6M trajectories across diverse geometries (\cite{ettinger2021large}) \\
      Lyft Level 5 & 170k instances; 1 city & Vehicle interactions; trajectory-based safety studies & 25s per instance, 10 Hz & Yes & Limited geographic diversity; 10 object classes (\cite{lyft2019}) \\
      KITTI & 15k scenes; 1 city & Early surrogate measures, safety benchmarks & 1–2 min sequences, 10 Hz & No & Classic full-stack dataset, still widely cited (\cite{kitti}) \\
      DAIR-V2X & 71k frames & Infrastructure-based safety monitoring & -- & Yes & First large-scale real-world V2X dataset (\cite{Dair-v2x}) \\
      OPV2V & 33k samples; CARLA simulation & Cooperative perception, proactive safety design & 10 Hz & No & 18k V2X frames, sim environment (\cite{Opv2v}) \\
      DeepAccident & 285k samples; sim dataset & Rare crash/near-crash events, proactive safety & 10 Hz & Yes & Includes 57k V2X scenarios (\cite{deepaccident}) \\
      Shift & 2.5M frames; 4850 segments; 8 cities & Long-term safety and mobility analysis & 10 Hz & No & Corridor-scale benchmarking dataset (\citep{shift}) \\
      \bottomrule
    \end{tabularx}
    \begin{tablenotes}[flushleft]
      \scriptsize
      \item[a] \textbf{Trajectory coverage}: Number of driving scenarios and geographic spread.
      \item[b] \textbf{Safety relevance}: Potential application for crash risk, near-miss, or surrogate safety studies.
    \end{tablenotes}
  \end{threeparttable}
\end{table}

Argoverse-2 was selected after benchmarking open-source AV datasets (Table \ref{tab:1}). Waymo Open offers greater overall mileage but does not explicitly target interaction-dense segments at the same level, Lyft Level-5 is limited to a single city, and nuScenes samples at 2 Hz, reducing sensitivity to rapid interactions. Argoverse-2 offers a practical balance of sampling rate, multi-agent density, HD-map detail, and interaction-centric scenario selection, making it well-suited for geometry-aware near-miss extraction and corridor-scale COR estimation. This study used vehicle trajectories and HD-map details, which provide multiple agent tracks labeled by object type (e.g., vehicle, pedestrian, cyclist). To operationalize the proposed framework, we processed these modules through a four-stage pipeline: (i) extraction of multi-agent trajectories and necessary information, (ii) coordinate transformation from local map space to global geographic space, (iii) integration with HD-map boundary features to define V–I exposure, and (iv) selection of a high-risk urban corridor for detailed analysis.

\subsection{Motion forecasting dataset}
Each Argoverse-2 motion-forecasting scenario has a three-level structure: scenario metadata, agent tracks, and frame-level states. Scenarios are 11 s long and provide a birds-eye-view 2D centroid and heading for each tracked object at 10 Hz (110 time steps), enabling fine-grained interaction modeling. Scenario metadata includes a unique scenario ID, recording city, and a focal agent (typically the ego AV). Each scenario contains multiple agent tracks with object type (e.g., vehicle, pedestrian, cyclist), scoring category (focal/scored/unscored), and a time-ordered sequence of states. Frame-level states provide position, velocity components, and heading, which form the basis for computing 2D-TTC. We directly use these trajectories as provided by Argoverse-2; tracks are produced by Argo’s internal multi-sensor fusion and map-aligned ego poses, which help keep trajectories tightly registered to the HD map.

To implement the proposed framework, we developed a Python batch-processing pipeline using the Argoverse-2 SDK. The pipeline parses scenarios and extracts tracks for all agents (not only the focal vehicle) to preserve the multi-agent interaction context. For each track, we retrieve frame-level kinematic attributes, including \((x, y)\) position, heading, velocity components, and timestamps. For the present analysis, we retain only motor-vehicle agents and filter out pedestrians and cyclists; the same pipeline can be extended to incorporate vulnerable road users in future work.

Because positions are reported in city-specific local coordinate frames, all coordinates are transformed to a globally referenced WGS84 system. This is done by projecting each city’s local origin into the appropriate UTM zone (e.g., Zone 17 for Miami, Pittsburgh, and Detroit), offsetting local \((x, y)\) values, and inverse-projecting to latitude/longitude using \texttt{pyproj}. Batch scripts traverse scenario folders, extract attributes and convert coordinates, and export unified CSV files containing both the original city-frame coordinates and the converted lat/lon values. To verify that this transformation preserves trajectory geometry, we conducted a visual geospatial validation for the Biscayne Boulevard corridor by overlaying converted trajectories and HD-map features on satellite imagery (Fig.~\ref{fig:3}). Vehicle paths closely follow mapped lane centerlines and intersection geometry, indicating residual registration error is small relative to lane width and is acceptable for conflict localization.

In addition to the motion-forecasting trajectories, we use the Argoverse-2 sensor annotation files $(\texttt{annotations.feather})$, which provide 3D bounding-box dimensions for each tracked object at each timestamp (including \texttt{length\_m} and \texttt{width\_m}). The motion-forecasting tracks are joined to these annotations using the shared scenario identifier, \texttt{timestamp\_ns}, and \texttt{track\_uuid}. For each vehicle track, the corresponding \texttt{length\_m} and \texttt{width\_m} values are used as the plan-view dimensions of the oriented rectangular footprint in the geometry-aware 2D-TTC computation.

The processed files form the input for V–V and V–I near-miss detection. For each valid scenario, we evaluate V-V interactions by computing pairwise relative spacing and approach rates across vehicle pairs, while V–I proximity is evaluated relative to HD-map-derived lane edges/curb boundaries. Scenarios without substantive interaction (e.g., sustained straight-lane cruising without turning, merging, or lateral maneuvers) are excluded to emphasize behaviorally informative, high-interaction conditions consistent with proactive extreme-conflict modeling.

\subsection{HD map}

The Argoverse-2 HD map dataset provides vectorized representations of roadway geometry and drivable space, enabling infrastructure-aware risk estimation. These maps are distributed as JSON files in local city coordinates and contain lane centerlines, left/right boundaries, crosswalk polygons, and drivable area definitions.

Lane-level geometry was extracted using the official Argoverse-2 API. The \texttt{get\_all\_lane\_segments()} method returns centerline polylines along with left/right boundaries that define the navigable road space. Additional features, such as crosswalk polygons and drivable area masks, were also retrieved to support validation and visualization, though only lane boundaries and centerlines were directly used in near-miss detection.

\begin{figure}[h]
    \centering
    \setlength{\abovecaptionskip}{0pt}
    \includegraphics[width=0.9\textwidth]{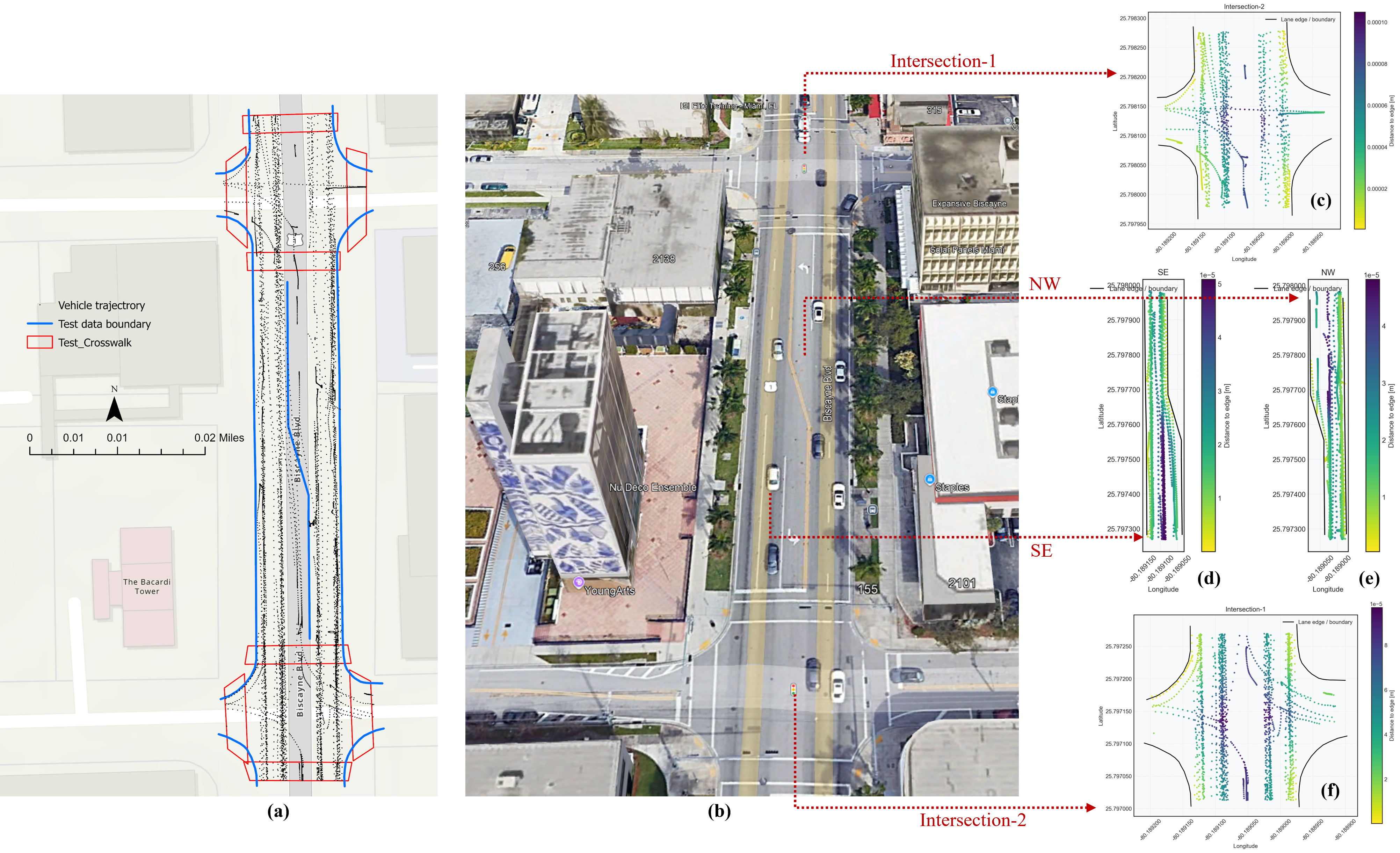}
    \caption{AV data transformed to real-world coordinates and aligned: (a) trajectories with lane boundaries and crosswalks, (b) validation with satellite imagery, and (c–f) directional traffic patterns highlighting interaction hotspots}

    \label{fig:3}
\end{figure}

Because all coordinates are expressed in city-specific Cartesian frames, global georeferencing was required to align HD maps with trajectory data. For each city, the WGS84 geodetic origin was projected into its respective UTM zone (e.g., Zone 17 for Miami and Pittsburgh, Zone 14 for Austin, Zone 10 for Palo Alto, Zone 18 for Washington, D.C.) using the \texttt{pyproj} library. Each local \((x, y)\) coordinate was then offset by this origin and inverse-projected into latitude–longitude, ensuring spatial consistency across cities. Once transformed, all HD map components: lane boundaries, centerlines, and crosswalks—were indexed by their global coordinates and converted into shapefiles or GeoPandas geometries. This geospatial transformation enabled efficient spatial querying, collision proximity analysis, and visual overlay with satellite imagery and trajectory data. In particular, the lane boundaries were instrumental in detecting V–I near-misses by allowing for the computation of lateral vehicle encroachments beyond the legal driving space. From the processed dataset, a high-density urban corridor in Miami was present to showcase the transformation (see Fig.~\ref{fig:3}). This illustrates the transformation of AV data into real-world coordinates, its alignment with HD maps and satellite imagery, and the resulting traffic patterns that highlight interaction hotspots across intersections and midblocks.

\subsection{Trajectory preprocessing and derivation of vehicle dynamics}
Before computing 2D-TTC, a series of preprocessing steps was applied to ensure the precision, continuity, and physical plausibility of vehicle motion data from the Argoverse-2 dataset. These steps included extracting raw trajectory data, smoothing and filtering the trajectories, and deriving dynamic variables such as speed, acceleration, and steering angle. First, all multi-agent trajectories were transformed from local city-specific frames to global WGS84 coordinates, enabling consistent alignment with HD map geometries. Raw trajectories were then examined for noise caused by sensor jitter, occlusion, or missed detections—issues more common in human-driven vehicles than in AV tracks. To suppress such noise while preserving curvature and lane-change dynamics, cubic B-spline smoothing was applied to the position coordinates (\citep{eilers1996flexible,choi2024safe}). This method effectively removed local fluctuations while retaining key geometric features (see Fig. \ ref {fig:4}). Stationary or irrelevant agents, such as parked vehicles and bicycles, were filtered out to reduce computational load and focus only on interaction-relevant entities.

\begin{figure}[h]
    \centering
    \includegraphics[width=0.9\textwidth]{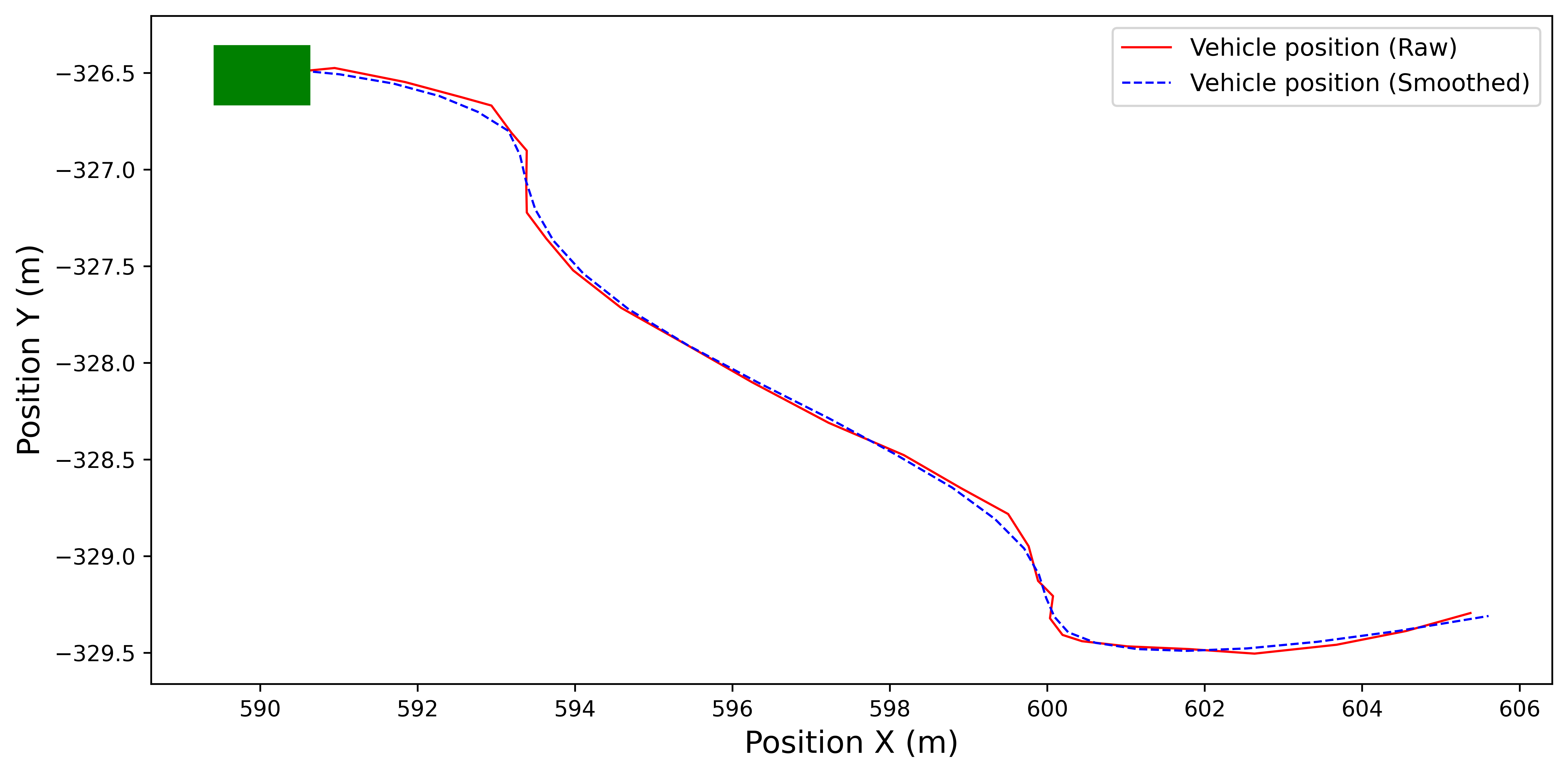}
    \caption{Effect of B-spline smoothing on a sample trajectory, showing removal of noise while preserving geometric fidelity}
    \label{fig:4}
\end{figure}

Dynamic control inputs were then computed. Instantaneous speed $v_t$ was derived as the Euclidean norm of velocity components:

\begin{equation}
v_t = \sqrt{v_{x,t}^2 + v_{y,t}^2}
\label{eq34}
\end{equation}

Residual noise in $v_t$ was suppressed with a Savitzky–Golay filter (\citep{chen2004simple}) using a second-order polynomial and a 110-frame window (11s). Longitudinal acceleration (\(a_t\)) was obtained by first-order numerical differentiation:

\begin{equation}
a_t = \frac{v_{t+1} - v_t}{\Delta t}
\label{eq35}
\end{equation}

where \(\Delta t = 0.01\)s reflects the 10 Hz sampling rate. Heading $\theta_t$ was computed from directional vectors of each track and was unwrapped to maintain continuity across discontinuities at \(\pm \pi\). The yaw rate \(\dot{\delta}_t\) was derived from the kinematic bicycle model:

\begin{equation}
\dot{\theta}_t = \frac{\theta_{t+1} - \theta_t}{\Delta t}
\label{eq36}
\end{equation}

\begin{equation}
\delta_t = \tan^{-1}\left( \frac{L \cdot \dot{\theta}_t}{v_t} \right)
\label{eq37}
\end{equation}

The final dataset included per-frame values of speed, acceleration, heading, yaw rate, and steering angle for each track, providing stable and interpretable control inputs. These preprocessed trajectories formed the basis for subsequent near-miss detection using Algorithm \ref{alg:collision} and the HBSGRP-UGEV model.

\subsection{Study area}\label{4.2}

 Under realistic and heterogeneous urban traffic conditions, we selected a high-volume study corridor along Biscayne Boulevard in Miami, Florida. This arterial links Downtown Miami to the Upper East Side and northern city limits and is characterized by dense access activity, frequent turning and lane-changing maneuvers, and diverse traffic control. The corridor was chosen because it has high scenario density in Argoverse-2 and high-fidelity HD map coverage, and because its mapped curb and median boundaries provide explicit roadway-edge geometry needed to consistently define V–I near-misses.

\begin{figure}[h]
    \centering
    \includegraphics[width=0.75\textwidth]{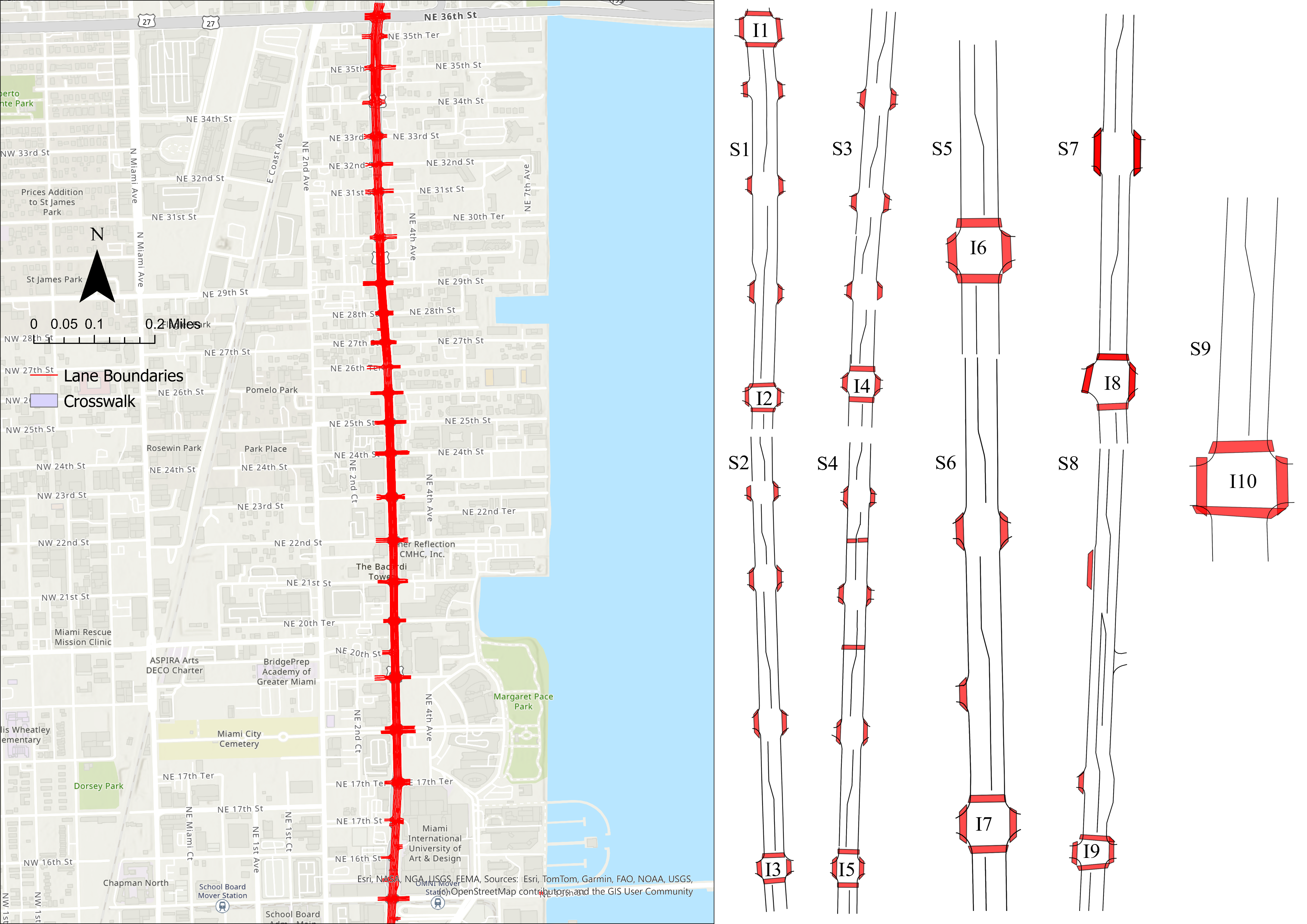}
    \caption{Segmented study corridor along Biscayne Boulevard, Miami, with lane boundaries and crosswalks. Insets show direction-specific segments (S1–S9) and intersections (I1–I10)}
    \label{fig:5}
\end{figure}

As shown in Fig.~\ref{fig:5}, the study area was extracted in ArcGIS Pro based on scenario density, HD map fidelity, and maneuver diversity. The resulting network forms a continuous corridor with bidirectional operations, multiple intersections, and midblock sections. For analytical consistency, the corridor was subdivided into 10 intersections (I1–I10) and 18 directional segments organized into nine segment pairs (S1–S9). Segment limits were defined midblock to preserve operational continuity, while intersection extents were delineated to capture conflict-prone subareas (turning paths, merges, and channelization transitions). These subregions define the hierarchical modeling structure and enable direction-aware inference.

Table~\ref{tab:4} summarizes the 28 study subregions. Intersections generally include fewer scenarios and unique vehicles than adjacent segments (e.g., I2: 47 scenarios/220 vehicles; I4: 96 scenarios/568 vehicles), reflecting shorter dwell time within intersection boundaries. By contrast, segments accumulate substantially more scenarios and vehicles due to sustained travel (e.g., S7 and S8 each include several hundred scenarios and $>$4000 unique vehicles)

\begin{table}[h!]
\centering
\caption{Profile of study locations along Biscayne Boulevard}
\label{tab:4}
\small
\renewcommand{\arraystretch}{1.0}
\resizebox{\textwidth}{!}{%
\begin{tabular}{clccccccc}
\hline
\textbf{ID} & \textbf{Location} & \textbf{Direction} & \textbf{Scenarios} & \textbf{Unique vehicles} & \textbf{Lane no} & \textbf{Lane width (m)} & \textbf{Driveway/Minor intersection} & \textbf{Median type} \\
\hline
\textit{Segment}\\
\hline
S1 & (NE 36–33 St)            & NB / SB   & 131 / 111  & 1706 / 813    & 3 / 3 & 3.5  & 3               & Raised \\

S2 & (NE 33–29 St)            & NB / SB   & 151 / 155  & 1532 / 1148   & 3 / 3 & 3.64  & 3             & Raised \\

S3 & (NE 29–26 St)            & NB / SB   & 191 / 186  & 1721 / 1295   & 3 / 3 & 3.36  & 4             & Flush paved \\

S4 & (NE 26–22 St)            & NB / SB   & 176 / 199  & 1542 / 1581   & 3 / 3 & 3.51  & 4             & Flush paved \\

S5 & (NE 22–21 St)            & NB / SB   & 127 / 123  & 705 / 643     & 3 / 2 & 3.3  & No             & Flush paved \\

S6 & (NE 21–19 St)            & NB / SB   & 235 / 245  & 1588 / 1861   & 3 / 3 & 3.62  & 1             & Flush paved \\

S7 & (NE 19–17 St)            & NB / SB   & 308 / 332  & 1704 / 2410   & 2 / 2 & 3.12  & 1              & Flush paved \\

S8 & (NE 17–15 St)            & NB / SB   & 294 / 314  & 1955 / 2406   & 3 / 3 & 3.57  & 2             & Flush paved \\

S9 & (NE 15–13 St)            & NB / SB   & 232 / 216  & 1565 / 1217   & 3 / 2 & 3.43  & 1             & Flush paved \\
\hline
\textit{Intersection}\\
\hline
I1 & NE 36 St                 & --        & 97         & 743           & --    & --   & --             & --\\
I2 & NE 33 St                 & --        & 47         & 220           & --    & --   & --              & -- \\
I3 & NE 29 St                 & --        & 132        & 699           & --    & --   & --             & -- \\
I4 & NE 26 St                 & --        & 96         & 568           & --    & --   & --      & -- \\
I5 & NE 22 St                 & --        & 106        & 640           & --    & --   & --             & -- \\
I6 & NE 21 St                 & --        & 134        & 647           & --    & --   & --             & -- \\
I7 & NE 19 St                 & --        & 214        & 1332          & --    & --   & --             & -- \\
I8 & NE 17 St                 & --        & 205        & 1328          & --    & --   &  --   & -- \\
I9 & NE 15 St                 & --        & 252        & 1332          & --    & --   & --             & -- \\
I10 & NE 13 St                & --        & 232        & 1139          & --    & --   & --             & -- \\

\hline
\end{tabular}%
}
\end{table}

Lane configurations range from 2 to 3 lanes per direction, with average lane widths of 3.1 m (S7) to 3.6 m (S6). Access density also varies across segments (e.g., multiple driveways/minor access points in S3–S4 versus few or none in S5–S6), creating heterogeneous exposure to lateral disturbances and side-entry maneuvers. Median treatments alternate between raised and flush-paved sections, which can affect turning opportunities and the potential for edge encroachment.

Building on the spatial segmentation and subregion profiles above, we next quantify time-varying safety conditions using the proposed 2D-TTC pipeline (Algorithm \ref{alg:collision}). Because this study adopts the BM sampling strategy, there is no global consensus on the optimal block duration. Prior BM–GEV studies based on video data often employ long, time-based blocks (e.g., several minutes) and select block sizes based on goodness-of-fit criteria. In contrast, AV trajectory datasets consist of short scenarios (e.g., Argoverse-2: 17 s; Waymo: 20 s; KITTI: 1–2 min) and high-frequency interaction windows. Consequently, this study adopts an interaction-window BM strategy, consistent with real-time near-miss modeling in our prior work (\citep{anis2025real}), in which blocks are defined by unique interaction units (\citep{ali2022assessing}). Specifically, each continuously observed interaction window, either a V–V or a V–I interaction within a scenario, is treated as a block, and the block extreme is defined as the minimum 2D-TTC over the 11 s (110-frame) window. For each interaction window, the near-miss process is evaluated at every frame by forward-simulating motion using an RK4-based kinematic projection and subsequently computing 2D-TTC. Thus, each retained extreme summarizes 110 high-frequency kinematic states and 110 interaction assessments, providing substantial within-block information despite the short scenario duration.

\begin{figure}[h]
    \centering
    \setlength{\abovecaptionskip}{0pt}
    \subcaptionbox{V-V\label{fig:I_VV}}
    {\includegraphics[width=0.30\textwidth]{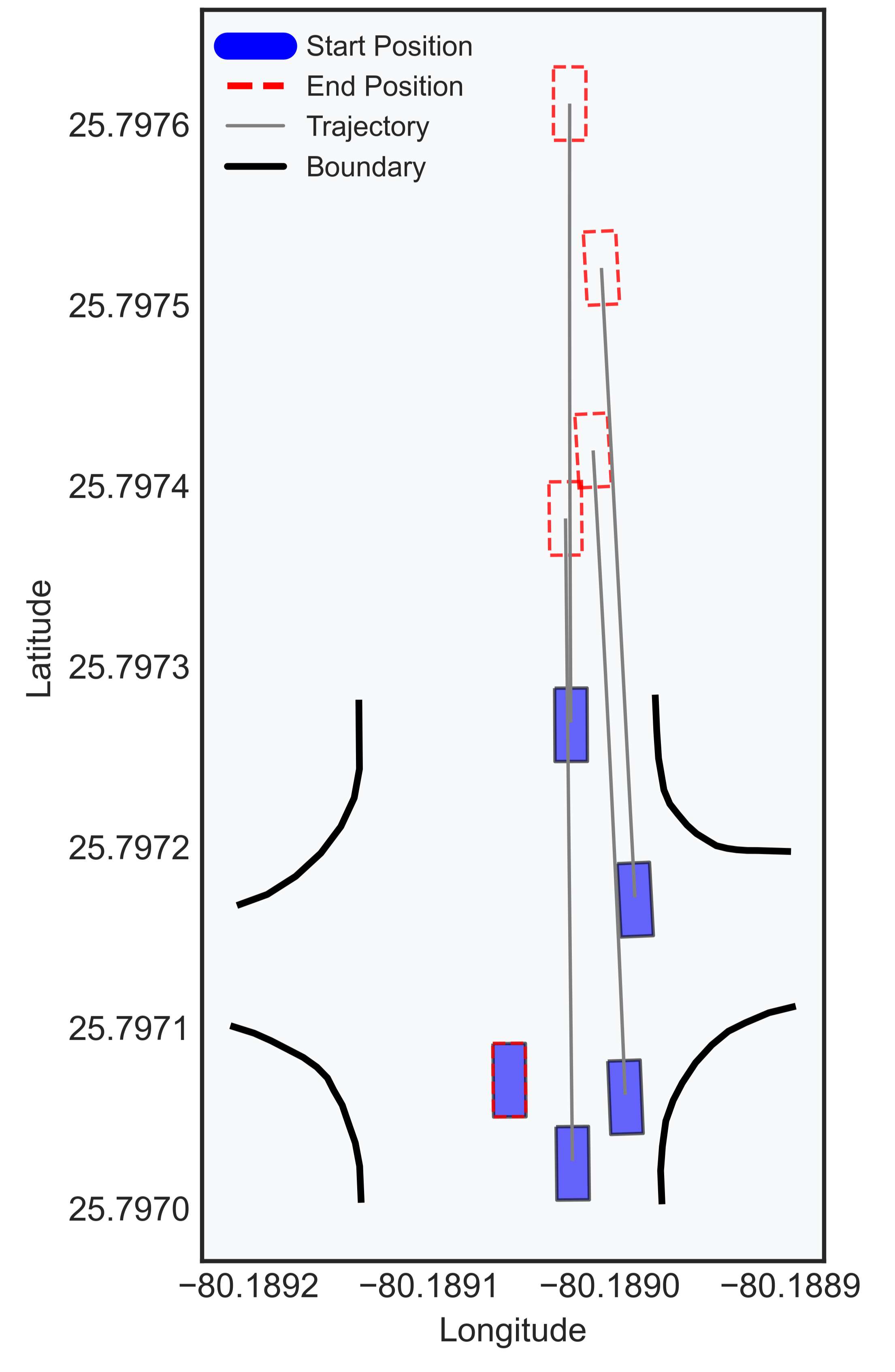}}
    \hspace{5mm}
    \subcaptionbox{V-I\label{fig:I_VI}}
    {\includegraphics[width=0.20\textwidth]{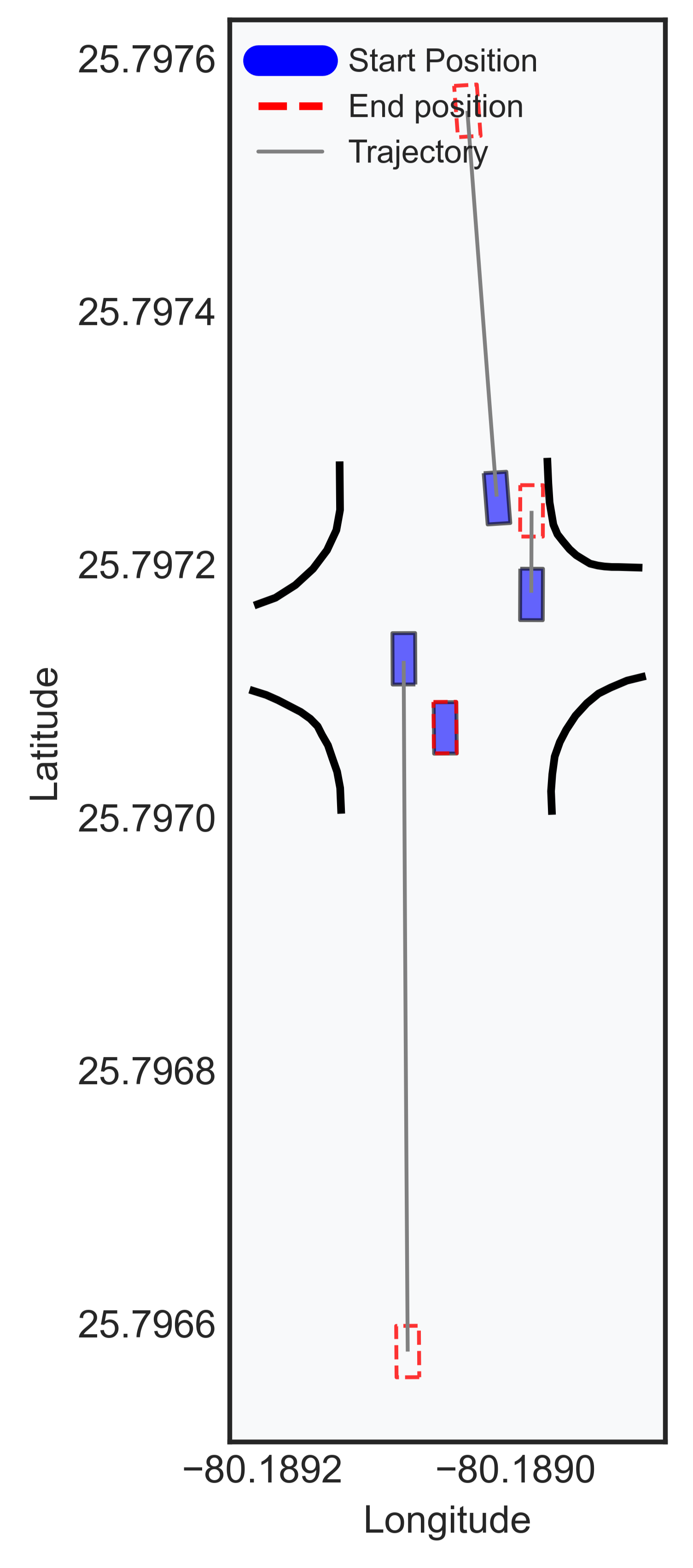}}
    \caption{Representative intersection-level near-miss detection using the 2D-TTC framework: (a) V–V and (b) V–I}
    \label{fig:6}
\end{figure}

Because multiple vehicles are present in a single 11s scenario, multiple interaction windows occur simultaneously. We therefore distinguish between (i) candidate near-miss detected across vehicles and frames within a scenario and (ii) the single extreme block retained for EVT per interaction window. Defining blocks by interaction windows also avoids repeatedly selecting adjacent frame-wise minima from the same persistent interaction, which would overweight long-lasting conflicts and amplify temporal dependence. For each retained block, extreme near-miss events are identified by forward-simulating vehicle trajectories under the geometric and traffic conditions summarized in Table~\ref{tab:4}. Trajectories are projected over a 3s horizon, and spatial overlaps with surrounding traffic or roadway boundaries are evaluated as potential near-miss events. To focus the BM sample on genuinely unsafe edge-encroachment behavior, we retain V–I encounters only when the interaction window contains short 2D-TTC values with moderate-to-high approach speeds, filtering out routine low-speed turning proximity that can appear as boundary approach under constant-input projection. Representative outcomes from this process are shown in Fig.~\ref{fig:6}, illustrating the spatial dynamics of predicted V–V and V–I near-misses within the Miami study corridor. These event visualizations provide the empirical basis for block-level 2D-TTC extraction across the 10 intersections (I1–I10), enabling localized evaluation of crash-prone conditions. Finally, all microscopic covariates (e.g., relative speed, relative acceleration/deceleration, jerk, heading/steering differences) are extracted from the same interaction window and aligned to the extreme event, either evaluated at the frame where the extreme 2D-TTC occurs.

\begin{table}[h!]
\centering
\caption{Summary statistics for V–V and V–I near-misses at Intersections}
\label{tab:5}
\renewcommand{\arraystretch}{1}
\resizebox{\textwidth}{!}{%
\begin{tabular}{lcccccccc}
\hline
\textbf{Metric} & \multicolumn{4}{c}{\textbf{V–V}} & \multicolumn{4}{c}{\textbf{V–I}} \\
\cline{2-5} \cline{6-9}
& Avg & Min & Max & Std & Avg & Min & Max & Std \\
\hline
\multicolumn{2}{l}{Continuous variables} \\
\hline
TTC (s) & 1.617 & 0.11 & 2.99 & 0.863 & 1.304 & 0.10 & 2.99 & 0.889 \\
Relative speed (m/s) & 5.257 & 0.00 & 23.724 & 4.446 & 8.753 & 0.00 & 22.276 & 4.035 \\
Relative acc./dec. (m/s$^2$) & -0.535 & -13.706 & 10.343 & 2.227 & 0.161 & -5.925 & 13.199 & 0.702 \\
Relative distance (m) & 11.673 & 2.419 & 46.857 & 6.844 & 1.520 & 0.60 & 3.80 & 1.110 \\
Jerk ($\mathrm{m/s^3}$) & -5.550 & -13.609 & 9.315 & 2.652 & -7.900 & -10.800 & 2.800 & 14.700 \\
Heading difference (rad) & 1.077 & 0.00 & 6.251 & 1.410 & -0.045 & -3.138 & 3.141 & 1.647 \\
Steering difference (rad)  & 0.409 & 0.00 & 3.142 & 0.666 & -0.006 & -1.571 & 1.571 & 0.130 \\
Traffic volume (veh) & 13.462 & 2.00 & 31.00 & 5.969 & 10.461 & 1.00 & 31.00 & 5.392 \\
Vehicle length (m) & 4.78   &  3.18 &   6.92 & 0.49 & 5.1 &   3.95 & 12.65&  1.30\\
Vehicle width (m) & 2.22   &  1.76 &   3.16 & 0.30 & 2.04 &   1.36 & 3.08&  0.23\\
\hline
\multicolumn{9}{l}{Categorical variable: Near-miss location (Proportion \%)} \\
\hline
Through & & & 72.29  & & & 0.00 & & \\
Left Turn & & &  25.99 & & & 46.08  & & \\
Right Turn & & &8.90  & & & 53.92  & & \\
\hline
\end{tabular}
}
\end{table}

Descriptive statistics for intersection-level V–V and V–I near-misses are reported in Table \ref{tab:5}. The continuous indicators include 2D-TTC and interaction dynamics (relative speed, relative acceleration/deceleration, relative distance, jerk, and heading/steering differences), along with vehicle dimensions and exposure (volume). Mean 2D-TTC values are approximately 1–2s (V–V: 1.62s; V–I: 1.30s), while the most safety-critical cases lie in the lower tail, with minima near 0.10–0.11s. Table further quantifies the lower-tail structure of 2D-TTC, showing that crash-prone near-miss risk is concentrated in a narrow extreme range. To ensure that extremely small 2DTTC values from geometry-oriented rectangle footprints (e.g., spurious corner overlap), the corresponding scenarios were manually reviewed by visualizing the time-evolving oriented bounding boxes and headings; the critical cases reflected plausible relative motion and geometry rather than numerical overlap errors. V–I near-misses tend to exhibit higher relative speeds and smaller distance gaps, consistent with abrupt boundary approaches, whereas V–V near-misses show larger heading/steering deviations and higher jerk variability, consistent with unstable maneuvers (e.g., abrupt turns or evasive swerves). Categorical summaries further highlight differences: most V–V events occur during through movements, whereas V–I near-misses are concentrated in lateral/turning maneuvers, consistent with edge approach under turning and merge dynamics.

\begin{figure}[h]
    \centering
    \setlength{\abovecaptionskip}{0pt}
    \subcaptionbox{V-V \label{fig:7vv}}
    {\rotatebox{270}{\includegraphics[width=0.25\textwidth]{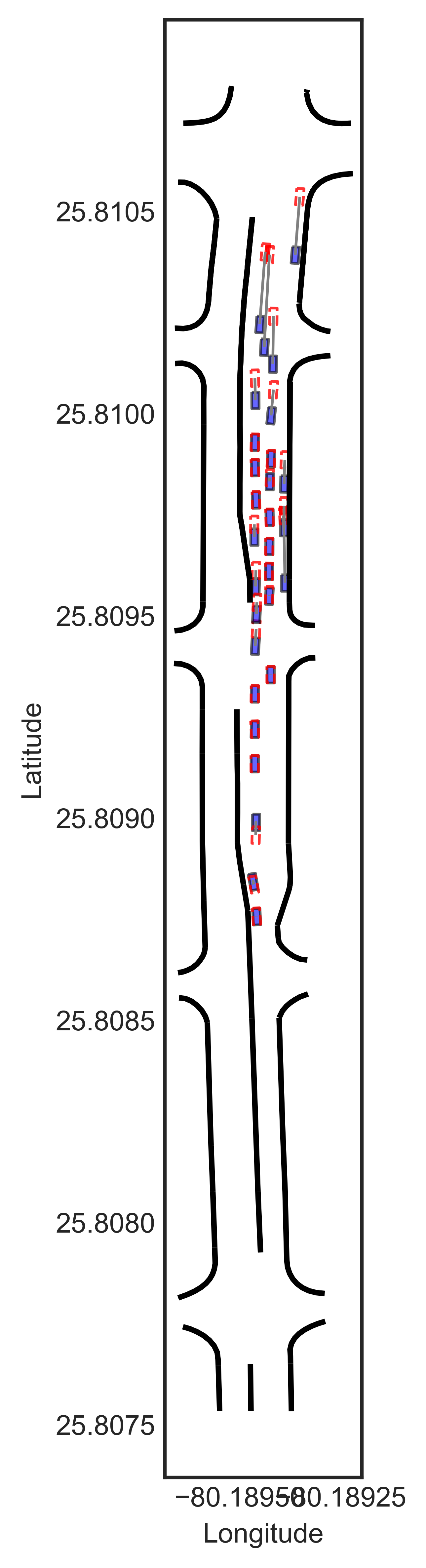}}}
    \hspace{5mm}
    \subcaptionbox{V-I \label{fig:S_VI}}
    {\rotatebox{270}{\includegraphics[width=0.25\textwidth]{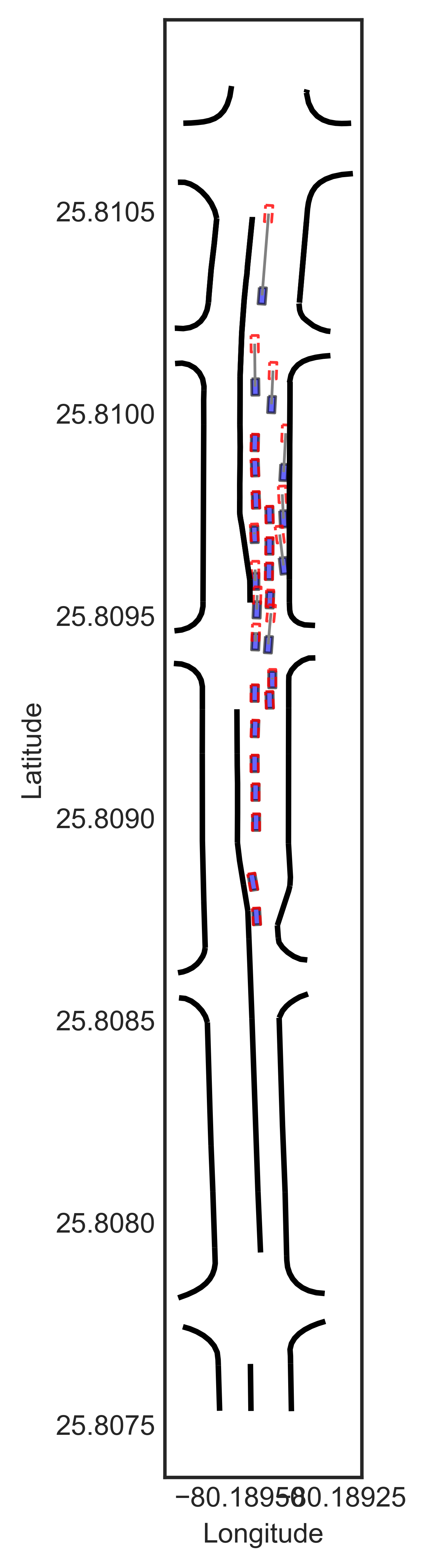}}}
    \caption{Representative segment-level near-miss detection using the 2D-TTC framework: (a) V–V and (b) V–I}
    \label{fig:7}
\end{figure}

The analysis was also conducted for directional segments (S1–S9) using the same 2D-TTC pipeline (See Fig. \ref{fig:7}). Table~\ref{tab:6} reports segment-level descriptive statistics. For V–V near-misses, mean TTC values are longer than at intersections (1.83 s), suggesting larger headways in midblock traffic, while very small TTC minima and high relative-speed maxima indicate occasional aggressive overtaking and rapid convergence. Relative distances span a wider range than at intersections, reflecting more dispersed longitudinal interactions in segment settings.

\begin{table}[h!]
\centering
\caption{Summary statistics for V–V and V–I near-misses at Segments}
\label{tab:6}
\renewcommand{\arraystretch}{1}
\resizebox{\textwidth}{!}{%
\begin{tabular}{lcccccccc}
\hline
Metric & \multicolumn{4}{c}{V–V} & \multicolumn{4}{c}{V–I} \\
\cline{2-5} \cline{6-9}
& Avg & Min & Max & Std & Avg & Min & Max & Std \\
\hline
\multicolumn{2}{l}{Continuous variables} \\
\hline
TTC (s) & 1.829 & 0.11 & 2.99 & 0.805 & 1.272 & 0.10 & 2.99 & 0.912 \\
Relative speed ($\mathrm{m/s}$) & 4.513 & 0.00 & 33.909 & 4.278 & 10.262 &  0.000  &28.067  &4.212 \\
Relative acc./dec. (m/s$^2$) & -0.871   &-23.014 &   18.336&    3.139 & 0.164& -14.067&  23.343&  1.400 \\
Relative distance (m) & 15.361&    1.931&   141.548&   12.773 & 1.720 & 0.32 & 6.56 & 1.02 \\
Jerk ($\mathrm{m/s^3}$) & -7.121 & -17.502 & 8.423 & 1.552 & -5.70 & -12.72 & 3.556 & 4.77 \\
Heading difference (rad) & 0.368 &    0.000 &    6.256 &   0.990 & -0.032&  -3.102&   3.138&  1.586 \\
Steering difference (rad) & 0.577  &   0.000&     3.142&    0.756 & -0.001&  -1.571 &  1.571 & 0.167 \\
Traffic volume (veh) & 16.801     &1.000   & 65.000   &11.009 &  12.816   &1.000  &65.000  &7.906 \\
Vehicle Length (m) & 4.74   &  3.0 &   7.85 & 1.11 & 4.79 &   2.95 & 8.65&  0.87\\
Vehicle width (m) & 2.12   &  1.36 &   3.08 & 0.23 & 2.22 &   1.76 & 4&  0.18\\
Driveway intensity ($\mathrm{m^{-1}}$)   &0.009     &0.000     &0.015    &0.005& 0.009     &0.000     &0.015    &0.005\\
Lane width (m) & 3.28   &  2.99 &   3.96 &  0.14 & 3.27 &  2.99 &   3.96& 0.141\\
Lane number & 2.73   &  2 &   4 &  0.578 & 2.51&   2.0&   4.0&  0.52\\
\hline
\multicolumn{9}{l}{Categorical variable:  (Proportion \%)} \\
\hline
& Through && 58.55 & & &  0.00 & &  \\
Collision location& Left Turn && 9.52 & & &  46.08 & &  \\
& Right Turn && 31.93 &  & & 53.92 & &  \\
Median type& Divided && 23.60 & & &  13.98& &  \\
& Undivided && 76.4 & & &  86.02 & &  \\
\hline            
\end{tabular}
}
\end{table}

For V–I near-misses, the mean TTC was shorter with greater relative speeds, underscoring the abrupt nature of vehicle–infrastructure encounters. Compared with intersections, segment-level V–I near-misses tend to occur at longer approach distances but exhibit greater variability in jerk and heading, consistent with lateral drift or corrective steering as vehicles move toward lane edges, curbs, or medians. These events frequently cluster near driveways and minor access points, where side-entry activity and path adjustments increase the likelihood of encroachment. Because the V–I definition is restricted to same-carriageway curbs/medians and the retained encounters are filtered by 2D-TTC, the resulting V–I sample reflects elevated boundary-encroachment risk rather than generic turning or lane-change exposure. Segment-level access exposure is captured by driveway intensity, defined as the density of driveways and minor access points per unit segment length. The categorical breakdown reinforces these patterns. For V–V events, nearly 59\% of near-misses were associated with through movements, while about 32\% and 10\% were linked to right and left turns, respectively. In contrast, V–I near-misses were dominated by turning maneuvers, with more than half (53.9\%) linked to right turns and 46\% to left turns, and no near-misses detected during through movements. This distribution highlights the heightened role of lateral maneuvers in producing infrastructure-related risks. Furthermore, median type also influenced outcomes: 76\% of V–V near-misses and 86\% of V–I near-misses occurred along undivided sections, underscoring the vulnerability of open medians and flush-paved designs in segment environments.

Taken together, these segment-level results indicate that, while intersections concentrate tightly constrained V–V conflicts, segments exhibit distinct risk mechanisms associated with higher-speed interactions and boundary encroachments. This supports treating intersections and segments as functionally different safety domains when modeling near-miss dynamics and designing countermeasures.

\begin{figure}[h]
    \centering
    \setlength{\abovecaptionskip}{0pt}
    \subcaptionbox{Intersection\label{a}}
    {\includegraphics[width=0.6\textwidth]{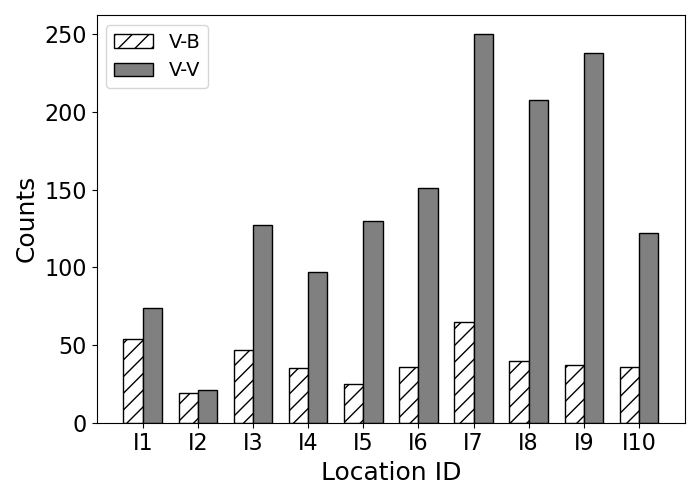}}
    \hspace{5mm}
    \subcaptionbox{Segment: NB\label{b}}
    {\includegraphics[width=0.45\textwidth]{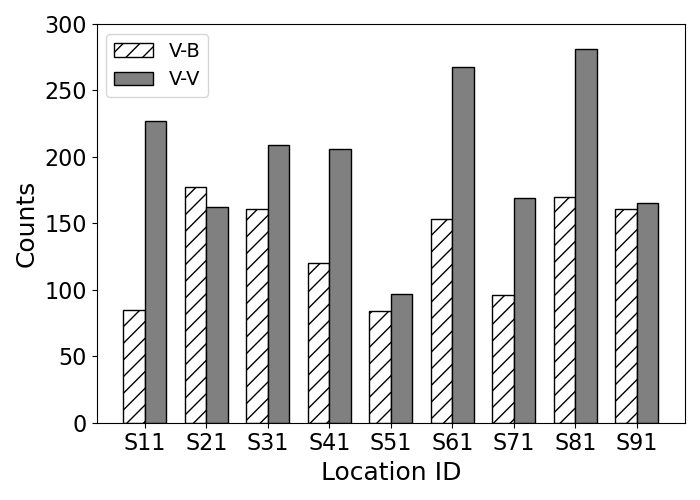}}
    \hspace{5mm}
    \subcaptionbox{Segment: SB\label{c}}
    {\includegraphics[width=0.45\textwidth]{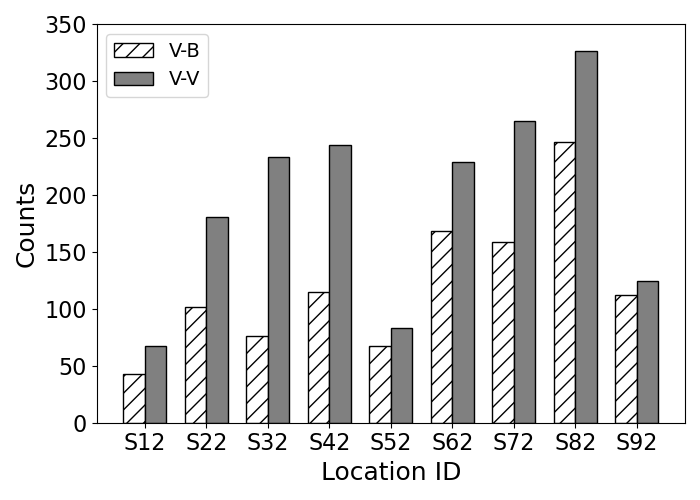}}
    \caption{BM near-miss frequencies by location type and direction}
    \label{fig:8}
\end{figure}

After extracting block extremes per valid interaction window, the BM extremes are aggregated by location and summarized in Fig.~\ref{fig:8}. The figure reports extreme near-miss frequencies across 28 locations. To contextualize exposure, Tables~\ref{tab:5}–\ref{tab:6} also report frame-level volumes, which varied across scenarios. Because the ego-centric LiDAR has a limited sensing range, the observable traffic volume varies across frames and interaction windows and is therefore used as a local exposure measure to contextualize 2DTTC-based extremes. The study extends prior EVT-based near-miss work by jointly modeling V–V and V–I extremes under the HBSGRP–UGEV framework and incorporating roadway geometry and infrastructure context alongside vehicle dynamics. In addition to kinematic predictors emphasized in earlier studies (\citep{desai2021correlating, jun2007relationships, tak2015development}), we include geometric covariates, such as lane configuration and boundary context; for example, steering-related measures capture abrupt lateral maneuvers associated with boundary incursions (\citep{gilbert2021multi}). Fig.~\ref{fig:8} also reveals clear spatial patterns: opposite-direction segments (S82, S72, S62) show the highest V–I extreme frequencies, while primary-direction segments (S11, S21, S81) exhibit elevated V–V frequencies consistent with midblock longitudinal interactions. Among intersections, I7–I9 emerge as hotspots for both near-miss types, reflecting the compounded operational complexity of signal control, turning movements, and interaction density.

Overall, the distribution of block extremes supports the BM approach in capturing rare but safety-critical events. Nearly all sites meet or exceed the commonly cited minimum of 30 observations for GEV-based inference (\citep{zheng2014freeway}), except for I2, which provides a statistically adequate basis for HBSGRP–UGEV estimation. The observed heterogeneity across directions and facility types further motivates the inclusion of direction and location-specific effects to enable context-sensitive, corridor-scale crash-risk estimation.

\section{Modeling estimation result}\label{5}

This study develops a set of Bayesian hierarchical UGEV models to estimate corridor-wide crash occurrence risk (COR) from extreme 2D-TTC near-miss events along Biscayne Boulevard in Miami. Extreme near-misses are extracted using a BM strategy; each V–V or V–I defines a block extreme used for EVT inference. Because extreme events are sparse at individual sites, direct site-by-site estimation can yield unstable posteriors and wide credible intervals. To improve statistical stability while preserving spatial structure, we adopt a corridor-wide grouped formulation. Each site defines one group $k$, and all block extremes observed at that site are assigned to the corresponding group. This enables partial pooling through corridor-level hyperparameters while allowing selected effects to vary across the 28 groups.

The HBSGRP–UGEV model is the primary analytical framework, with a simpler fixed-effect hierarchical model (HBSFP) also estimated for comparison. For each near-miss type (V–V and V–I), two models were therefore developed: (1) HBSFP and (2) HBSGRP. The grouped random-parameters formulation in HBSGRP allows selected coefficients to vary across spatial clusters, thereby capturing latent heterogeneity in interaction dynamics and roadway characteristics. To compare alternative model specifications, we relied on several Bayesian information criteria commonly used in roadway safety research. The Deviance Information Criterion (DIC) (\citep{spiegelhalter2002bayesian, el2012measuringa, el2012measuringb}) is frequently reported as a summary measure of model fit, with lower values indicating superior predictive fit (\citep{zheng2019bayesian}). However, prior studies have demonstrated that its numerical value can be affected by the model's parameterization (e.g., \citep{geedipally2012negative,geedipally2014caution}). Align with earlier safety studies, a difference greater than five DIC units was taken to indicate a meaningfully better fit between competing models (\citep{el2012measuringa, el2012measuringb}). To strengthen the assessment of predictive performance and reduce sensitivity to parameterization,  additionally report the Widely Applicable Information Criterion (WAIC) and the Leave-One-Out Cross-validation Information Criterion (LOOIC), which have been recommended as more robust alternatives to DIC for Bayesian model comparison (\citep{vehtari2017practical}) and have seen growing use in recent roadway safety applications (\citep{dzinyela2024negative, islam2023grouped, khodadadi2023evaluating}).

Estimation was conducted using MultiBUGS, a parallelized extension of OpenBUGS tailored for high-dimensional Bayesian inference. Its support for block updates and parallel execution makes it well-suited for fitting hierarchical models with grouped random effects and extreme-value behavior within complex data structures. Parameters were estimated by Markov Chain Monte Carlo (MCMC) simulation; two parallel MCMC chains of 50000 iterations were run, with the first 20000 discarded as burn-in and the remaining 30000 retained for posterior inference. Convergence was assessed both visually (trace plots) and quantitatively (Brooks–Gelman–Rubin diagnostic), with all monitored parameters achieving BGR values below 1.1, confirming reliable convergence (\citealp{gelman1992inference, brooks1998general, el2009urban}). This procedure yielded robust posterior summaries of means, standard deviations, and 95\% credible intervals, even under correlation structures induced by spatial clustering and random effects.

The modeling framework incorporated a comprehensive set of covariates into the location ($\mu$) and scale ($\sigma$) parameters of the nonstationary UGEV distribution to capture contextual and behavioral variability in COR. For both V–V and V–I interactions, dynamic covariates included relative (speed, distance, acceleration, deceleration, headway, steering angle), jerk, and traffic volume, thereby capturing multiple dimensions of vehicle dynamics. Geometric and infrastructure-related covariates were also included: lane number, lane width, driveway intensity, median type, and vehicle lane position. 

All covariates were specified in both the $\mu$ and $\sigma$ components, enabling the models to capture variation in both the central tendency and dispersion of extreme 2D-TTC. In contrast, the shape parameter ($\xi$) was held constant across all locations. This decision reflects a tradeoff between flexibility and stability: although $\xi$ governs tail heaviness in the UGEV distribution, it is notoriously difficult to estimate from sparse BM samples. Allowing $\xi$ to vary with covariates often induces instability and overfitting (\citep{coles2001introduction, cooley2006bayesian}), whereas fixing it improves convergence and yields more interpretable posterior distributions.

\subsection{Posterior results and model performance}

Tables~\ref{tab:7}–\ref{tab:10} report the posterior summaries for both HBSFP and HBSGRP models, estimated separately for V–V and V–I near-miss events at intersections and segments. Across all cases, the HBSGRP structure consistently outperforms the HBSFP baseline, indicating a better ability to capture unobserved heterogeneity and location-level variability. For the HBSGRP models, the DIC values are 508.9 (V–V) and 3179.6 (V–I) at intersections and 4189.0 (V–V) and 6550.0 (V–I) at segments. These correspond to relative reductions of 3.41\% and 0.54\% at intersections, and 7.49\% and 3.11\% at segments, compared with their HBSFP counterparts. Such improvements exceed the widely accepted threshold of five DIC units, indicating substantively better model fit (\citealp{el2012measuringa, el2012measuringb}). To further evaluate robustness, WAIC and LOOIC were also computed, both of which produced rankings consistent with DIC, thereby reinforcing the HBSGRP's superior performance. By allowing selected coefficients to vary across spatial groups, the HBSGRP models capture latent heterogeneity in both the central tendency and dispersion, a capability absent in fixed-effect structures. Collectively, these results demonstrate that grouped random parameters models provide a more flexible and accurate framework for modeling COR dynamics in complex urban traffic environments.

\subsubsection{Intersection-level} \label{Int-HBSGRP-UGEV}
The intersection-level HBSGRP–UGEV model was designed to capture heterogeneity in extreme 2D-TTC from V–V interactions at signalized intersections. In an initial specification, all covariates were allowed to vary across site groups; however, the final random-parameter set was determined using a theory-first rationale and then empirically confirmed and parsimonious. Specifically, random effects were retained for the intercept and two varying-dynamics covariates, relative (speed and distance). These quantities directly govern the short-horizon evolution of 2D-TTC and are expected to vary across intersections with signal phasing, approach geometry, queue discharge conditions, and driver compliance. In contrast, other vehicle dynamics (e.g., relative acceleration/deceleration, jerk, heading/steering differences) are typically more episodic and noise-sensitive at high frequency; once closing dynamics are accounted for, their between-site variation was not consistently identifiable and contributed limited additional explanatory power.

\begin{figure}[htbp]
    \centering

    \begin{subfigure}[b]{0.45\textwidth}
        \includegraphics[width=\linewidth]{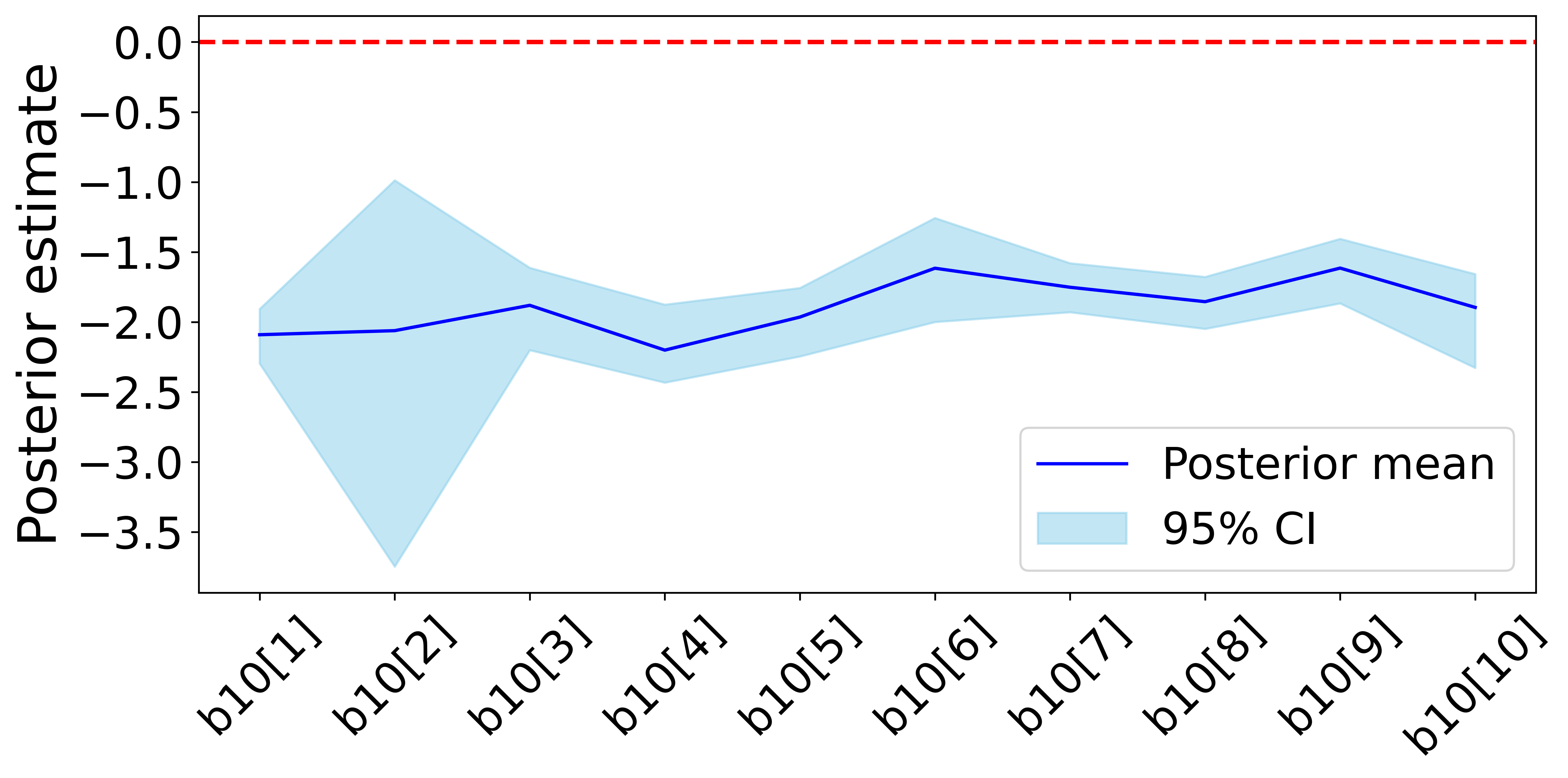}
        \caption{$\beta_{\mu,0[1,10]}$}
        \label{fig:b10}
    \end{subfigure}
    \hspace{0.02\textwidth} 
    \begin{subfigure}[b]{0.45\textwidth}
        \includegraphics[width=\linewidth]{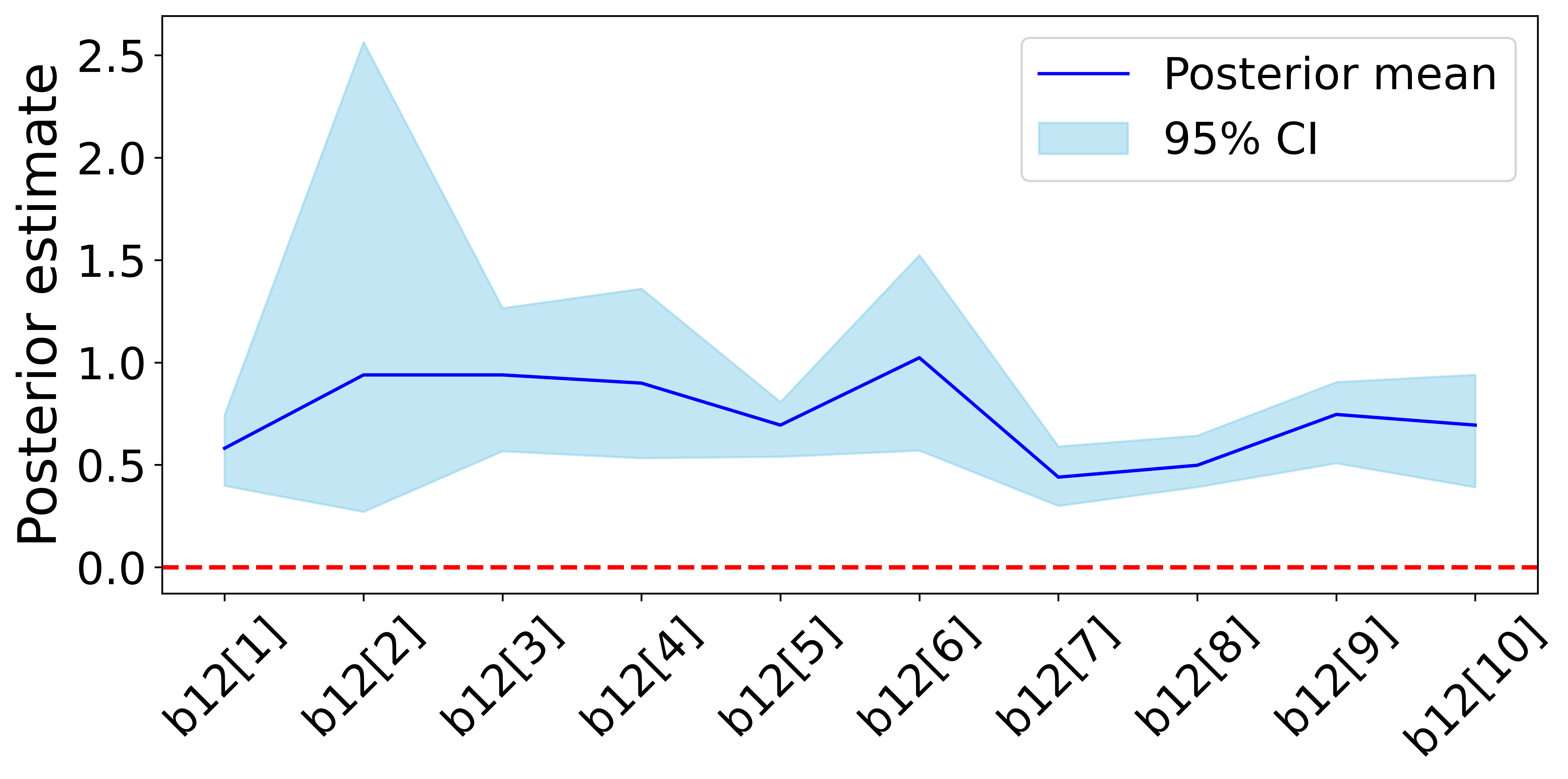}
        \caption{$\gamma_{\mu,1[1,10]}$}
        \label{fig:b12}
    \end{subfigure}

    \vspace{0.1cm}
    \begin{subfigure}[b]{0.45\textwidth}
        \includegraphics[width=\linewidth]{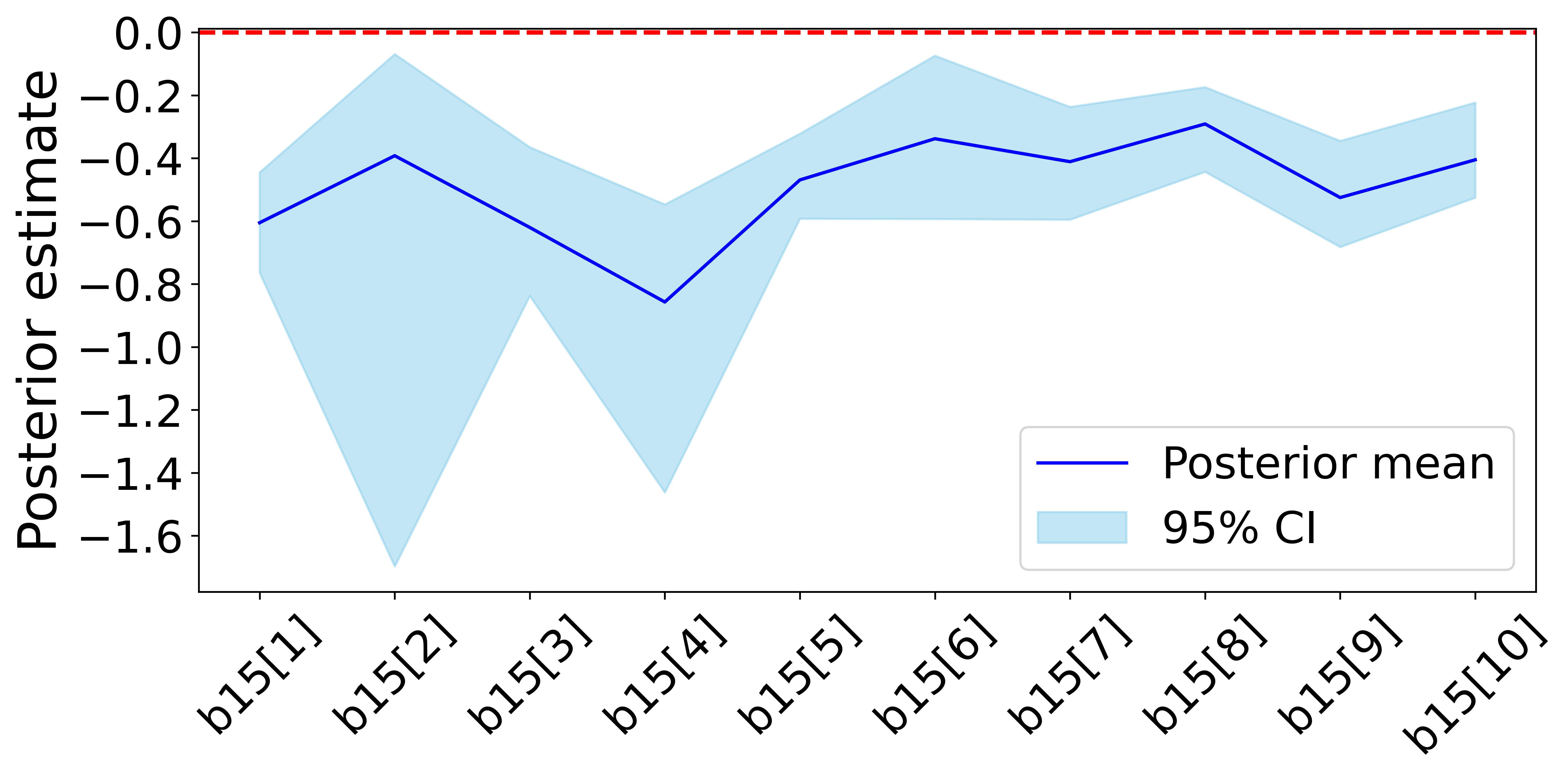}
        \caption{$\gamma_{\mu,2[1,10]}$}
        \label{fig:b15}
    \end{subfigure}
    \hspace{0.02\textwidth} 
    \begin{subfigure}[b]{0.45\textwidth}
        \includegraphics[width=\linewidth]{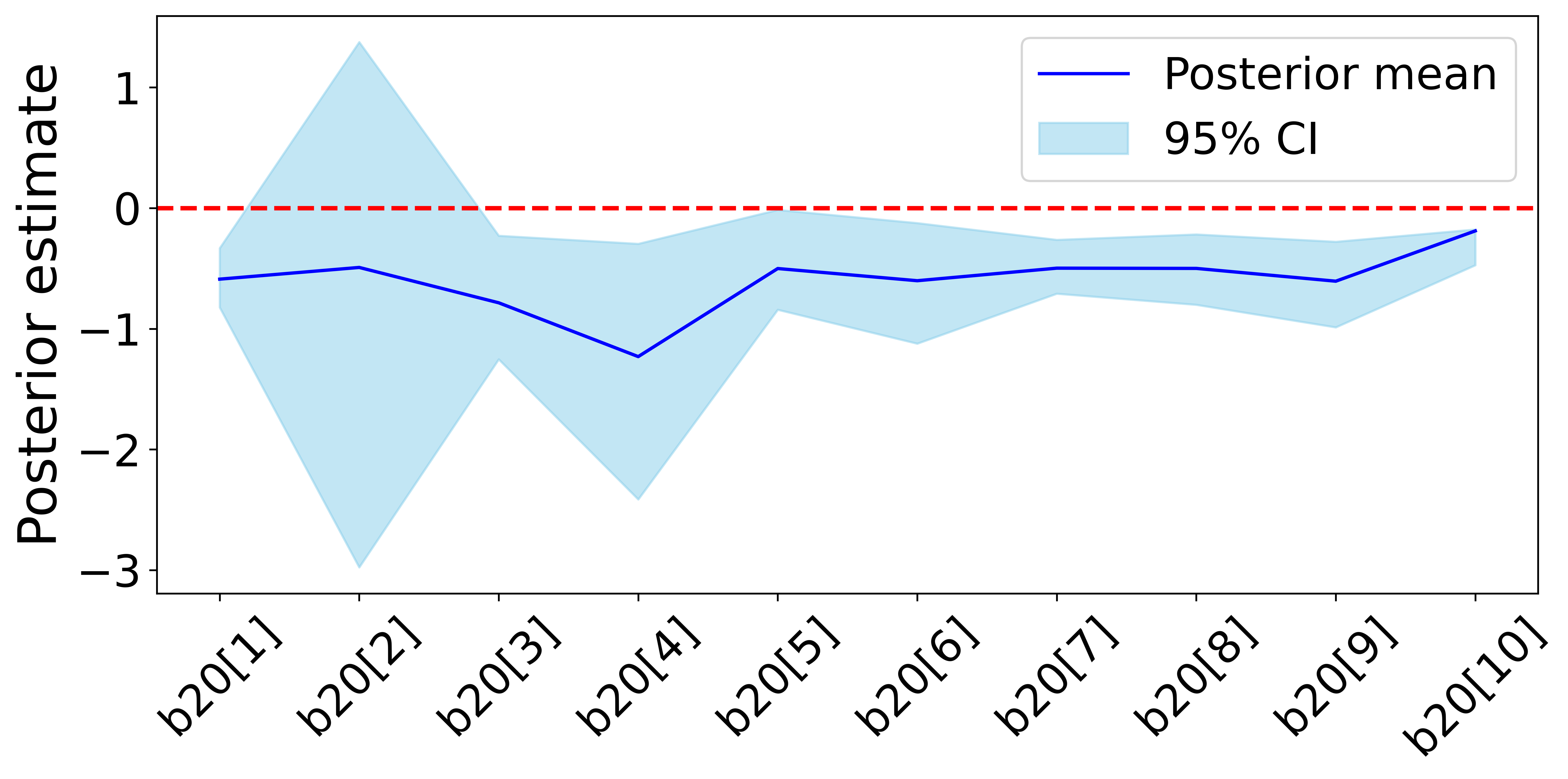}
        \caption{$\beta_{\sigma,0[1,10]}$}
        \label{fig:b20}
    \end{subfigure}

    \vspace{0.1cm}
    \begin{subfigure}[b]{0.47\textwidth}
        \includegraphics[width=\linewidth]{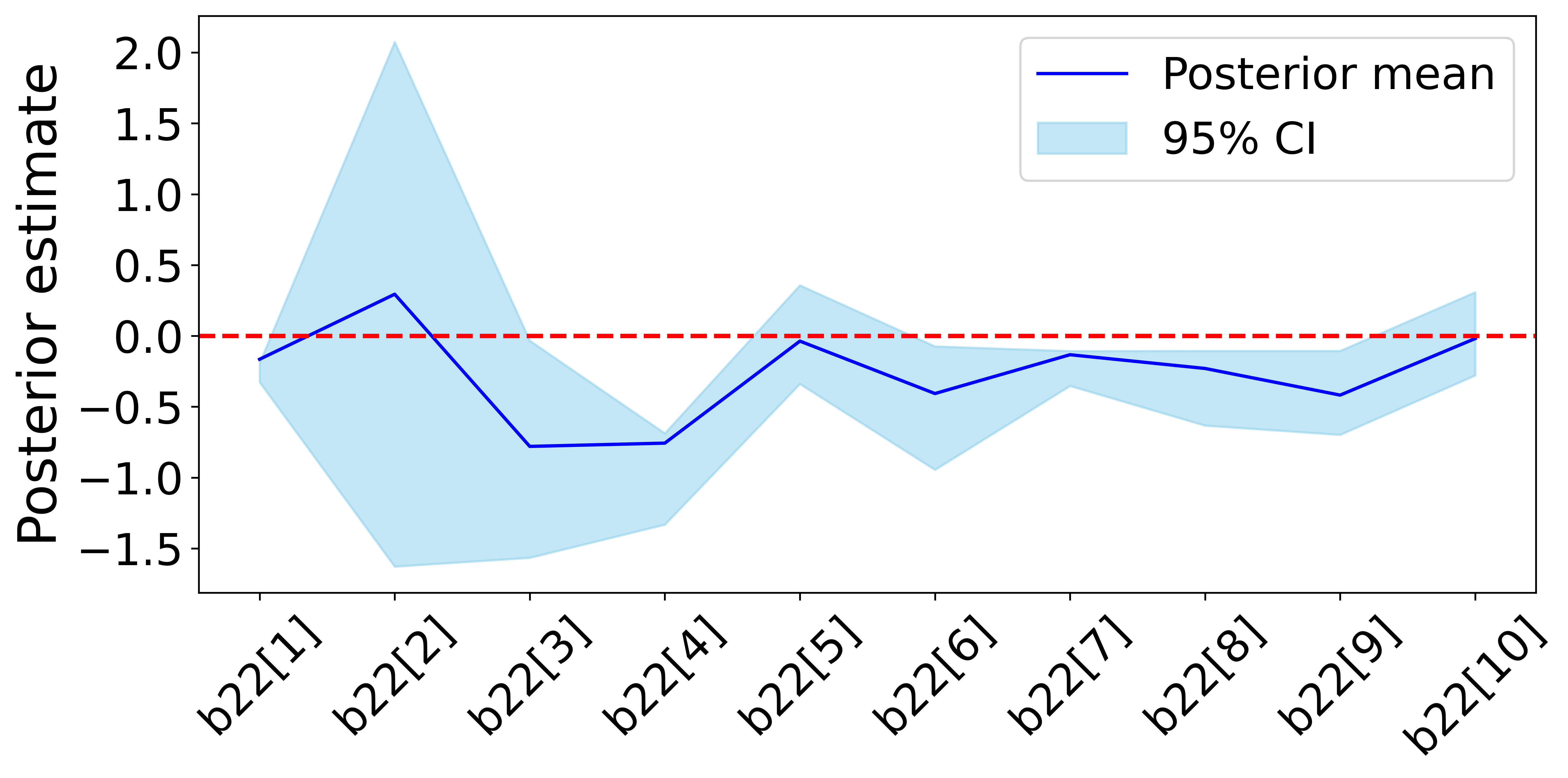}
        \caption{$\gamma_{\sigma,1[1,10]}$}
        \label{fig:b22}
    \end{subfigure}
    \hspace{0.02\textwidth} 
    \begin{subfigure}[b]{0.45\textwidth}
        \includegraphics[width=\linewidth]{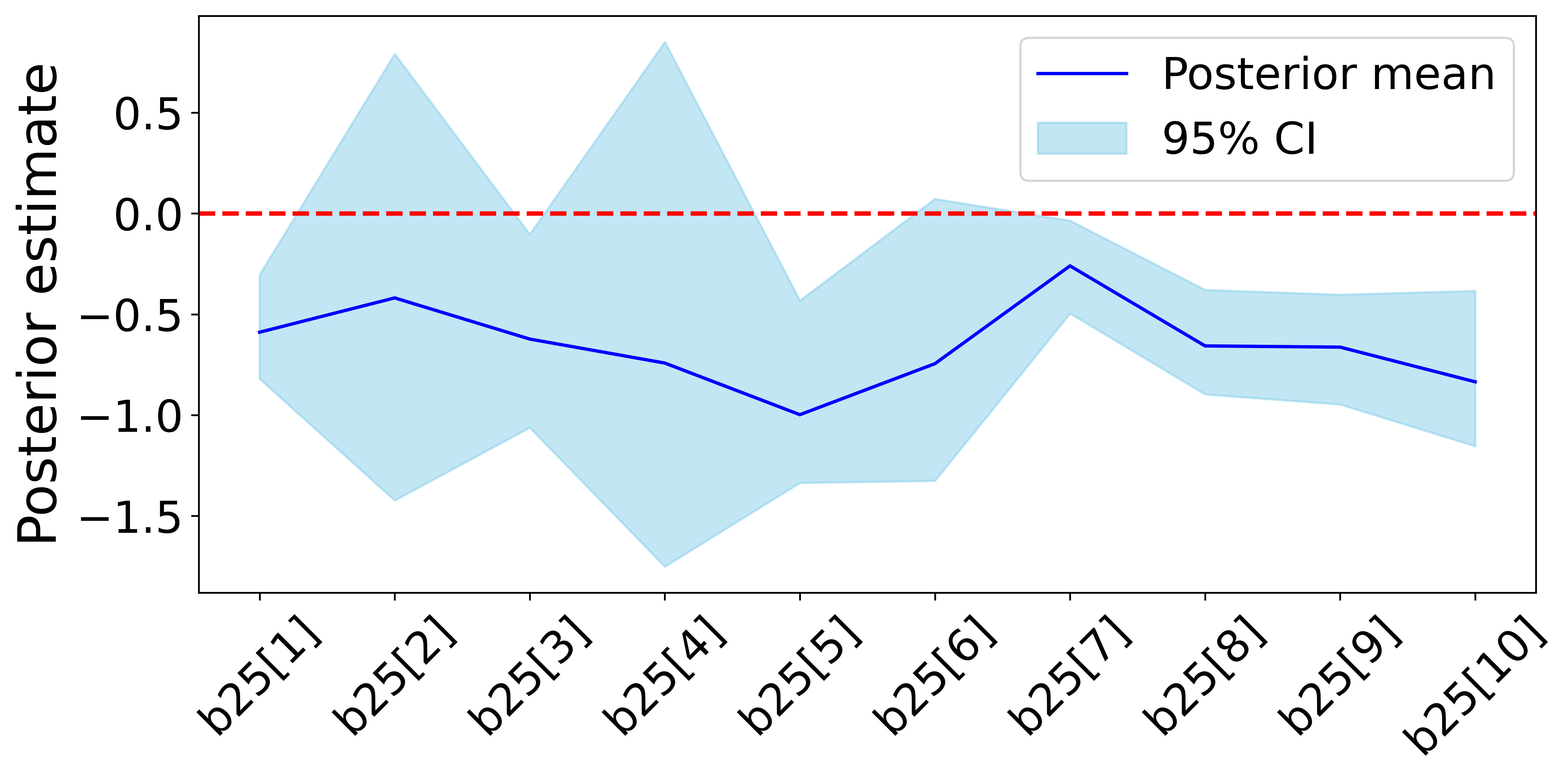}
        \caption{$\gamma_{\sigma,2[1,10]}$}
        \label{fig:b25}
    \end{subfigure}

    \vspace{0.1cm}
    \begin{subfigure}[b]{0.45\textwidth}
        \includegraphics[width=\linewidth]{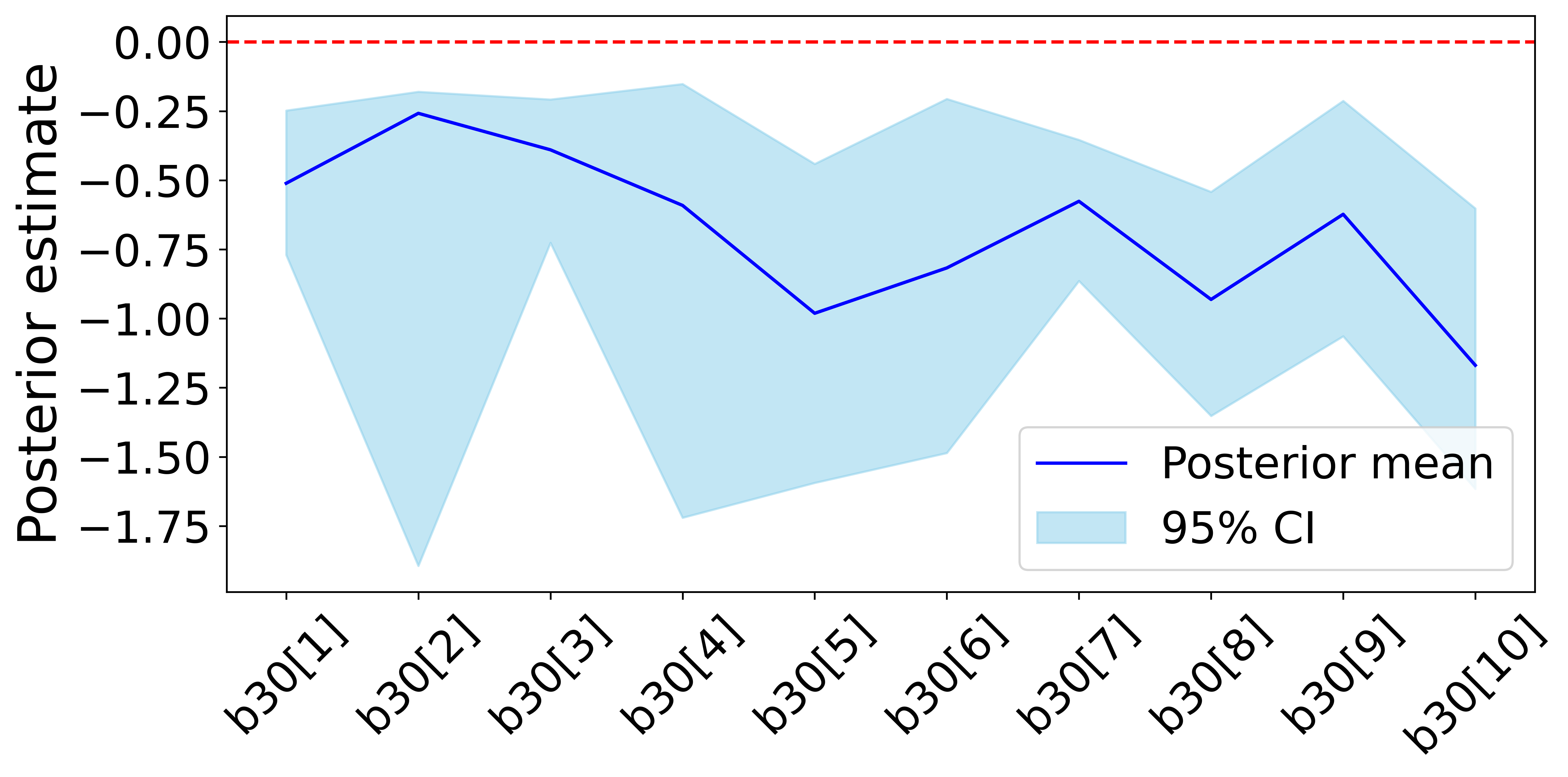}
        \caption{$\beta_{\xi,0[1,10]}$}
        \label{fig:b30}
    \end{subfigure}

    \caption{Posterior estimates for V-V near-misses at Intersections}
    \label{fig:9}
\end{figure}

To avoid overfitting and preserve interpretability, restricted random effects to this minimal, theoretically motivated subset and relied on hierarchical partial pooling to shrink site-specific coefficients toward corridor-level means when information was limited. Candidate random-effects structures were compared using posterior stability (non-degenerate random-effect variances), predictive information criteria, and posterior predictive checks; covariates with random-effect variances concentrated near zero or unstable posteriors were retained as fixed effects (\citep{anis2025pedestrian}). The resulting specification captures intersection-to-intersection variability in closing behavior while maintaining a parsimonious, generalizable structure; group-specific random effects are illustrated in Fig.~\ref{fig:9}, with posterior summaries reported in Table~\ref{tab:7}.

\begin{table}[htbp]
\centering
\caption{Posterior estimates for V-V near misses at Intersections}
\label{tab:7}
\scriptsize
\begin{threeparttable}
\resizebox{\textwidth}{!}{
\begin{tabular}{l l l c c c c c c}
\hline
\textbf{Model Parameters} & \textbf{Hyperparameter} & \textbf{Covariate} & \multicolumn{3}{c}{\textbf{HBSFP}} & \multicolumn{3}{c}{\textbf{HBSGRP}} \\
\cline{4-9}
 & & & Mean & SD\tnote{a} & 95\% CRI\tnote{b} & Mean & SD\tnote{a} & 95\% CRI\tnote{b} \\
\hline
\multicolumn{9}{l}{\textit{Location Parameter} ($\mu_{k,i}$)} \\
\hline
$\beta_{\mu,0}$ & Fixed & Intercept & -1.870 & 0.039 & $[-1.917, -1.811]^{\dagger}$ & - & - & -\\
$\beta_{\mu,0,k}$ & Random & Intercept & - & - & -&-1.893&	0.186&	$[-2.309,	-1.572]^{\dagger}$ \\
$\gamma_{\mu,1,k}$ & Random & Relative speed    & - & - & -& 0.745&	0.181&	$[0.393,	1.134]^{\dagger}$\\
$\gamma_{\mu,2,k}$ &Random&  Relative distance & - & - & -& -0.491	&0.157&	$[-0.819, -0.280]^{\dagger}$\\
$\beta_{\mu,1}$ & Fixed& Relative speed & 0.639 & 0.050 & $[0.246, 0.739]^{\dagger}$ & - & - & - \\
$\beta_{\mu,2}$ & Fixed&  Relative acceleration  & 1.064&0.074&$[0.126, 2.018]^{\dagger}$& -0.155&	0.032&	$[-0.221,	-0.093]^{\dagger}$\\
$\beta_{\mu,3}$ & Fixed&Relative deceleration & -0.013 & 0.037 & [-0.088, 0.059]&0.140&	0.063&	$[0.004, 0.238]^{\dagger}$\\
$\beta_{\mu,4}$ &Fixed&  Relative distance & -0.442 & 0.036 & $[-0.505, -0.418]^{\dagger}$ &-&-&-\\
$\beta_{\mu,5}$ &Fixed& Jerk & -0.012 & 0.006 & $[-0.021, -0.002]^{\dagger}$ & -0.027 & 0.015 & $[-0.046, -0.006]^{\dagger}$ \\
$\beta_{\mu,6}$ & Fixed& Heading difference & 0.024 & 0.026 & [-0.029, 0.072]& 0.028&	0.039&	[-0.052,	0.097]\\
$\beta_{\mu,7}$ & Fixed& Steering difference &  -0.004 & 0.049 & [-0.096, 0.092]& 0.059&	0.026&	$[0.003,	0.107]^{\dagger}$\\
$\beta_{\mu,8}$ & Fixed&  Volume & -0.003 & 0.005 & [-0.012, 0.006]&-0.018	&0.035&	[-0.085, 0.051]\\
$\beta_{\mu,9}$ & Fixed& Turn movement (Left =1, else=0)  &0.086 & 0.026 & [-0.102, 0.242]&0.073&	0.011&	$[0.042,	0.205]^{\dagger}$\\
\hline
\multicolumn{9}{l}{\textit{Scale Parameter} ($\log \sigma_{k,i}$)} \\
\hline
$\beta_{\sigma,0}$ &Fixed&Intercept & -0.571 & 0.058 & $[-0.680, -0.455]^{\dagger}$& - & - & -\\
$\beta_{\sigma,0,k}$ & Random & Intercept & - & - & -& -0.599&	0.311&	$[-1.240,	-0.056]^{\dagger}$  \\
$\gamma_{\sigma,1,k}$ & Random &  Relative speed  & - & - & -&-0.265	&0.254&	$[-0.810,	-0.043]^{\dagger}$\\
$\gamma_{\sigma,2,k}$ &Random &  Relative distance & - & - & -&-0.653	&0.297&	$[-1.121,	-0.033]^{\dagger}$\\
$\beta_{\sigma,1}$ & Fixed&  Relative speed  & 0.012 & 0.084 & [-0.139, 0.183]&-&-&-\\
$\beta_{\sigma,2}$ & Fixed& Relative acceleration &-1.686 & 0.865 & $[-3.066, -0.627]^{\dagger}$&-0.226&	0.066&	$[-0.360,	-0.108]^{\dagger}$\\
$\beta_{\sigma,3}$ &Fixed& Relative deceleration & -0.218 & 0.051 & $[-0.321, -0.120]^{\dagger}$&-0.253&	0.053&	$[-0.356,	-0.151]^{\dagger}$\\
$\beta_{\sigma,4}$ & Fixed& Relative distance & -0.062 & 0.008 & $[-0.078, -0.046]^{\dagger}$ & - & - & - \\
$\beta_{\sigma,5}$ & Fixed& Jerk & 0.019 & 0.009 & $[0.005, 0.030]^{\dagger}$ & 0.012 & 0.003 & $[0.007, 0.018]^{\dagger}$\\
$\beta_{\sigma,6}$ & Fixed& Heading difference &-0.063&	0.071&	[-0.212,	0.060]& -0.055 & 0.014 & $[-0.133, -0.026]^{\dagger}$\\
$\beta_{\sigma,7}$ & Fixed& Steering difference & 0.063 & 0.065 & [-0.061, 0.191]&-0.053&	0.040	&$[-0.117,	-0.011]^{\dagger}$\\
$\beta_{\sigma,8}$ & Fixed& Volume &-0.001 & 0.007 & [-0.015, 0.013]&-0.004	&0.044	&[-0.090, 0.084]\\
$\beta_{\sigma,9}$ & Fixed&  Turn movement(Left =1, else=0)& 0.042 & 0.120 & [-0.183, 0.285] & 0.217	&0.123&	$[0.025, 0.471]^{\dagger}$\\
\hline
\multicolumn{9}{l}{\textit{Shape Parameter} ($\xi_{k,i}$)} \\
\hline
$\beta_{\xi,0}$ & Fixed & Intercept & -0.305 & 0.051 & $[-0.413, -0.208]^{\dagger}$& - & - & -\\
$\beta_{\xi,0,k}$ & Random & Intercept & - & - & -& -0.684&	0.253&	$[-1.309,-0.314]^{\dagger}$\\
\hline
\multicolumn{9}{l}{\textit{Model Fit}} \\
\hline
DIC &  &  & \multicolumn{3}{c}{526.9} & \multicolumn{3}{c}{508.9} \\
WAIC &  &  & \multicolumn{3}{c}{539.4} & \multicolumn{3}{c}{521.5} \\
LOOIC &  &  & \multicolumn{3}{c}{540.1} & \multicolumn{3}{c}{522.3} \\
\hline
\end{tabular}
}
\begin{tablenotes}
\item[HBSFP] Hierarchical Bayesian Structure Fixed Parameter
\item[HBSGRP] Hierarchical Bayesian Structure Grouped Random Parameters
\item[a] Standard deviation
\item[b] 95\% Bayesian credible interval 
\item[-] Covariate not included in the model 
\item[$\dag$] Indicates statistical significance at the 95\% level (interval excludes 0)
\end{tablenotes}
\end{threeparttable}
\end{table}

Consistent reductions in DIC, WAIC, and LOOIC confirm the superior fit of the HBSGRP model, highlighting the added value of allowing speed and spacing effects to vary across intersections. The direction and magnitude of these coefficients provide insights into the mechanisms underlying COR. Relative speed and distance emerge as the dominant dynamics, influencing both $\mu$ and $\sigma$ parameters. Relative speed exhibits a positive association with the location component and a negative association with the scale component (Fig.~\ref{fig:b12}, Fig.~\ref{fig:b22}), whereas relative distance exhibits a negative association with the location component and a negative association with the scale component (Fig.~\ref{fig:b15}, Fig.~\ref{fig:b25}). Taken together, these changes imply that higher closing-rate conditions and tighter spacing shift the fitted extreme distribution toward more critical tail behavior while simultaneously concentrating the distribution. Fig.~\ref{fig:9} illustrates the spatial heterogeneity across intersections, and the corresponding group-level posterior intervals are provided. While most locations show consistent trends, site I2 exhibits wider credible intervals for the location effects and less stable scale-related posteriors, attributable to its smaller BM sample size and the resulting weaker posterior precision.

Beyond random effects, several covariates were modeled as fixed effects because of their stable contributions across intersections. These effects can be interpreted in terms of driver behavior. These factors highlight the roles of longitudinal dynamics, lateral control, maneuver type, and exposure in shaping near-miss outcomes. Table~\ref{tab:7} shows that relative acceleration is negatively associated with both the location and scale components, implying lower predicted exceedance probability and a more concentrated extreme distribution under higher-acceleration conditions. By contrast, relative deceleration is positively associated with the location component and negatively associated with the scale component, implying higher predicted exceedance probability with a more concentrated tail under braking-dominant interactions. Jerk is negatively associated with the location component but positively associated with the scale component, implying lower central extreme levels but greater dispersion in tail behavior, i.e., more variability in predicted exceedance probability across comparable contexts.

Lateral control variables also contribute meaningfully at intersections. Steering difference shows a positive association with the location component and a negative association with the scale component, while heading difference exhibits the same sign pattern. Taken together, these coupled changes imply that larger lateral corrections and trajectory misalignment shift the fitted extreme distribution toward more crash-prone tail behavior while concentrating it, resulting in higher exceedance probability under unstable lateral maneuvers in constrained intersection geometry. The left-turn indicator is significant, with positive associations in both the location and scale components, indicating that left-turn maneuvers shift the extreme 2D-TTC distribution toward more crash-prone behavior while increasing dispersion, consistent with the added conflict complexity of opposing flows and heterogeneous gap-acceptance. Traffic volume is not statistically significant in either component; it is retained as an exposure control, and its weak effect suggests that block-level extremes are dominated by instantaneous interaction states rather than by aggregate demand within the short scenario windows.

Finally, the estimated shape parameter ($\xi$) is negative and significant across all intersections (Fig.~\ref{fig:b30}), indicating a bounded upper tail of the UGEV distribution. This supports the Weibull domain as the appropriate tail form for modeling extreme near-misses in signalized urban settings. Importantly, group-level variability in $\xi$ suggests that intersection design and control strategies influence not only the frequency but also the extremal nature of near-miss outcomes.

\begin{figure}[htbp]
    \centering

    \begin{subfigure}[b]{0.47\textwidth}
        \centering
        \includegraphics[width=\linewidth]{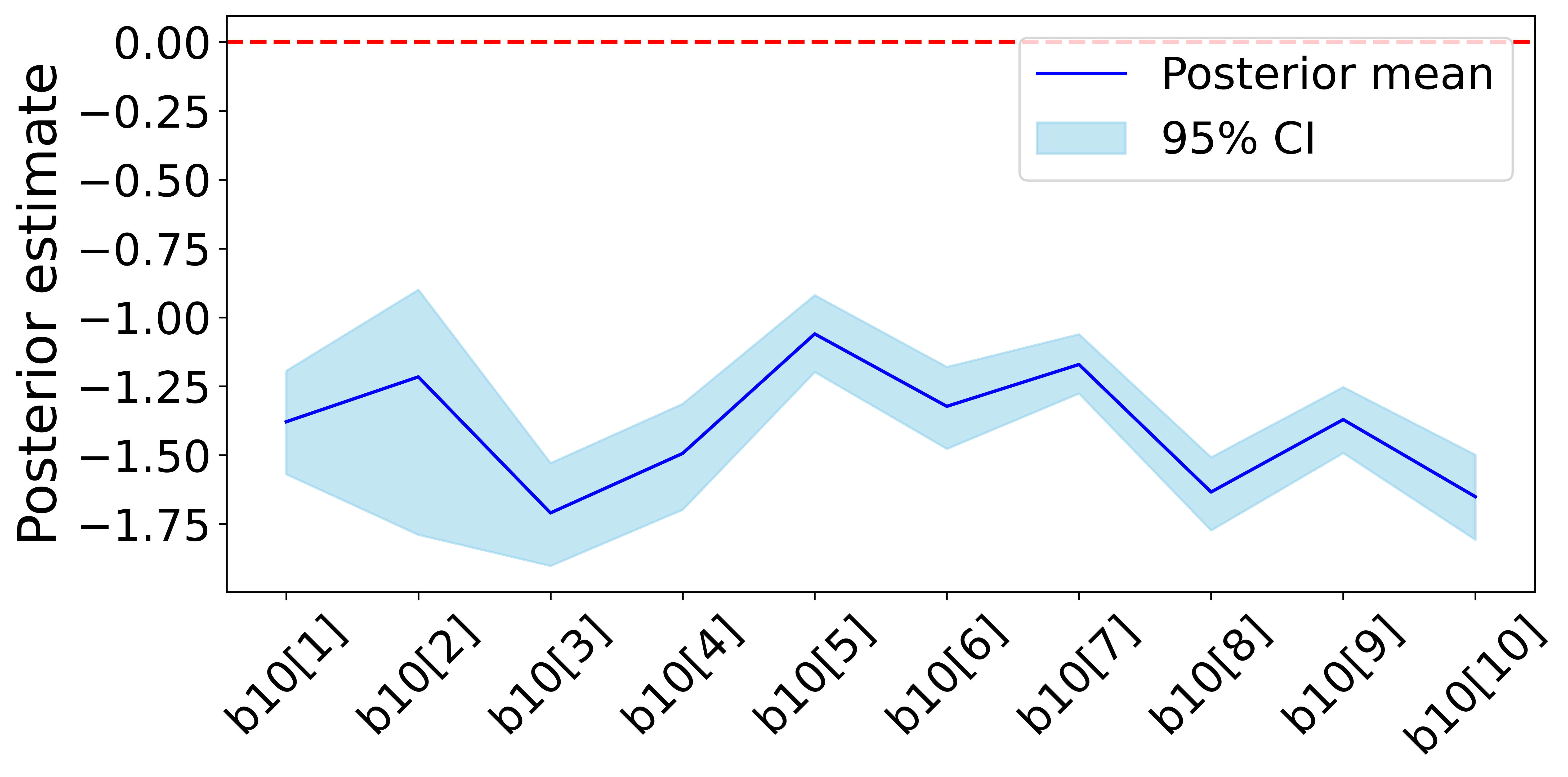}
        \caption{$\beta_{\mu,0[1,10]}$}
        \label{fig:10a}
    \end{subfigure}
    \hspace{0.02\textwidth} 
    \begin{subfigure}[b]{0.47\textwidth}
        \centering
        \includegraphics[width=\linewidth]{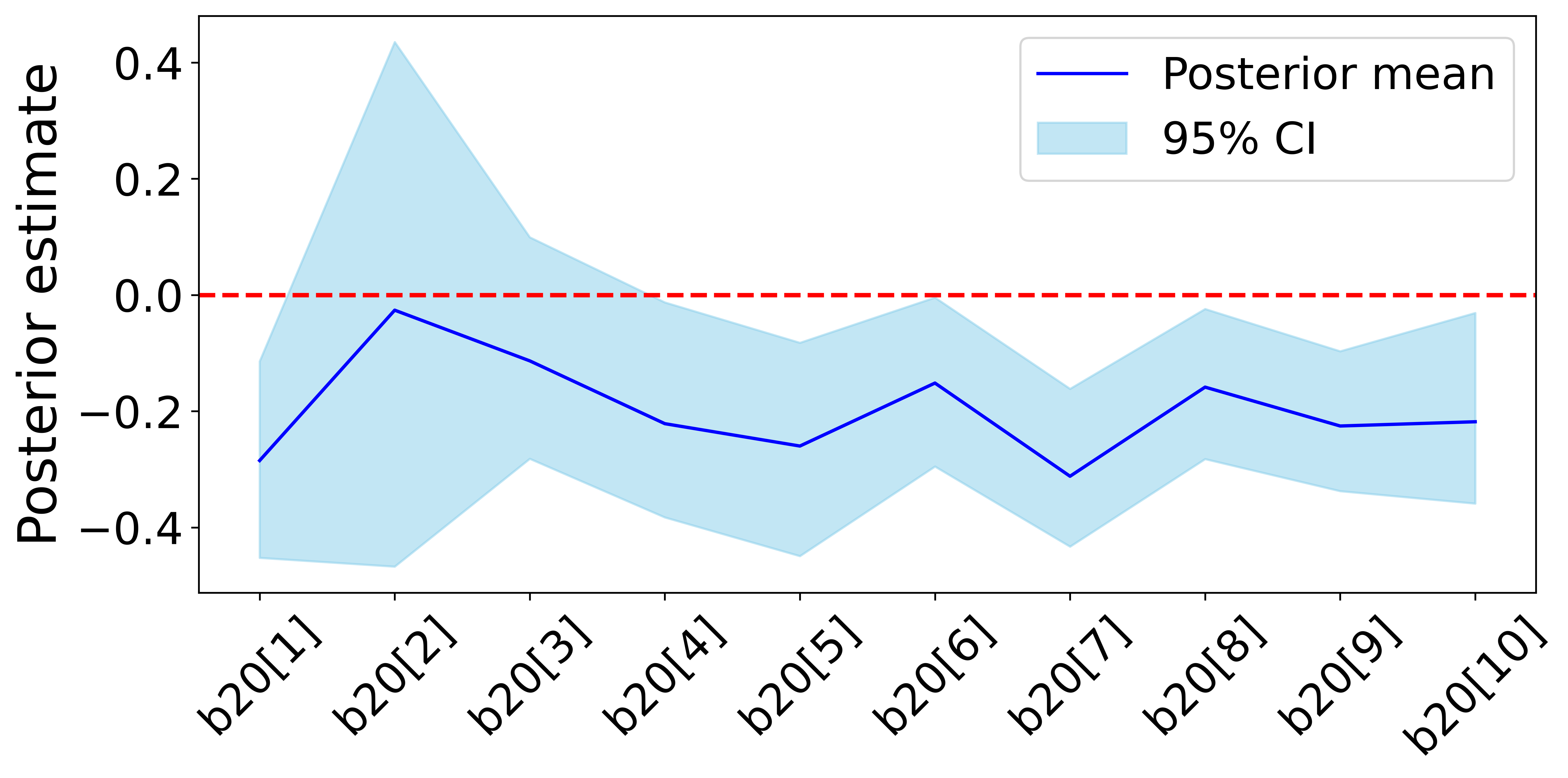}
        \caption{$\beta_{\sigma,0[1,10]}$}
        \label{fig:10b}
    \end{subfigure}

    \vspace{0.1cm}
    \begin{subfigure}[b]{0.5\textwidth}
        \centering
        \includegraphics[width=\linewidth]{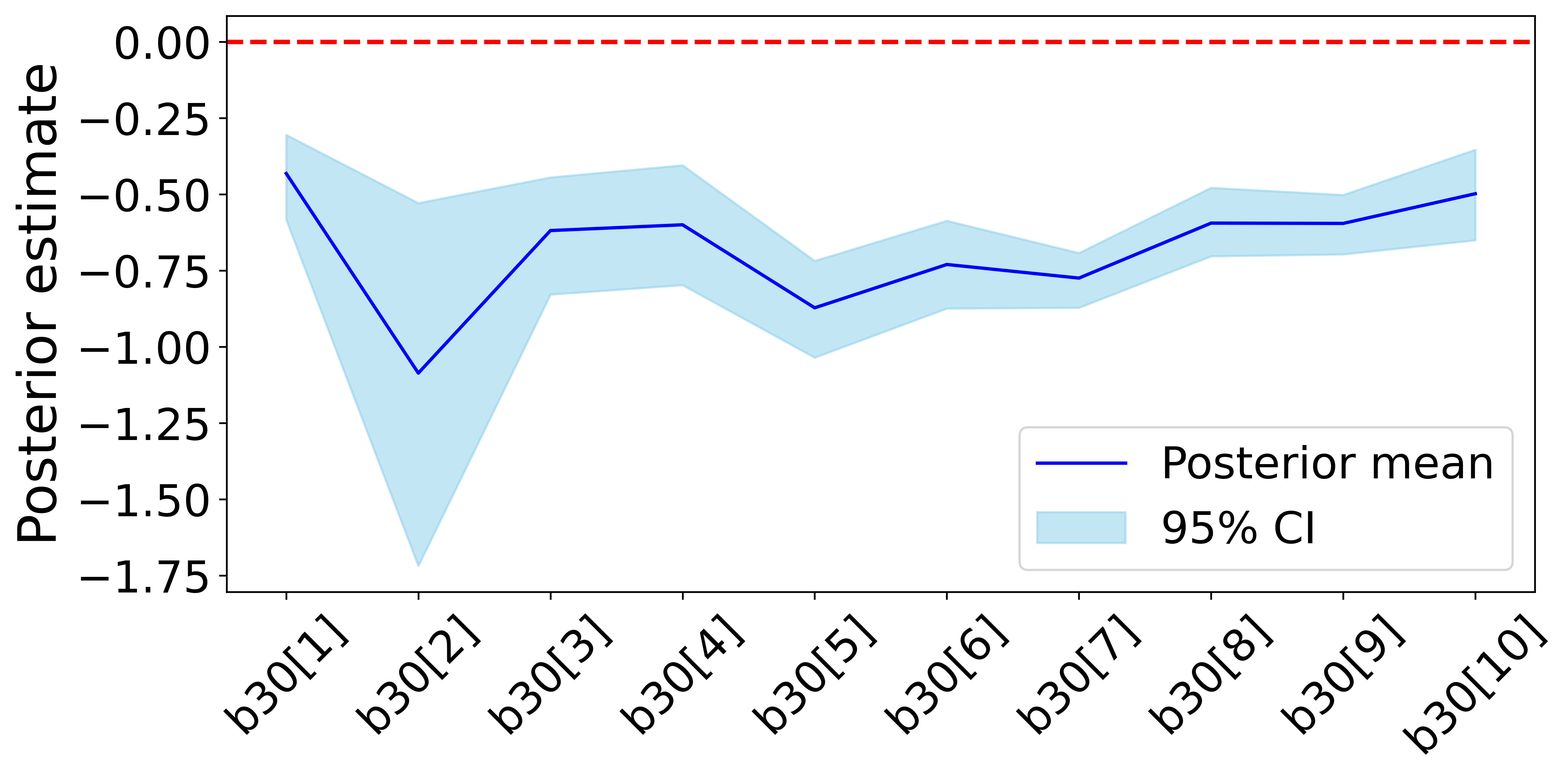}
        \caption{$\beta_{\xi,0[1,10]}$}
        \label{fig:10c}
    \end{subfigure}

    \caption{Posterior estimates for V-I near-misses at Intersections}
    \label{fig:10}
\end{figure}

The intersection-level HBSGRP model for V–I near-misses specified the intercept as a grouped random parameters in both the location and scale components (Fig.~\ref{fig:10}) to capture baseline heterogeneity across intersections. Relative speed was modeled as a grouped random effect in the location component, and relative distance was modeled as a grouped random effect in the scale component. Although these effects are weakly identified at some sites (credible intervals include zero), they were retained because they are mechanistically relevant to boundary-encroachment dynamics and because they improve predictive fit relative to the fixed-parameter specification. All remaining covariates were treated as fixed effects. Fit statistics (Table~\ref{tab:8}) show modest but consistent improvements in DIC, WAIC, and LOOIC versus HBSFP, indicating smaller gains from grouped random parameters for V–I than for V–V at intersections.

\begin{table}[htbp]
\centering
\caption{Posterior estimates for V-I near misses at Intersections}
\label{tab:8}
\scriptsize
\begin{threeparttable}
\resizebox{\textwidth}{!}{
\begin{tabular}{l l l c c c c c c}
\hline
\textbf{Model Parameters} & \textbf{Hyperparameter} & \textbf{Covariate} & \multicolumn{3}{c}{\textbf{HBSFP}} & \multicolumn{3}{c}{\textbf{HBSGRP}} \\
\cline{4-9}
 & & & Mean & SD\tnote{a} & 95\% CRI\tnote{b} & Mean & SD\tnote{a} & 95\% CRI\tnote{b} \\
\hline
\multicolumn{9}{l}{\textit{Location Parameter} ($\mu_{k,i}$)} \\
\hline
$\beta_{\mu,0}$ & Fixed & Intercept & -1.392 & 0.039 & $[-1.461, -1.299]^{\dagger}$ & - & - & -\\
$\beta_{\mu,0,k}$ & Random & Intercept & - & - & -&-1.401&	0.092&	$[-1.598,	-1.236]^{\dagger}$ \\
$\gamma_{\mu,1,k}$ & Random & Relative speed    & - & - & -& 0.025&	0.020&	[-0.012,	0.065]\\
$\beta_{\mu,1}$ & Fixed& Relative speed &-0.012 & 0.022 & [-0.054, 0.031] & - & - & - \\
$\beta_{\mu,2}$ & Fixed&  Relative acceleration  & -0.020 & 0.023 & [-0.067, 0.022] & -0.020&	0.024&	[-0.068,	0.029]\\
$\beta_{\mu,3}$ & Fixed&Relative deceleration & -0.066 & 0.013 & $[-0.075, -0.055]^{\dagger}$&-0.026 & 0.011 & $[-0.071, -0.009]^{\dagger}$\\
$\beta_{\mu,4}$ &Fixed&  Relative distance & -0.357 & 0.035 & $[-0.427, -0.290]^{\dagger}$ &-0.359&	0.026&	$[-0.405,	-0.306]^{\dagger}$\\
$\beta_{\mu,5}$ &Fixed& Jerk & 0.270 & 0.355 & [-0.457, 0.556] & -0.021 & 0.005 & [-0.026, 0.01] \\
$\beta_{\mu,6}$ & Fixed& Heading difference &0.040 & 0.022 & $[0.005, 0.084]^{\dagger}$& 0.062	&0.026&	$[0.010,	0.127]^{\dagger}$\\
$\beta_{\mu,7}$ & Fixed& Steering difference &  0.009 & 0.018 & [-0.027, 0.045]& 0.038&	0.022&	$[0.014, 0.074]^{\dagger}$\\
$\beta_{\mu,8}$ & Fixed&  Volume & -0.017 & 0.021 & [-0.061, 0.023]&-0.023&	0.023&	[-0.069,	0.027]\\
$\beta_{\mu,9}$ & Fixed& Turn movement (Left =1, else=0)  &-0.129 & 0.080 & $[-0.265, -0.061]^{\dagger}$&-0.579&	0.310&	$[-1.231,	-0.220]^{\dagger}$\\
\hline
\multicolumn{9}{l}{\textit{Scale Parameter} ($\log \sigma_{k,i}$)} \\
\hline
$\beta_{\sigma,0}$ &Fixed&Intercept & -0.100 & 0.038 & $[-0.167, -0.009]^{\dagger}$& - & - & -\\
$\beta_{\sigma,0,k}$ & Random & Intercept & - & - & -& -0.197&	0.098&	$[-0.374,	-0.001]^{\dagger}$  \\
$\gamma_{\sigma,1,k}$ &Random &  Relative distance & - & - & -&0.042&	0.034&	[-0.028,	0.107]\\
$\beta_{\sigma,1}$ & Fixed&  Relative speed  & 0.036 & 0.019 & $[0.013, 0.063]^{\dagger}$&0.040&	0.019&	$[0.001,	0.076]^{\dagger}$\\
$\beta_{\sigma,2}$ & Fixed& Relative acceleration &0.012 & 0.015 & [-0.015, 0.042]&0.046&	0.011&	$[0.009,	0.084]^{\dagger}$\\
$\beta_{\sigma,3}$ &Fixed& Relative deceleration & -0.147 & 0.056 & $[-0.285, -0.027]^{\dagger}$ &-0.177 & 0.056 & $[-0.289, -0.047]^{\dagger}$\\
$\beta_{\sigma,4}$ & Fixed& Relative distance & 0.209 & 0.038 & $[0.140, 0.282]^{\dagger}$ & - & - & - \\
$\beta_{\sigma,5}$ & Fixed& Jerk & 0.611 & 0.133 & $[0.227, 0.886]^{\dagger}$ & 0.109 & 0.033 & $[0.016, 0.156]^{\dagger}$\\
$\beta_{\sigma,6}$ & Fixed& Heading difference & -0.023 & 0.018 & $[-0.057, -0.004]^{\dagger}$&-0.090&	0.021&	$[-0.130,	-0.049]^{\dagger}$\\
$\beta_{\sigma,7}$ & Fixed& Steering difference & -0.012 & 0.018 & [-0.045, 0.023]&-0.024&	0.020&	$[-0.062,	-0.017]^{\dagger}$\\
$\beta_{\sigma,8}$ & Fixed& Volume &0.004 & 0.016 & [-0.028, 0.034]&0.029&	0.018&	$[0.003,	0.064]^{\dagger}$\\
$\beta_{\sigma,9}$ & Fixed&  Turn movement (Left =1, else=0) & 0.100 & 0.053 & $[0.019, 0.197]^{\dagger}$ & 0.264&	0.049&	$[0.163,	0.349]^{\dagger}$\\
\hline
\multicolumn{9}{l}{\textit{Shape Parameter} ($\xi_{k,i}$)} \\
\hline
$\beta_{\xi,0}$ & Fixed & Intercept & -0.677 & 0.042 & $[-0.749, -0.594]^{\dagger}$& - & - & -\\
$\beta_{\xi,0,k}$ & Random & Intercept & - & - & -& -0.680&	0.096&	$[-0.876,	-0.501]^{\dagger}$\\
\hline
\multicolumn{9}{l}{\textit{Model Fit}} \\
\hline
DIC &  &  & \multicolumn{3}{c}{3197} & \multicolumn{3}{c}{3179.6} \\
WAIC &  &  & \multicolumn{3}{c}{3202} & \multicolumn{3}{c}{3194.4} \\
LOOIC &  &  & \multicolumn{3}{c}{3212.87} & \multicolumn{3}{c}{3209.8} \\
\hline
\end{tabular}
}
\begin{tablenotes}
\item[HBSFP] Hierarchical Bayesian Structure Fixed Parameter
\item[HBSGRP] Hierarchical Bayesian Structure Grouped Random Parameters
\item[a] Standard deviation
\item[b] 95\% Bayesian credible interval 
\item[-] Covariate not included in the model 
\item[$\dag$] Indicates statistical significance at the 95\% level (interval excludes 0)
\end{tablenotes}
\end{threeparttable}
\end{table}

Relative deceleration shows a negative association with the location parameter and a negative association with the scale parameter, indicating that stronger braking near roadway boundaries is associated with a systematic shift in extreme behavior and a concentration of the distribution. Relative distance shows a negative association with the location parameter, whereas its contribution to the scale component is not statistically significant at the pooled level, suggesting that reduced clearance robustly shifts the distribution toward more critical extremes across intersections, whereas its influence on dispersion is more site-dependent and uncertain. Lateral control indicators show a consistent coupled structure: both heading and steering differences exhibit positive associations with the location parameter and negative associations with the scale parameter, indicating that greater lateral misalignment is associated with shifted extreme behavior accompanied by reduced dispersion. The left-turn indicator shows a negative association with the location parameter and a positive association with the scale parameter, suggesting that left-turn contexts shift the location component while broadening dispersion, which supports threshold-dependent risk responses when exceedance probabilities are evaluated across severity levels. In contrast, speed and throttle-related dynamics act primarily through dispersion: relative speed and relative acceleration show positive associations with the scale parameter, while their pooled associations with the location component are weak, indicating that approach-speed and acceleration behavior near infrastructure is more strongly linked to variability in extremes than to a uniform location shift.

The shape parameter remains negative and significant across intersections (\ref{fig:10c}), supporting the Weibull domain for V–I near misses and confirming that these events have a bounded upper tail. This indicates a physical and behavioral ceiling on severity, constrained by roadway geometry and driver response capacity. Spatial variation in $\xi$ further underscores the influence of intersection design and infrastructure placement on the extremal properties of near-miss distributions. Taken together, the intersection-level results suggest that V–I near misses are more strongly shaped by clearance, braking, and lateral maneuvers than by speed differentials. In contrast, V–V near misses are dominated by high speeds, which yield deterministic severity, whereas V–I outcomes exhibit greater variability driven by the interplay among infrastructure geometry, driver avoidance strategies, and intersection control.

\subsubsection{Segment-level \label{sec}}

The segment-level HBSGRP model for V–V near-misses specifies grouped random parameters to capture heterogeneity across directional segments (Table~\ref{tab:9}). Random effects were assigned to the location and scale intercepts and to three time variant covariates: relative speed, relative deceleration, and relative distance, allowing their contributions to vary by segment . All remaining covariates (relative acceleration, jerk, heading and steering differences, traffic volume, lane attributes, and driveway/access density) were treated as fixed effects to maintain a parsimonious specification when additional between-segment variation was not reliably identifiable. The HBSGRP model provides a better fit than the HBSFP alternative, supporting partial pooling for modeling segment-level variation in extreme near-miss risk.

\begin{table}[htbp]
\centering
\caption{Posterior estimates V-V near misses at Segments}
\label{tab:9}
\scriptsize
\begin{threeparttable}
\resizebox{\textwidth}{!}{
\begin{tabular}{l l l c c c c c c}
\hline
\textbf{Model Parameters} & \textbf{Hyperparameter} & \textbf{Covariate} & \multicolumn{3}{c}{\textbf{HBSFP}} & \multicolumn{3}{c}{\textbf{HBSGRP}} \\
\cline{4-9}
 & & & Mean & SD\tnote{a} & 95\% CRI\tnote{b} & Mean & SD\tnote{a} & 95\% CRI\tnote{b} \\
\hline
\multicolumn{9}{l}{\textit{Location Parameter} ($\mu_{k,i}$)} \\
\hline
$\beta_{\mu,0}$ & Fixed & Intercept & -2.117&	0.032&	$[-2.181,	-2.025]^{\dagger}$ & - & - & -\\
$\beta_{\mu,0,k}$ & Random & Intercept & - & - & -&-1.047 & 0.171 & $[-1.369, -0.772]^{\dagger}$ \\
$\gamma_{\mu,1,k}$ & Random & Relative speed    & - & - & -&0.406&	0.068& $[0.286,	0.543]^{\dagger}$\\
$\gamma_{\mu,2,k}$ & Random & Relative deceleration    & - & - & -&-0.082&	0.016&	$[-0.197,	-0.057]^{\dagger}$\\
$\gamma_{\mu,3,k}$ &Random&  Relative distance & -&-&-&-0.600&	0.082&	$[-0.779,	-0.454]^{\dagger}$\\
$\beta_{\mu,1}$ & Fixed& Relative speed & 0.446&	0.037&	$[0.376,	0.519]^{\dagger}$ & - & - & - \\
$\beta_{\mu,2}$ & Fixed&  Relative acceleration  & -0.068	&0.017	&$[-0.154, -0.034]^{\dagger}$&-0.079&	0.037&	$[-0.163, -0.032]^{\dagger}$\\
$\beta_{\mu,3}$ & Fixed&Relative deceleration & -0.074	&0.018&	$[-0.108, -0.040]^{\dagger}$&-&-&-\\
$\beta_{\mu,4}$ &Fixed&  Relative distance & -0.669&	0.063&	$[-0.801,	-0.552]^{\dagger}$ &-&-&-\\
$\beta_{\mu,5}$ &Fixed& Jerk & -0.004&	0.001&	$[-0.007, -0.002]^{\dagger}$ &-0.156	& 0.043&	$[-0.234,	-0.101]^{\dagger}$ \\
$\beta_{\mu,6}$ & Fixed& Heading difference & 0.043&	0.016&	$[0.011,	0.073]^{\dagger}$&0.044&	0.008&	$[0.029, 0.059]^{\dagger}$\\
$\beta_{\mu,7}$ & Fixed& Steering difference &  0.042	&0.012&	$[0.018,	0.066]^{\dagger}$&0.026&	0.010&	$[0.007,	0.046]^{\dagger}$\\
$\beta_{\mu,8}$ & Fixed&  Volume & 0.035&	0.013&	$[0.009,	0.060]^{\dagger}$&0.019&	0.008&	$[0.002,	0.038]^{\dagger}$\\
$\beta_{\mu,9}$ & Fixed& Lane no  &-0.006&	0.011&	[-0.027,	0.015]& 0.009	&0.011&	[-0.011,	0.032]\\
$\beta_{\mu,10}$ & Fixed& Lane width & 0.003&	0.001&	$[0.001,	0.007]^{\dagger}$&0.017&	0.009&	$[0.002,	0.031]^{\dagger}$\\
$\beta_{\mu,11}$ & Fixed& Driveway density & -0.001&	0.015&	[-0.029, 0.027]&0.489&	0.114&	$[0.256,	0.578]^{\dagger}$\\
$\beta_{\mu,12}$ & Fixed& Median (Undivided=1, else=0) & -0.004&	0.033&	[-0.070,	0.062]&-0.412&	0.155&	$[-0.711,	-0.070]^{\dagger}$\\
$\beta_{\mu,13}$ & Fixed& Vehicle position (Left lane=1, else=0) &0.029&	0.038&	[-0.048,	0.103]&0.055	&0.006&	$[0.012,	0.105]^{\dagger}$\\
\hline
\multicolumn{9}{l}{\textit{Scale Parameter} ($\log \sigma_{k,i}$)} \\
\hline
$\beta_{\mu,0}$ &Fixed&Intercept & -0.622	&0.045	&$[-0.711,	-0.533]^{\dagger}$& - & - & -\\
$\beta_{\mu,0,k}$ & Random & Intercept & - & - & -&-0.253 & 0.047 & $[-0.952, -0.156]^{\dagger}$ \\
$\gamma_{\sigma,1,k}$ & Random & Relative speed    & - & - & -&-0.055&	0.026& $[-0.107,	-0.003]^{\dagger}$\\
$\gamma_{\sigma,2,k}$ & Random & Relative deceleration    & - & - & -&-0.119&	0.018&	$[-0.154,	-0.084]^{\dagger}$\\
$\gamma_{\sigma,3,k}$ &Random&  Relative distance & -&-&-&-0.304&	0.117&	$[-0.520,	-0.084]^{\dagger}$\\
$\beta_{\sigma,1}$ & Fixed&  Relative speed  & -0.116&	0.050&	$[-0.217, -0.023]^{\dagger}$&-&-&-\\
$\beta_{\sigma,2}$ & Fixed& Relative acceleration &-0.611&	0.097&	$[-0.817, -0.439]^{\dagger}$&-0.024	&0.173&	[-0.277,	0.298]\\
$\beta_{\sigma,3}$ &Fixed& Relative deceleration & -0.060&	0.019&	$[-0.097,	-0.024]^{\dagger}$&-&-&-\\
$\beta_{\sigma,4}$ & Fixed& Relative distance & 0.003&	0.064&	[-0.121,	0.130] & - & - & - \\
$\beta_{\sigma,5}$ & Fixed& Jerk & 0.006&	0.000&	$[0.004,	0.008]^{\dagger}$ & 0.357&0.147	&$[0.061, 0.611]^{\dagger}$\\
$\beta_{\sigma,6}$ & Fixed& Heading difference & 0.065	&0.017&	$[0.032,	0.099]^{\dagger}$& 0.033&	0.011&	$[0.017,0.068]^{\dagger}$\\
$\beta_{\sigma,7}$ & Fixed& Steering difference & 0.011	&0.016	&[-0.020,	0.043]& -0.001&	0.019&	[-0.031, 0.029]\\
$\beta_{\sigma,8}$ & Fixed& Volume &-0.011	&0.016&	[-0.043,	0.020]&-0.012	&0.015&	[-0.030,	0.029]\\
$\beta_{\sigma,9}$ & Fixed& Lane no  &-0.031&	0.014&	$[-0.049,	-0.005]^{\dagger}$&-0.028&	0.017&	$[-0.061,	-0.002]^{\dagger}$\\
$\beta_{\sigma,10}$ & Fixed& Lane width & -0.006&	0.003&	$[-0.012,	-0.000]^{\dagger}$&-0.015&	0.009&	$[-0.044,	-0.001]^{\dagger}$\\
$\beta_{\sigma,11}$ & Fixed& Driveway density & 0.040	&0.018	&$[0.004,	0.075]^{\dagger}$&0.284&	0.215&	$[0.019, 0.648]^{\dagger}$\\
$\beta_{\sigma,12}$ & Fixed& Median (Undivided=1, else=0)  & 0.076	&0.042&	[-0.006,	0.158]&-0.205&	0.095&	$[-0.369,	-0.058]^{\dagger}$\\
$\beta_{\sigma,13}$ & Fixed& Vehicle position (Left lane=1, else=0) &-0.049	&0.052&	[-0.156,	0.051]&-0.109&	0.052&	$[-0.209,	-0.006]^{\dagger}$\\
\hline
\multicolumn{9}{l}{\textit{Shape Parameter} ($\xi_{k,i}$)} \\
\hline
$\beta_{\xi,0}$ & Fixed & Intercept & -0.116&	0.030&	$[-0.178, -0.059]^{\dagger}$& - & - & -\\
$\beta_{\xi,0,k}$ & Random & Intercept & - & - & -&-0.185&	0.144&	$[-0.436, -0.163]^{\dagger}$\\
\hline
\multicolumn{9}{l}{\textit{Model Fit}} \\
\hline
DIC &  &  & \multicolumn{3}{c}{4528} & \multicolumn{3}{c}{4189} \\
WAIC &  &  & \multicolumn{3}{c}{4594} & \multicolumn{3}{c}{4276} \\
LOOIC &  &  & \multicolumn{3}{c}{4608.25} & \multicolumn{3}{c}{4278} \\
\hline
\end{tabular}
}
\begin{tablenotes}
\item[HBSFP] Hierarchical Bayesian Spatial Fixed Parameter
\item[HBSGRP] Hierarchical Bayesian Spatial Grouped Random Parameters
\item[a] Standard deviation
\item[b] 95\% Bayesian credible interval 
\item[-] Covariate not included in the model 
\item[$\dag$] Indicates statistical significance at the 95\% level (interval excludes 0)
\end{tablenotes}
\end{threeparttable}
\end{table}

For the segment-level V–V model, the grouped random effects indicate that closing dynamics shift the fitted extreme distribution through paired changes in the location and scale components, and that these joint shifts vary across segment contexts. Relative speed exhibits a positive association with the location component and a negative association with the scale component, implying that as closing speed increases, the fitted distribution moves toward more critical extremes while simultaneously becoming more concentrated around those extremes. The posterior variation across segments indicates that this coupled shift is strongest on segments characterized by more uninterrupted progression, whereas access-rich segments show wider posterior intervals, consistent with more heterogeneous interaction patterns. Relative distance shows a negative association with the location component and a negative association with the scale component, indicating that tighter spacing is linked to more critical tail behavior and less dispersion in extreme outcomes. Relative deceleration contributes through a coupled location/scale pattern whose magnitude is segment-dependent, consistent with the fact that braking can represent either conflict dissipation or crisis response depending on prevailing speeds and spacing; this heterogeneity is reflected in the segment-specific posterior intervals. In segments with fewer block-maxima observations, the intercept posteriors are less precise, but the corridor-level pattern remains: closing-rate and spacing jointly govern how the extreme distribution shifts and concentrates, which is the mechanism through which COR changes.

Among the fixed effects, relative acceleration and jerk have negative associations with the location component, whereas jerk has a positive association with the scale component, suggesting that aggressive longitudinal control is associated with a shift in the external distribution and increased dispersion in outcomes. Heading difference has positive associations with both the location and scale components, whereas steering difference has a positive location association but a scale association whose interval overlaps zero, indicating that lateral misalignment tends to shift the external distribution, with less consistent evidence that it systematically changes dispersion. Exposure and geometry covariates also show coupled behavior: volume and lane width have positive location associations, while lane number and lane width have negative scale associations, suggesting that cross-sectional capacity primarily manifests through changes in dispersion rather than a consistent shift in the center of extremes. Driveway density shows positive associations at both the location and scale levels, indicating that access activity is linked to both a shift and a widening of the extreme distribution. The median type (undivided) shows negative associations in both location and scale, and left-lane position shows a positive location association and a negative scale association, indicating that cross-section and lane placement shape COR through simultaneous changes in the fitted center and the dispersion of extreme near-miss outcomes. Overall, segment-level COR is best interpreted as the result of these coupled location–scale shifts, with grouped vehicle dynamics providing the primary source of spatial heterogeneity (Table~\ref{tab:9}).

Finally, the shape parameter ($\xi$) is negative and statistically significant across segment groups, indicating a bounded upper tail for extreme near-miss behavior. Overall, longitudinal dynamics (speed, braking, and spacing) drive segment-level extremes, while roadway features (median type, lane width, and driveway density) primarily modulate variability. The improved fit of the HBSGRP model shows that allowing key dynamics to vary by segment is important for capturing midblock heterogeneity that fixed-parameter models miss.

\begin{figure}[htbp]
    \centering

    \begin{subfigure}[b]{0.45\textwidth}
        \includegraphics[width=\linewidth]{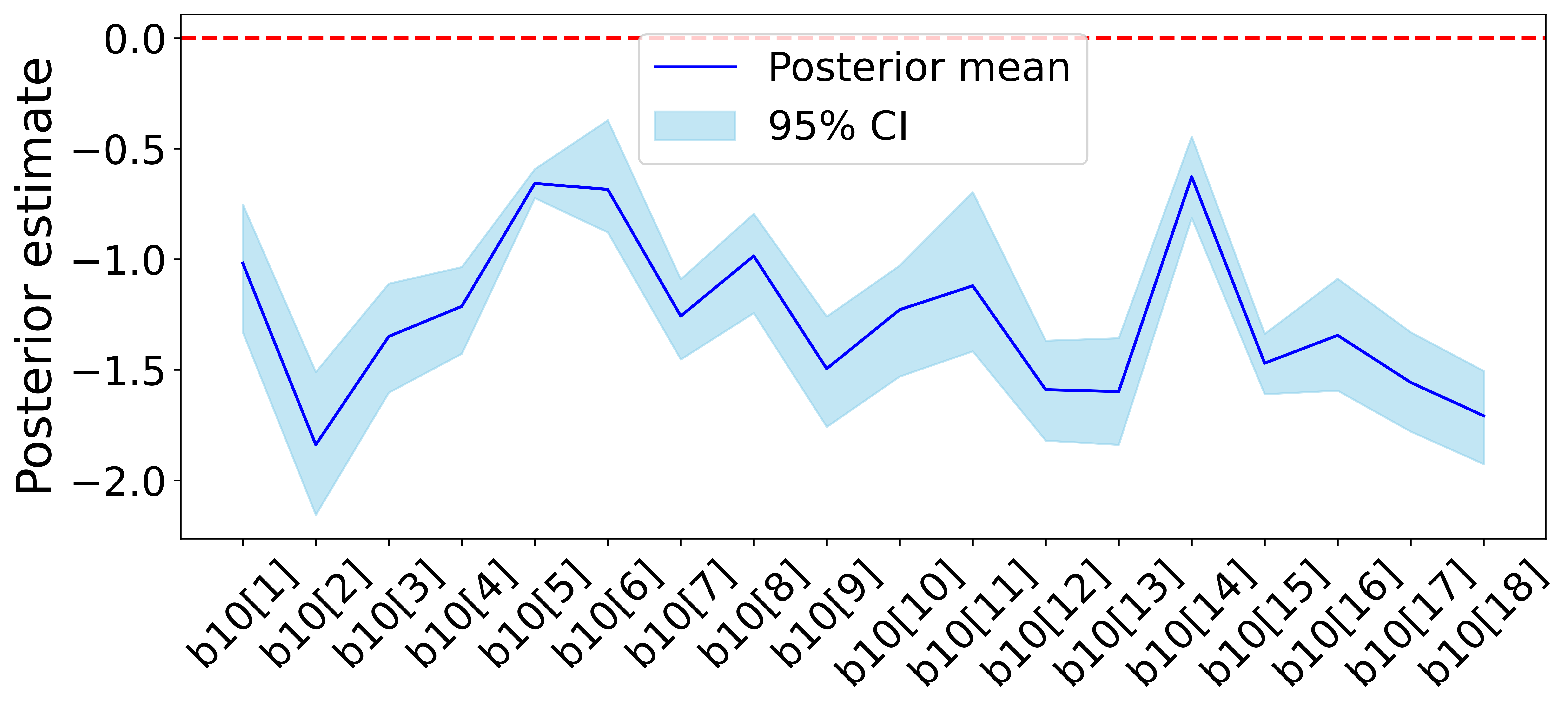}
        \caption{$\beta_{\mu,0[1,18]}$}
        \label{fig:12a}
    \end{subfigure}
    \hfill
    \begin{subfigure}[b]{0.45\textwidth}
        \includegraphics[width=\linewidth]{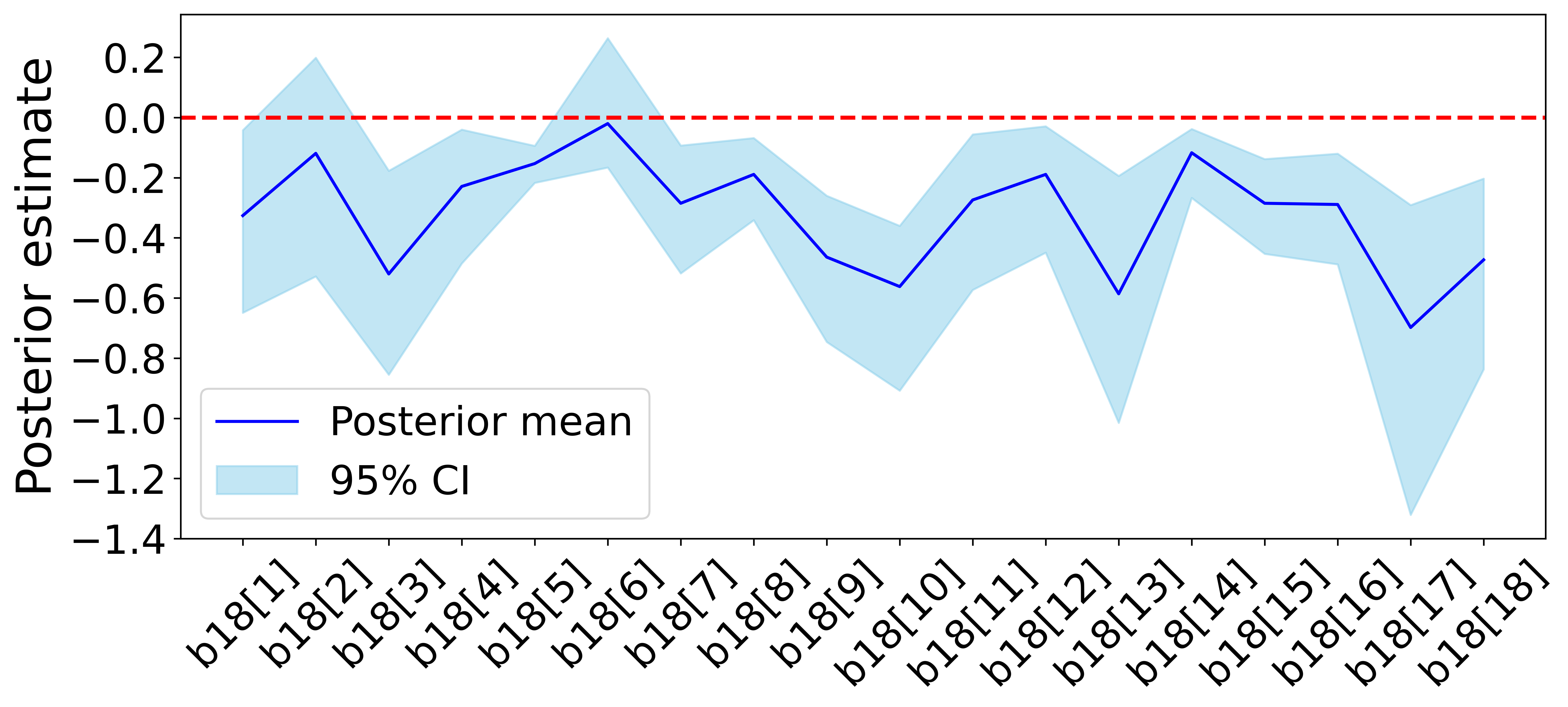}
        \caption{$\gamma_{\mu,1[1,18]}$}
        \label{fig:12b}
    \end{subfigure}

    \vspace{0.1cm}
    \begin{subfigure}[b]{0.45\textwidth}
        \includegraphics[width=\linewidth]{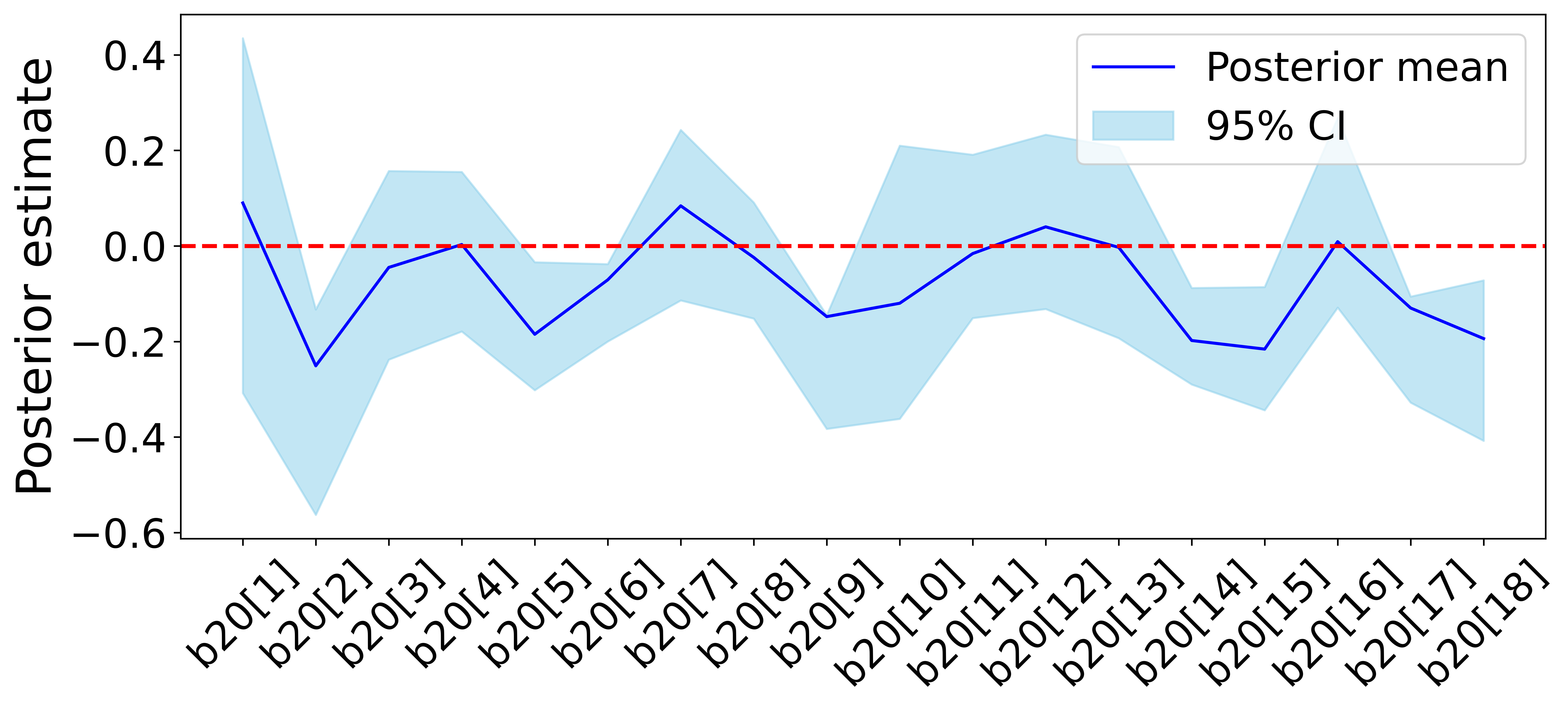}
        \caption{$\beta_{\sigma,0[1,18]}$}
        \label{fig:12c}
    \end{subfigure}
    \hfill
    \begin{subfigure}[b]{0.45\textwidth}
        \includegraphics[width=\linewidth]{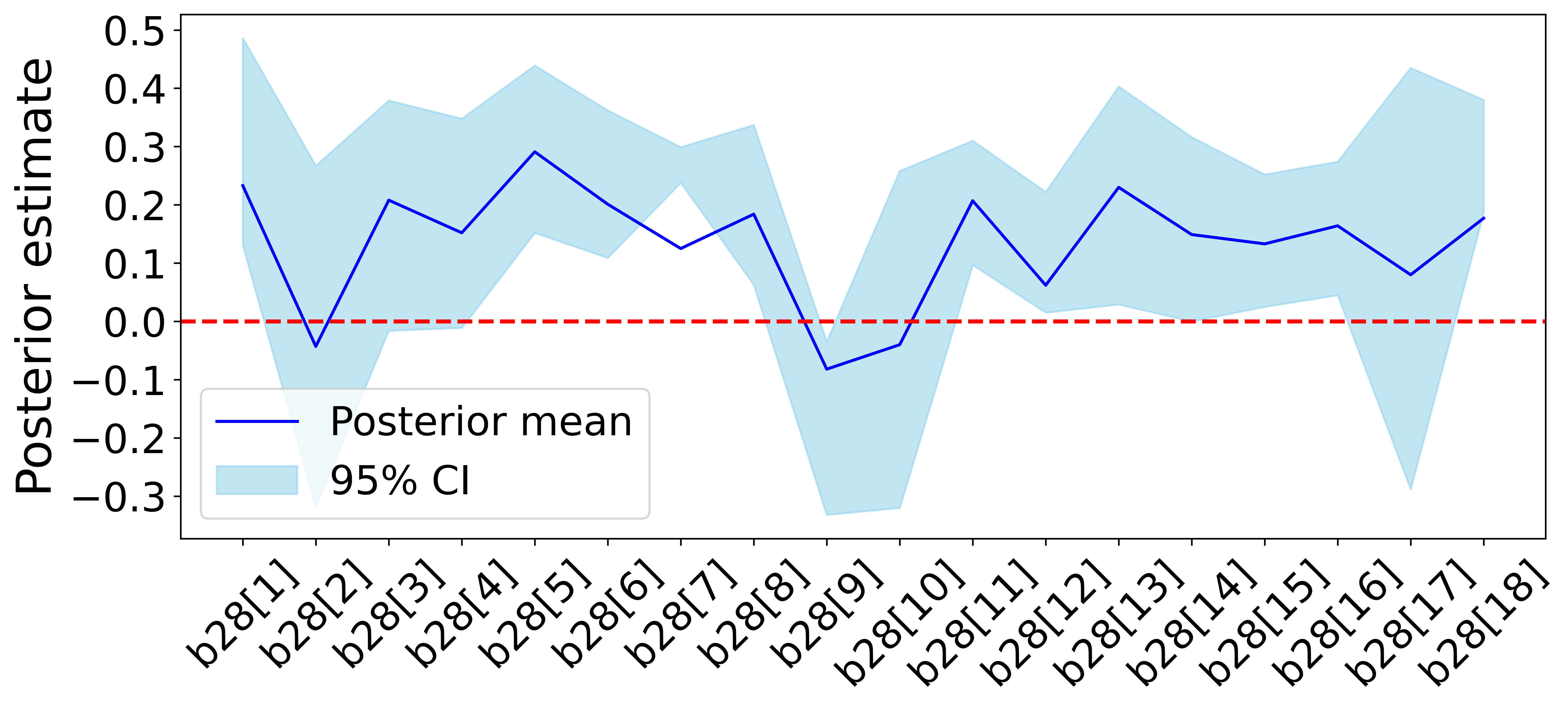}
        \caption{$\gamma_{\sigma,1[1,18]}$}
        \label{fig:12d}
    \end{subfigure}

    \vspace{0.1cm}
    \begin{subfigure}[b]{0.45\textwidth}
        \includegraphics[width=\linewidth]{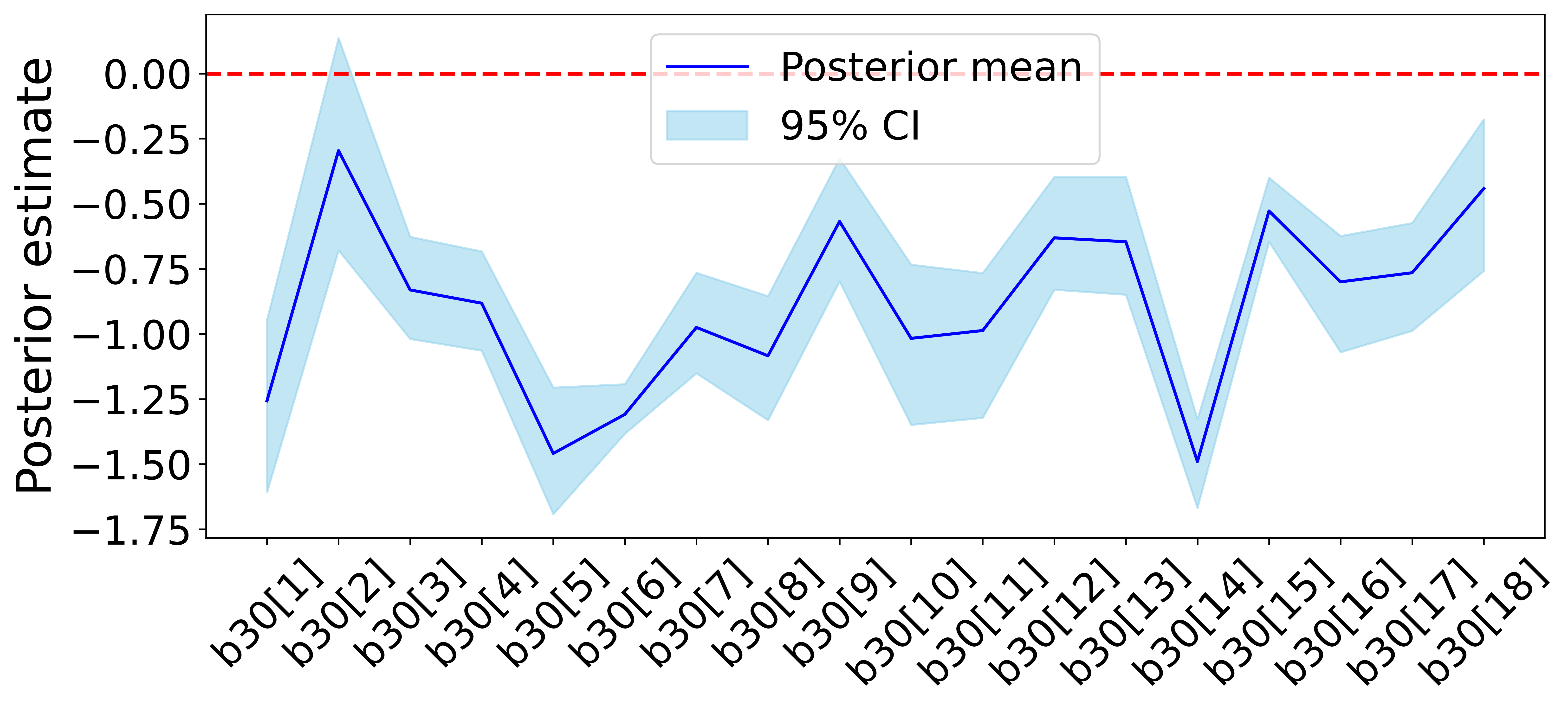}
        \caption{$\beta_{\xi,0[1,18]}$}
        \label{fig:12e}
    \end{subfigure}

    \caption{Posterior estimates V-I near misses at Segments}
    \label{fig:12}
\end{figure}

The HBSGRP model for segment-level V–I near-misses offers insights into how longitudinal dynamics and roadway design jointly shape the severity and variability of extreme 2D-TTC outcomes (Table~\ref{tab:10}, Fig.~\ref{fig:12}). In particular, the location intercept is random and significant, and the relative distance is random in both components, with a negative association in the location component and a positive association in the scale component. Implying that reduced clearance shifts extremes toward more critical outcomes while increasing dispersion, and that this sensitivity varies by segment. All other covariates, including relative speed, acceleration, deceleration, jerk, heading difference, steering angle difference, traffic volume, lane attributes, driveway density, median type, and left-lane presence, are modeled as fixed effects to represent their more consistent contributions across segment environments.

\begin{table}[htbp]
\centering
\caption{Posterior estimates V-I near misses at Segments}
\label{tab:10}
\scriptsize
\begin{threeparttable}
\resizebox{\textwidth}{!}{
\begin{tabular}{l l l c c c c c c}
\hline
\textbf{Model Parameters} & \textbf{Hyperparameter} & \textbf{Covariate} & \multicolumn{3}{c}{\textbf{HBSFP}} & \multicolumn{3}{c}{\textbf{HBSGRP}} \\
\cline{4-9}
 & & & Mean & SD\tnote{a} & 95\% CRI\tnote{b} & Mean & SD\tnote{a} & 95\% CRI\tnote{b} \\
\hline
\multicolumn{9}{l}{\textit{Location Parameter} ($\mu_{k,i}$)} \\
\hline
$\beta_{\mu,0}$ & Fixed & Intercept & -1.287&	0.037&	$[-1.352,	-1.22]^{\dagger}$ & - & - & -\\
$\beta_{\mu,0,k}$ & Random & Intercept & - & - & -&-1.263&	0.122&	$[-1.494,	-1.037]^{\dagger}$ \\
$\gamma_{\mu,1,k}$ &Random&  Relative distance & - & - & -&-0.317&	0.148&	$[-0.632,	-0.104]^{\dagger}$\\
$\beta_{\mu,1}$ & Fixed& Relative speed & -0.027	&0.018&	$[-0.061,	-0.008]^{\dagger}$ & 0.009	&0.001&	$[0.006,	0.027]^{\dagger}$\\
$\beta_{\mu,2}$ & Fixed&  Relative acceleration  & 8.36E-04&	0.019&	[-0.045,	0.036]&0.025&	0.013&	$[0.002,	0.057]^{\dagger}$\\
$\beta_{\mu,3}$ & Fixed&Relative deceleration & -0.039	&0.008&	$[-0.098,	-0.004]^{\dagger}$&-0.032	&0.008&	$[-0.087, -0.014]^{\dagger}$\\
$\beta_{\mu,4}$ &Fixed&  Relative distance & -0.257&	0.025&	$[-0.304,	-0.207]^{\dagger}$ &-&-&-\\
$\beta_{\mu,5}$ &Fixed& Jerk & -0.124 & 0.069 & [-0.224, 0.289 ] &-0.074 & 0.039 & [-0.124, 0.202] \\
$\beta_{\mu,6}$ & Fixed& Heading difference &  0.025&	0.025&	[-0.025,	0.069]&0.027&	0.014&	$[0.001,	0.049]^{\dagger}$\\
$\beta_{\mu,7}$ & Fixed& Steering difference &  0.033&	0.017&	$[0.004,	0.065]^{\dagger}$&0.041&	0.016&	$[0.003,	0.066]^{\dagger}$\\
$\beta_{\mu,8}$ & Fixed&  Volume &-0.002&	0.007&	[-0.016,0.010]& 0.043&	0.021&$[0.015, 0.084]^{\dagger}$\\
$\beta_{\mu,9}$ & Fixed& Lane no  &-0.001&	0.012&	[-0.019,	0.016]&0.042&	0.021&	$[0.004,	0.078]^{\dagger}$\\
$\beta_{\mu,10}$ & Fixed& Lane width & -0.027&	0.018&	$[-0.061,	-0.008]^{\dagger}$&0.013	&0.009&	$[0.000,	0.032]^{\dagger}$\\
$\beta_{\mu,11}$ & Fixed& Driveway density & -0.007&	0.024&	[-0.057,	0.042]&-0.237&	0.105&	$[-0.403,	-0.020]^{\dagger}$\\
$\beta_{\mu,12}$ & Fixed& Median (Undivided=1, else=0)  & -0.644&	0.102&	$[-0.773, -0.513]^{\dagger}$&-0.083&	0.059&	$[-0.196,	-0.019]^{\dagger}$\\
$\beta_{\mu,13}$ & Fixed& Vehicle position (Left lane=1, else=0) &0.057&	0.033&	$[0.028,	0.091]^{\dagger}$&-0.064&	0.014&	$[-0.082,	-0.038]^{\dagger}$\\
\hline
\multicolumn{9}{l}{\textit{Scale Parameter} ($\log \sigma_{k,i}$)} \\
\hline
$\beta_{\mu,0}$ &Fixed&Intercept &-0.495&	0.156&	$[-0.703,	-0.303]^{\dagger}$& - & - & -\\
$\beta_{\sigma,0,k}$ & Random & Intercept & - & - & -& -0.076&	0.104&	[-0.265,	0.083] \\
$\gamma_{\sigma,1,k}$ &Random &  Relative distance & - & - & -&0.135&	0.103&	$[0.011,	0.318]^{\dagger}$\\
$\beta_{\sigma,1}$ & Fixed&  Relative speed  & 0.039&	0.016&	$[0.012,	0.061]^{\dagger}$&0.013&	0.001&	$[0.004,	0.038]^{\dagger}$\\
$\beta_{\sigma,2}$ & Fixed& Relative acceleration & 0.002&	0.017&	[-0.030,	0.042]&-0.026&	0.014&	$[-0.064,	-0.001]^{\dagger}$\\
$\beta_{\sigma,3}$ &Fixed& Relative deceleration & -0.036	&0.013&	$[-0.075,	-0.014]^{\dagger}$&-0.048	&0.021&	$[-0.089,	-0.022]^{\dagger}$\\
$\beta_{\sigma,4}$ & Fixed& Relative distance & 0.167	&0.018&	$[0.136,	0.199]^{\dagger}$ & - & - & - \\
$\beta_{\sigma,5}$ & Fixed& Jerk & -0.108 & 0.054 & [0.452, -0.119] & -0.078 & 0.119 & [-0.312, 0.078]\\
$\beta_{\sigma,6}$ & Fixed& Heading difference & -0.019&	0.020&	[-0.053,	0.020]& -0.029&	0.017&	$[-0.049, -0.001]^{\dagger}$\\
$\beta_{\sigma,7}$ & Fixed& Realative steering angle & -0.033&	0.012&	$[-0.065,	-0.001]^{\dagger}$& -0.032	&0.024&	$[-0.059, -0.003]^{\dagger}$\\
$\beta_{\sigma,8}$ & Fixed& Volume &-0.007&	0.018&	[-0.039,	0.027]&0.001&	0.016&	[-0.026, 0.030]\\
$\beta_{\sigma,9}$ & Fixed& Lane no  & -0.028&	0.017&	$[-0.058, -0.006]^{\dagger}$&-0.021&	0.012&	$[-0.046, -0.002]^{\dagger}$\\
$\beta_{\sigma,10}$ & Fixed& Lane width & -0.010	&0.020	&[-0.044,	0.030]&0.018&	0.016&	$[0.010,	0.045]^{\dagger}$\\
$\beta_{\sigma,11}$ & Fixed& Driveway density & 0.016&	0.019&	[-0.021,	0.055]&-0.125&	0.028&	[-0.239,	0.018]\\
$\beta_{\sigma,12}$ & Fixed& Median (Undivided=1, else=0)  & 0.663	&0.166&	$[0.440,	0.880]^{\dagger}$&0.174&	0.090&	$[0.263,	0.553]^{\dagger}$\\
$\beta_{\sigma,13}$ & Fixed& Vehicle position (Left lane=1, else=0) &-0.046&	0.027&	$[-0.080,	-0.012]^{\dagger}$&0.055	&0.010&	$[0.039,	0.071]^{\dagger}$\\
\hline
\multicolumn{9}{l}{\textit{Shape Parameter} ($\xi_{k,i}$)} \\
\hline
$\beta_{\xi,0}$ & Fixed & Intercept &  -0.845&	0.028&	$[-0.897,	-0.788]^{\dagger}$g& - & - & -\\
$\beta_{\xi,0,k}$ & Random & Intercept & - & - & -&-0.887&	0.118&	$[-1.123,	-0.659]^{\dagger}$\\
\hline
\multicolumn{9}{l}{\textit{Model Fit}} \\
\hline
DIC &  &  & \multicolumn{3}{c}{6760} & \multicolumn{3}{c}{6550} \\
WAIC &  &  & \multicolumn{3}{c}{6775} & \multicolumn{3}{c}{6595} \\
LOOIC &  &  & \multicolumn{3}{c}{6791.2} & \multicolumn{3}{c}{6601} \\
\hline
\end{tabular}
}
\begin{tablenotes}
\item[HBSFP] Hierarchical Bayesian Spatial Fixed Parameter
\item[HBSGRP] Hierarchical Bayesian Spatial Grouped Random Parameters
\item[a] Standard deviation
\item[b] 95\% Bayesian credible interval 
\item[-] Covariate not included in the model 
\item[$\dag$] Indicates statistical significance at the 95\% level (interval excludes 0)
\end{tablenotes}
\end{threeparttable}
\end{table}

Regarding the fixed parameters, for instance Relative speed is positive in both components, indicating that higher-speed boundary interactions are associated with a systematic shift and greater dispersion. Relative acceleration is positive in the location component and negative in the scale component, whereas relative deceleration is negative in both components, indicating that braking-dominated boundary encounters are associated with a shift toward more critical extremes and a more concentrated distribution. Heading and steering differences are positive in the location component and negative in the scale component, suggesting that lateral deviations coincide with shifted extremes but with reduced dispersion.

Roadway and exposure variables show similar coupled behavior. Lane number is positive in the location component and negative in the scale component, while lane width is positive in both components, indicating that additional lanes align with reduced dispersion, whereas wider lanes align with greater dispersion. Driveway density is negative in the location component and does not show a stable scale association, indicating access-rich segments align with more critical extremes, with dispersion effects less consistent. Undivided medians are negative in the location component and positive in the scale component, linking undivided sections to more critical and more variable extremes. Vehicle position (left lane) is negative in the location component and positive in the scale component, indicating lane context shifts extremes while increasing dispersion.

The shape parameter is negative and significant across sites (Fig.~\ref{fig:12e}), indicating a bounded upper tail and supporting the Weibull domain for extreme 2D-TTC. This implies that, even at the segment level, physical and behavioral constraints limit how severe boundary-related extremes can become. Differences in this parameter across segments suggest that local geometry and operating conditions influence tail behavior.

Overall, segment results show that vehicle dynamics and roadway features jointly govern both the shift and spread of extremes. For V–V interactions, relative speed, relative deceleration, and relative distance are the primary drivers, jointly shifting the distribution towards more critical outcomes and concentrating it under high closing rate, hard braking, and tight spacing conditions. For V–I interactions, relative distance is the dominant dynamic driver, with its segment-specific variability reflecting how clearance to curbs/medians governs boundary-related extremes.

\section{Model validation}\label{6}
The trajectory-based near-miss dataset provides precise geolocation for each extracted extreme interaction (Eq.~\ref{eq:33}) because all vehicle states are map-referenced and time-stamped at high frequency. These geolocated extremes can be aggregated to corridor elements (intersections and directional segments) and compared spatially. In contrast, the available historical crash records for the study corridor are reported with imprecise geolocation; therefore, they cannot be matched to specific near-miss trajectories. Consequently, direct external validation that links predicted near-miss risk to individual crash records is not feasible with the available crash data. Validation is therefore designed to assess: (i) out-of-sample discrimination on unseen extreme blocks and (ii) spatial generalizability through site-by-site comparisons of expected versus observed extremes with uncertainty. Both HBSFP–UGEV and HBSGRP–UGEV were evaluated using these validation procedures; only the superior-performing HBSGRP–UGEV results are reported.

Model performance was evaluated using an out-of-sample event case--control design. The dataset was split at the block level into 70\% training and 30\% testing. For each severity threshold $\omega \in {-0.1,-0.2,\ldots,-0.9}$ sec, a test-set block was labeled as severe if its observed block maximum exceeded $\omega$. For each held-out block, the posterior exceedance probability was computed from the fitted HBSGRP--UGEV and HBSFP-UGEV model using Eq.~(\ref{eq:31}). Because the HBSGRP--UGEV specification consistently outperformed the HBSFP--UGEV baseline in predictive criteria and posterior stability, only HBSGRP--UGEV validations are reported here. Receiver operating characteristic (ROC) curves were constructed by varying the classification probability cutoff, and predictive performance was summarized using the area under the ROC curve (ROC-AUC), which is threshold-independent and suitable for rare-event classification.

\begin{figure}[h!]
    \centering
    \begin{subfigure}[t]{0.48\textwidth}
        \centering
        \includegraphics[width=\linewidth]{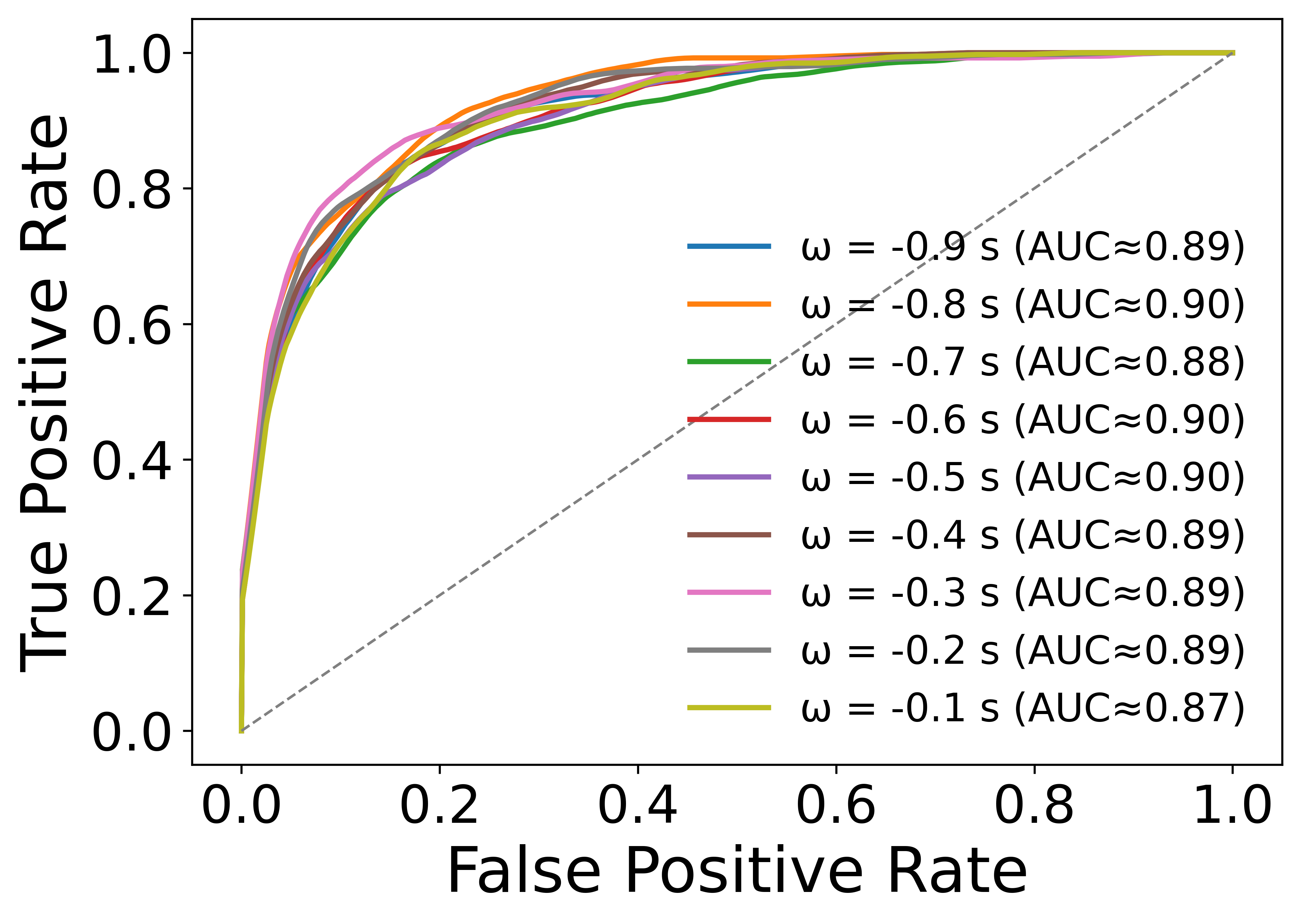}
        \caption{V–V Segment}
        \label{fig:roc_vv_segment}
    \end{subfigure}
    \hfill
    \begin{subfigure}[t]{0.48\textwidth}
        \centering
        \includegraphics[width=\linewidth]{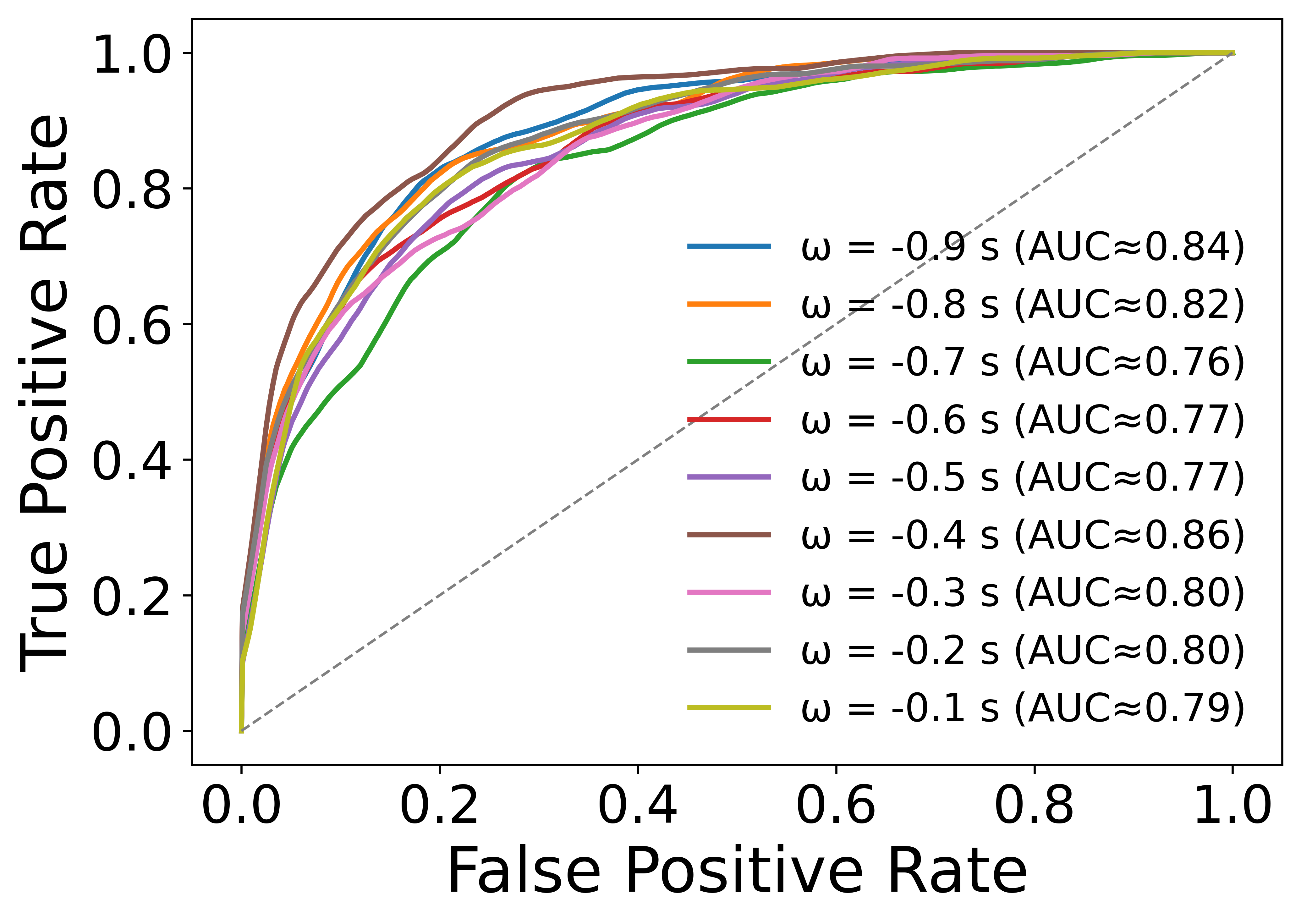}
        \caption{V–I Segment}
        \label{fig:roc_vb_segment}
    \end{subfigure}
    
    \vspace{0.1cm}
    
    \begin{subfigure}[t]{0.48\textwidth}
        \centering
        \includegraphics[width=\linewidth]{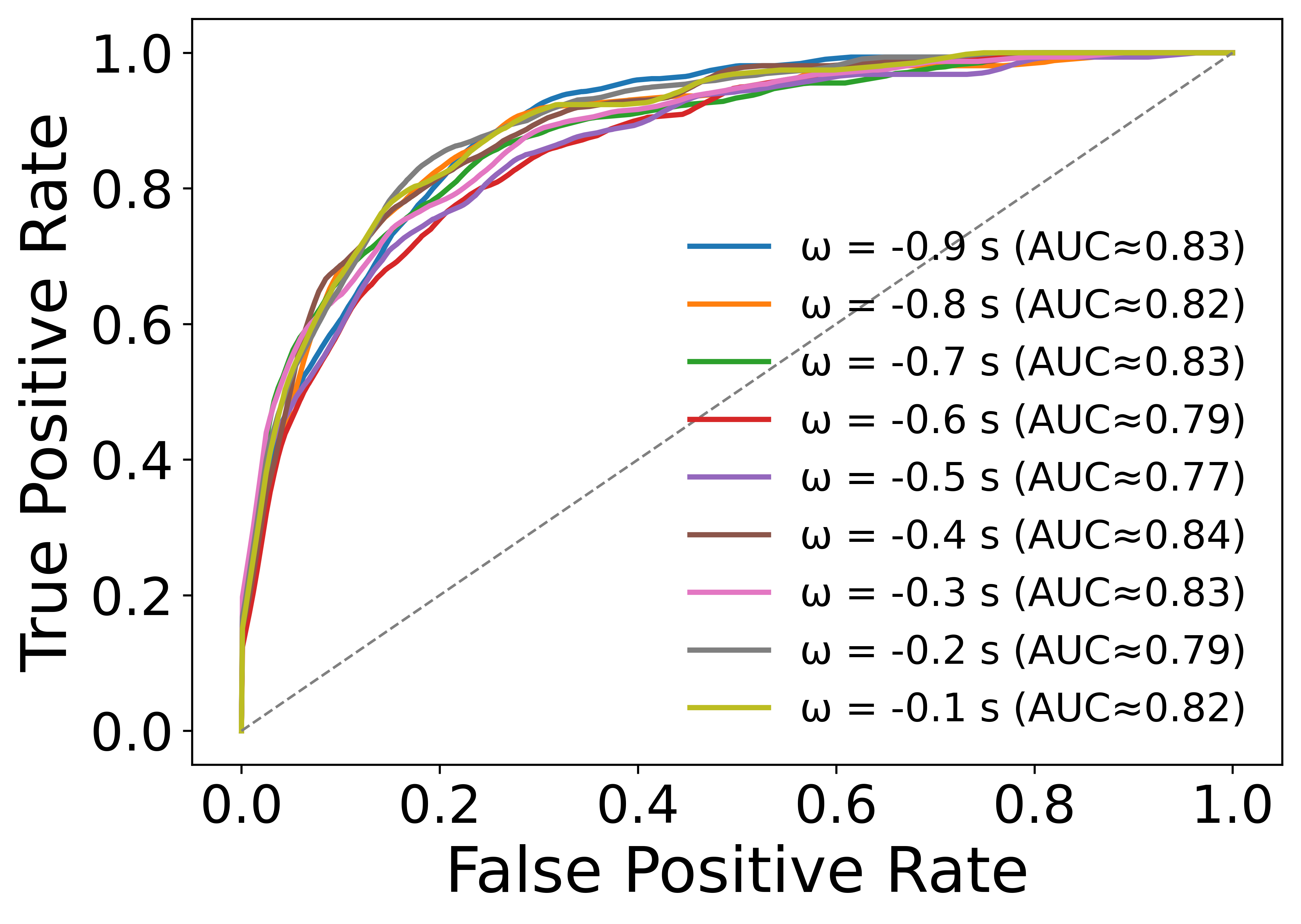}
        \caption{V–V Intersection}
        \label{fig:roc_vv_intersection}
    \end{subfigure}
    \hfill
    \begin{subfigure}[t]{0.48\textwidth}
        \centering
        \includegraphics[width=\linewidth]{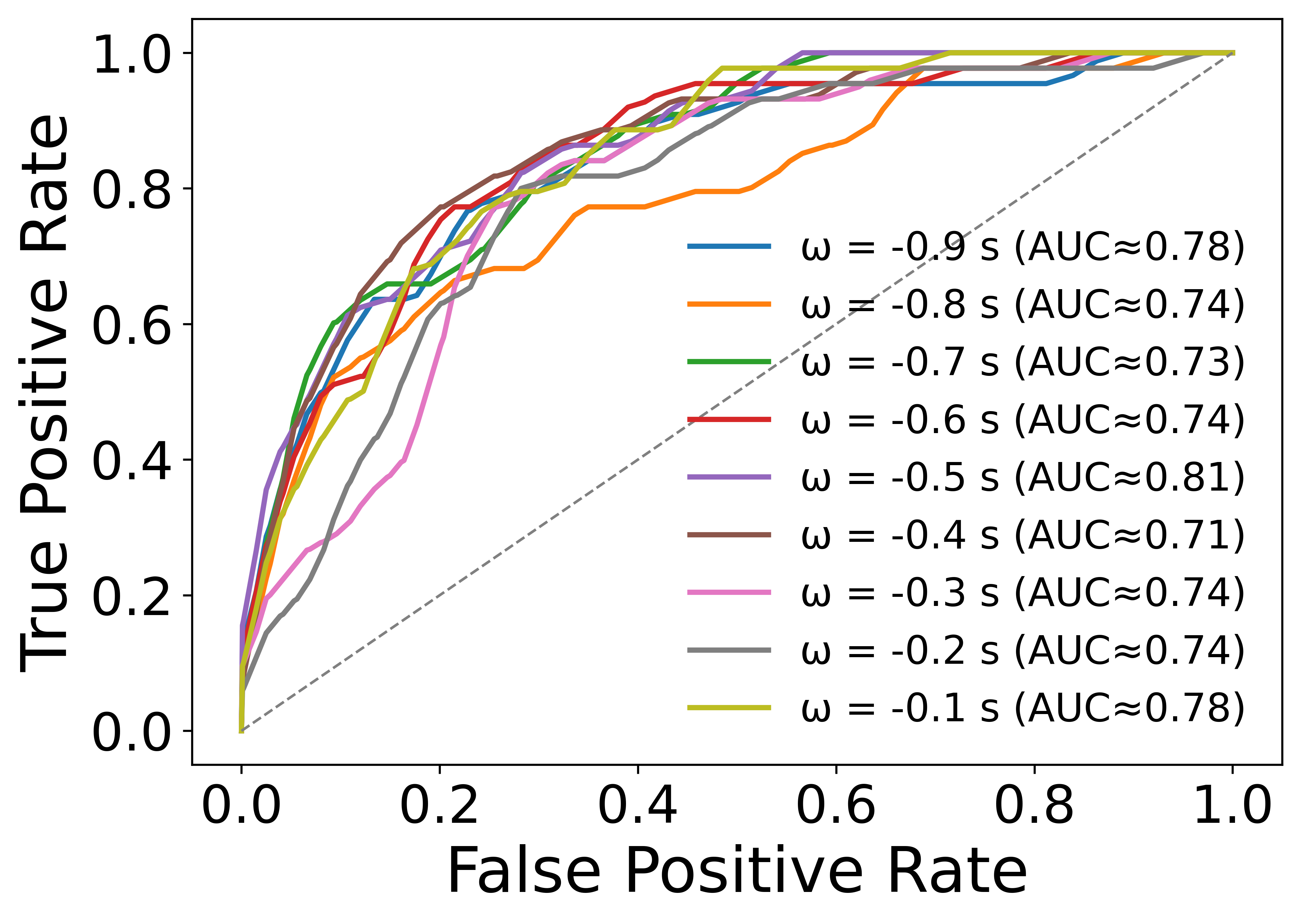}
        \caption{V–I Intersection}
        \label{fig:roc_vb_intersection}
    \end{subfigure}
    
    \caption{ROC curves for four HBSGRP–UGEV models across severity thresholds from $\omega = -0.1$s to $\omega = -0.9$s }
    \label{fig:13}
\end{figure}

ROC curves for the four HBSGRP--UGEV variants (V--V and V--I; segments and intersections) are shown in Fig.~\ref{fig:13}. The V--V segment model achieved the strongest identification (AUC $\approx 0.89$), followed by the V--V intersection model (AUC $\approx 0.82$). The V--I models exhibited lower identification, with AUC $\approx 0.79$ for segments and AUC $\approx 0.75$ for intersections. These results indicate that the framework more reliably distinguishes severe V--V near-misses, particularly along midblock segments, than infrastructure-related V--I near-misses in intersection environments, where turning trajectories, curb/median proximity, and signalized operations introduce greater variability.

While ROC--AUC assesses overall discrimination, it does not indicate whether the model reproduces spatial variation across corridor elements. To evaluate spatial generalizability, expected versus observed extremes were compared site by site, accounting for uncertainty. For each site $k$ and threshold $\omega$, the observed number of severe extremes in the test set was computed as the count of held-out blocks at site $k$ whose observed block maximum exceeded $\omega$. The corresponding expected number of severe extremes was computed by summing the model-based exceedance probabilities across the same held-out blocks. Uncertainty in expected counts was quantified using posterior predictive simulation, propagating posterior uncertainty in exceedance probabilities to replicated exceedance counts and summarizing the resulting predictive intervals. The expected-observed results indicate that the fitted models generally reproduce the spatial ordering of extremes across corridor elements, supporting the ability of the HBSGRP--UGEV framework to capture site-to-site variation rather than only aggregate frequency. The comparisons also reveal a systematic calibration pattern across thresholds: the model tends to overestimate exceedances at more severe thresholds (i.e., deeper tail levels) and underestimates them at less severe thresholds. This behavior suggests that the fitted tail response may be overly sensitive for the most extreme conditions and comparatively conservative for milder near-miss conditions.

Together, the hold-out ROC--AUC results (Fig.~\ref{fig:13}) and the site-by-site expected--observed comparisons with uncertainty provide complementary evidence of predictive performance: the former evaluates out-of-sample discrimination on unseen blocks, while the latter tests whether predicted extremes generalize spatially across corridor elements. The primary limitation remains the absence of event-level crash geolocation and timestamps, which preclude direct external validation of crash matches. Future work can strengthen external validation by leveraging crash datasets with event-level spatial--temporal precision and by implementing explicit spatial cross-validation designs (e.g., leave-one-site-out) when site-level sample sizes become sufficiently large.

\section{Summary and conclusions}\label{7}

This study addresses the challenge of estimating short-term crash occurrence risk (COR) in urban corridors, where conventional surrogate safety approaches often rely on simplified interaction geometry and provide limited support for corridor-scale inference under heterogeneous operations. Existing trajectory-based EVT studies have primarily emphasized vehicle–vehicle (V–V) interactions and rarely incorporate vehicle–infrastructure (V–I) conflicts within a unified corridor-scale framework, thereby limiting their applicability for proactive safety monitoring and infrastructure-relevant risk assessment. To bridge these gaps, we proposed a geometry-aware two-dimensional time-to-collision (2D-TTC) indicator that propagates oriented vehicle footprints and integrates HD-map roadway boundaries, enabling consistent extraction of extreme near-miss events from both V–V and V–I interactions.

Extreme near-misses were modeled using a nonstationary univariate generalized extreme value (UGEV) distribution embedded in a Hierarchical Bayesian Structure Grouped Random Parameters (HBSGRP) framework. The grouped structure supports partial pooling across intersections and directional segments while retaining location-specific heterogeneity, improving stability under sparse extreme samples. High-resolution trajectories from the Argoverse-2 dataset were combined with HD maps to construct interaction-window block maxima of 2D-TTC and to derive covariates representing (i) interaction dynamics (e.g., relative speed, acceleration/deceleration, jerk, and heading/steering differences, with exposure represented by traffic volume) and (ii) roadway context (e.g., lane width, driveway density, median type, turning movements, and lane position). Grouped random parameters were assigned to a small, theoretically motivated subset of high-impact dynamics to capture context-dependent effects without overfitting, and a baseline hierarchical Bayesian fixed-parameter model (HBSFP) was estimated for benchmarking.

Results indicate that the HBSGRP specification consistently improves predictive performance and posterior stability relative to HBSFP, as reflected by lower DIC, WAIC, and LOOIC values. Importantly, covariate effects were interpreted in terms of their implied impact on COR (i.e., on the fitted UGEV distribution and crash-level exceedance probability), rather than through individual GEV parameters in isolation. Under this distribution-level interpretation, closing-dynamics variables (relative speed, deceleration, and distance) emerged as the most influential drivers of crash-prone extreme risk for V–V interactions, particularly on segments, while V–I risk was more strongly tied to spacing/proximity conditions and boundary encroachment. At intersections, V–V near misses were dominated by relative speed and distance, while no dynamic variables were significant for V–I near misses. Roadway and maneuver context (e.g., steering/heading deviations, lane width, driveway density, median type, and turning movements) also contributed meaningfully to COR variation, reinforcing that corridor risk is jointly shaped by short-horizon behavior and geometric/infrastructure constraints.

Model credibility was supported through validation against held-out outcomes using ROC–AUC. Predictive discrimination was strongest for V–V segment models (AUC $\approx$ 0.89), followed by V–V intersection models (AUC $\approx$ 0.82), while V–I models achieved moderate performance (segments: $\approx$ 0.79; intersections: $\approx$ 0.75). These results suggest that V–V crash-prone extremes are more consistently explainable from interaction dynamics at the corridor scale, whereas V–I interactions—particularly at intersections—exhibit greater variability due to complex boundary approach behavior, turning paths, and map-defined infrastructure constraints.

While the proposed framework advances corridor-wide COR estimation, several limitations motivate future research. First, Argoverse-2 motion-forecasting scenarios are intentionally curated toward safety-critical, high-interaction episodes; therefore, the estimated COR should be interpreted as conditional on these traffic states. Second, the 2D-TTC projection assumes simplified short-horizon dynamics, which may underrepresent risk under extreme evasive maneuvers or highly nonlinear control responses. Third, V–V and V–I extremes are modeled as separate univariate processes under the interaction-window block definition; extending to a bivariate or multivariate hierarchical EVT formulation would require a common space–time block that yields paired extremes and sufficient information to identify joint tail dependence. Finally, the current study focuses on crash occurrence risk from extreme near-misses, without explicit severity stratification within the EVT layer. 

Several extensions can further strengthen the proposed framework. First, incorporating longer-duration and more representative trajectory streams from continuously operating AV fleets would allow benchmarking the estimated COR against network-wide baseline risk and assessing temporal stability beyond short interaction windows. Second, the current univariate EVT formulation could be extended to multivariate hierarchical EVT models that jointly represent V–V and V–I extremes, provided that a common space–time blocking strategy yields paired extremes with sufficient information to identify tail dependence. Third, integrating complementary physics-informed severity surrogates, such as post-impact delta-V or energy-based indicators, would enable joint modeling of crash occurrence and severity rather than occurrence alone. Fourth, extending the framework to include vulnerable road user (VRU) near-misses would improve relevance for urban safety applications. Finally, systematic comparison with POT–GPD-based formulations under comparable blocking and dependence controls would help clarify trade-offs between BM and POT approaches for short-horizon, high-frequency trajectory data. As trajectory data scale in coverage and duration, hierarchical EVT models such as HBSGRP–UGEV will gain additional statistical power, enabling more nuanced, multimodal safety assessments. Within this evolving data landscape, the proposed framework provides a scalable foundation for proactive corridor-wide monitoring, supporting earlier identification of crash-prone conditions and enabling data-driven traffic management consistent with Vision Zero goals.

\bigskip

\subparagraph{\textbf{Acknowledgement:}}
This research is funded by Federal Highway Administration (FHWA) Exploratory Advanced Research 693JJ323C000010. The results do not reflect FHWA's opinions.

\bigskip
\subparagraph{\textbf{Disclaimer:}}
The results presented in this document do not necessarily reflect those from the Federal Highway Administration.

\bigskip
\subparagraph{\textbf{Data availability:}}
Data and code will be made available on reasonable request.

\printcredits

\bibliographystyle{cas-model2-names}

\bibliography{ref}

\clearpage

\end{document}